\documentclass{tADP2e}
\usepackage{mathrsfs}
\usepackage{graphicx}
\usepackage{overpic}
\usepackage{epstopdf}
\usepackage{bm}
\usepackage{psfrag}
\usepackage{xcolor}
\usepackage{pict2e}
\usepackage{hyperref}
\usepackage{pifont}

\newcommand{\dL}{{\ensuremath{d_{\rm L}}}}

\newcommand{\eqnref}[1]{(\ref{#1})}
\newcommand{\Eqnref}[1]{Eq.~(\ref{#1})}
\newcommand{\Figref}[1]{Fig.~\ref{#1}}

\newcommand{\Secref}[1]{Section~\ref{#1}}

\newcommand{\lp}{\left(}
\newcommand{\rp}{\right)}
\newcommand{\obs}[1]{{#1}}
\renewcommand{\citep}[1]{\cite{#1}}

\newcommand\tr{{\mathrm{tr}}}
\newcommand{\ve}[1]{\ensuremath{\mbox{\boldmath$#1$}}}
\renewcommand{\dotsc}{\cdots}
\newcommand{\uxin}{{\partial_x^m  u}}
\renewcommand{\uxi}{{u_{x,m}}}

\newcommand{\ves}[1]{\ensuremath{\mbox{{\boldmath$\scriptstyle #1$}}}}
\newcommand{\ma}[1]{\ensuremath{\mathbb{#1}}}

\newcommand\nn{\nonumber}
\newcommand{\CC}[2]{\ensuremath{{\cal C}{{#1}_;}{#2}}}

\newcommand{\ku}{{\ensuremath{\rm Ku}}}
\newcommand{\st}{{\ensuremath{\rm St}}}

\newcommand{\Ordo}{\ensuremath{O}}

\newcommand{\DeltaL}{{\ensuremath{\Delta_{\rm L}}}}

\newcommand{\xd}{{x^{\rm d}(t)}}
\newcommand{\xdo}{{x^{\rm d}(0)}}
\newcommand{\xxd}{{\ve x^{\rm d}(t)}}
\newcommand{\RRR}{{\mathscr{R}}}
\newcommand{\AAA}{{\mathscr A}}
\newcommand{\VVV}{{\mathscr V}}

\begin{document}
% \jvol{00} \jnum{00} \jyear{2013} \jmonth{October}

% \articletype{GUIDE}
\title{Statistical models for spatial patterns of heavy particles in turbulence}
\author{K. Gustavsson$^{\rm a}$ \vspace{6pt} and
        B. Mehlig$^{\rm a,b}$$^{\ast}$\thanks{$^\ast$Corresponding author. Email:  Bernhard.Mehlig@physics.gu.se
\vspace{6pt}}\\\vspace{6pt}  $^{a}${\em{Department of Physics, University of Gothenburg, SE-41296 Gothenburg, Sweden}};\\ $^{b}${\em NORDITA, Roslagstullsbacken 23, 106 91 Stockholm, Sweden}
\\
%% \received{v4.1 released October 2013}
}
\maketitle
\begin{abstract}
The dynamics of heavy particles suspended in turbulent flows is of fundamental importance for a wide range of  questions in astrophysics, atmospheric physics, oceanography, and technology. Laboratory experiments and numerical simulations have demonstrated that heavy particles respond in intricate ways to  turbulent fluctuations of the carrying fluid: non-interacting particles may cluster together and form spatial patterns even though the fluid is incompressible, and the relative speeds of nearby particles can fluctuate strongly. Both phenomena depend sensitively on the parameters of the system. This parameter dependence is difficult to model from first principles since turbulence plays an essential role. Laboratory experiments are also very difficult, precisely since they must refer to a turbulent environment. But in recent years it has become clear that important aspects of the dynamics of heavy particles in turbulence can be understood in terms of statistical models where the turbulent fluctuations are approximated by Gaussian random functions with appropriate correlation functions. In this review we summarise how such statistical-model calculations have led to a detailed understanding of the factors that determine heavy-particle dynamics in turbulence. We concentrate on spatial clustering of heavy particles in turbulence. This is an important question because spatial clustering affects the collision rate between the particles and thus the long-term fate of the system.
\\[2mm]{\bf PACS:} 
 05.40.-a Fluctuation phenomena, random processes, noise, and Brownian motion; 
46.65.+g Random phenomena and media; 92.60.Mt Particles and aerosols; 47.55.Kf Particle-laden flows; 
47.27.-i Turbulent flows; 47.27.eb Turbulence - Statistical theories and models; 
47.27.Gs Isotropic turbulence; homogeneous turbulence; 05.40.Jc Brownian motion; 45.50.Tn Collisions
\\[2mm]{\bf Keywords:} Turbulent aerosols, Particles in turbulence, preferential concentration, spatial clustering, fractals, caustics, inertial particles, Stokes law, multiplicative amplification, Lyapunov exponents, eorrelation dimension.
\\[10mm]
\centerline{\bfseries Table of contents}\vspace{8pt}
\hbox to \textwidth{\hsize\textwidth\vbox{\hsize18pc
\hspace*{-12pt} {\ref{sec:intro}.}    Introduction\\
{\ref{sec:Model}.}    Statistical model\\
\hspace*{10pt}{\ref{sec:model}.}  Equation of motion\\
\hspace*{10pt}{\ref{sec:rvf}.}  Statistical model for the velocity field\\
\hspace*{10pt}{\ref{sec:dimlessvar}.}  Dimensionless numbers and variables\\
\hspace*{10pt}{\ref{sec:othermodels}.}  Other models\\
{\ref{sec:measures}.}    Spatial clustering\\
\hspace*{10pt}{\ref{sec:fractal}.}  Small-scale clustering\\
\hspace*{10pt}{\ref{sec:LSC}.}  Preferential sampling\\
\hspace*{10pt}{\ref{sec:clustering_caustics}.}  Clustering by caustics\\
{\ref{sec:mechanisms}.}    Clustering mechanisms\\
\hspace*{10pt}{\ref{sec:mulamp}.}   Small-scale clustering\\
\hspace*{10pt}{\ref{sec:prstr}.}   Preferential sampling \\}
\hspace{-24pt}\vbox{\noindent\hsize18pc
\hspace*{10pt}{\ref{sec:caustics}.}   Caustics\\
\hspace*{10pt}{\ref{sec:gs}.}   Gravitational settling\\
\hspace*{10pt}{\ref{sec:sum}.}   Summary\\
{\ref{sec:comparison}.}    Comparison with results of\\
\hspace*{8pt}        direct numerical simulations\\
\hspace*{10pt}{\ref{sec:compare}.}   Small-scale clustering \\
\hspace*{10pt}{\ref{sec:pv}.}   Preferential sampling\\
\hspace*{10pt}{\ref{sec:lambda_2}.}   Lyapunov spectrum\\
\hspace*{10pt}{\ref{sec:gravs}.}   Gravitational settling\\
{\ref{sec:Method}.}    Methods \\
\hspace*{10pt}{\ref{sec:mwna}.}   White-noise approximations\\
\hspace*{10pt}{\ref{sec:fkua}.}   Finite-$\ku$ approximations \\
{\ref{sec:conc}.}    Conclusions\\
      }}
\mbox{}\\[1mm]
Postprint version of the article published by Taylor \& Francis Group in 
Advances in Physics 65 (2016), 1, {\tt http://dx.doi.org/10.1080/00018732.2016.1164490}
\end{abstract}

%%%%%%%%%%%%%%%%%%%%%%%%%%%%%%%%%%

\section{Introduction}
\label{sec:intro}
The dynamics of particles moving in turbulent flows
is fundamental to understanding chemical and kinetic processes in
many areas in the Natural Sciences, and in technology. One
example is the  problem of rain initiation from turbulent cumulus
clouds.  It is thought that turbulence facilitates the {growth of rain droplets}~\citep{Sha03,Bodenschatz10,Dev12}.
This idea has a long history, but the mechanisms at work are not yet fully understood. The topic remains a subject of intensive research.

A second example is the dynamics {of} micron-sized dust grains in the turbulent gas surrounding a growing star. According to the standard model of planet formation {planetesimals of at least several metres in diameter form} through {collisional aggregation} of microscopic dust grains.
But fundamental questions remain unanswered, they are summarised in Ref.~\cite{Wil08}. See Ref.~\cite{Anders} for a recent review of the subject.

A third problem that requires the analysis of particle dynamics in turbulent flows is the dynamics of micro-organisms as well as inorganic matter in the turbulent ocean.  An important example is the formation, settling, and the disintegration of \lq marine snow\rq{} in the unsteady ocean. Marine snow refers to small tenuous aggregates of organic matter~\cite{Kio01} that are carried by or settle through the turbulence. Some micro-organisms swim actively. Also their motion is affected by turbulent velocity gradients, often in combination with additional torques due to shape or mass-density asymmetries \cite{Rei03,Den10,Gua12,Cen13,Zha14,Berglund2016}.

Fourth, technology provides many examples, too. Particle-laden flows are important in combustion processes~\cite{Kuo12}, dilute fibre suspensions~\citep{Lun11},  and the mixing of chemicals~\cite{Hes04}.
In many technical applications the dynamics of particles in pipe flows is of interest, and the question is
to understand the spatial patterns formed by the particles \cite{Pic05a,Sar11}, how they modify the flow~\cite{Mat03},
and how the particles deposit onto and re-suspend from the surface of the pipe \cite{Ree88,Bel15}.

The dynamics of particles in turbulent flows is often characterised by large fluctuations
in their spatial distribution and their relative velocities.
These fluctuations depend sensitively on the
parameters of the system, such as the length- and time-scales of the turbulent fluctuations, the particle size and density, as well as the fluid density, to name but a few.

In this review we discuss the spatial distribution of heavy small particles in incompressible turbulent flows
(so-called \lq turbulent aerosols\rq{}). Particles in turbulent aerosols may cluster together and form spatial patterns
even though direct interactions between the particles are negligible, and despite the fact that incompressible flows exhibit neither sinks nor sources.
For particles that simply follow the flow, spatial clustering
cannot occur in the steady state in incompressible velocity fields.
{Such} velocity fields have neither sinks nor sources, so {that} the spatial particle distribution must become and remain uniform at long times. Spatial  clustering in turbulent aerosols is thus an inertial effect. Finite inertia allows the particles to detach from the flow. The importance
of inertia is characterised by  the \lq Stokes number\rq{} $\st\equiv (\gamma \tau)^{-1}$. Here  $\gamma$ is the rate at which the relative {velocity} between a particle and the fluid is damped, and $\tau$ is the relevant correlation-time scale of the underlying flow.

Pattern formation by small-scale clustering in turbulent aerosols has been experimentally studied~\citep{Woo05,Saw08,Sal08,Mon10,Gib12,Saw12b}.
Earlier experimental results are reviewed in Refs.~\cite{War09,Bal10}.
Theoretical investigations have employed direct numerical simulations of turbulence to determine the dynamics of the suspended particles
and to analyse spatial clustering \citep{Wan93,Rea00,Hog01,Chu05,Pic05,Sim06,Bec06,Bec07,Cal08,Cal08b,Col09,Men10,Saw12a,Saw12b}.
These studies rely on the Stokes approximation for the force exerted by the fluid upon the particle.
This is a good approximation for heavy small particles, it is referred to as \lq one-way coupling\rq{} in the literature.
An {\em ab-initio approach} would require to solve the particle and fluid equations together (\lq two-way coupling\rq{}), taking into account the time-dependent boundary conditions of the fluid problem posed by the moving particle.
For certain time-independent linear flows this problem can be solved systematically by perturbation theory \citep{saffman1956,subramanian2005,einarsson2015a,einarsson2015b,candelier2015,rosen2015d}. 
 In general, however,  the problem must be solved numerically, yet direct numerical simulations of the entire problem in turbulence
are possible only for a single particle~\citep{Hom10}, or for very few.
At higher particle-number densities fluid-mediated and direct interactions between the particles may become important (\lq four-way coupling\rq{}~\cite{Elg94}). It is clear that the non-linear problem of describing spatial patterns of particles in turbulence is very difficult to solve in its most general form.

This review concerns turbulent aerosols where one-way coupling is a good approximation.
Different mechanisms have been suggested to explain spatial clustering in turbulent aerosols.
Maxey~\cite{Max87} argued that inertia allows the particles to centrifuge out of vortical flow regions.
This mechanism refers to an instantaneous positive
correlation between particle positions and straining regions in the
underlying turbulent flow.
Such \lq preferential sampling\rq{} of straining regions is often invoked
in discussing clustering in turbulent aerosols.
This effect causes inhomogeneities in the spatial distribution of particles on scales
that correspond to the size of the smallest turbulent eddies, and on larger scales.

Often the suspended particles are much smaller than the smallest eddies, and
it turns out that turbulent aerosols may exhibit significant small-scale spatial clustering, on scales
much smaller than the size of the smallest eddies.
In the steady state the particle positions form a fractal at each instance of time, substantially enhancing the probability of observing two particles very close together.
This affects how frequently the particles collide~\citep{Sun97,Bru98,And07,Bec05}.
To quantify this small-scale clustering
one must follow the dynamics of particles that are initially very close together and compute whether a small cloud of particles tends to collapse or diverge. This is determined by the
history of the fluid-velocity gradients experienced by the particles.
How the small cloud of particles contracts or expands is determined by a
random product of contraction and expansion factors evaluated along its path through the turbulent flow.
We refer to this mechanism as \lq multiplicative amplification\rq{} \citep{Wil03,Meh04,Dun05,Gus11a}.

There is a dichotomy here. Preferential sampling relies on instantaneous correlations between particle positions and local fluid-velocity gradients. Multiplicative amplification by contrast refers to the history of fluid-velocity gradients seen by the particles. How can these two points of view be reconciled?
The picture is further complicated by the fact that preferential sampling affects
the fluctuations of fluid-velocity gradients experienced by the particles. The resulting bias must be taken into account when computing  small-scale clustering by multiplicative amplification.  In the inertia-less case these fluctuations are referred to as \lq Lagrangian fluctuations\rq{} \cite{Snyder,TB2009}.  But when particle inertia matters the fluctuations along the particle paths differ from the Lagrangian fluctuations because the particles can detach from the flow.
The actual particle paths determine how the particles sample the gradients of the fluid velocity. The aggregate effect of these  velocity-gradient fluctuations determines whether the particles cluster or not.

This is a very difficult problem to solve {\em ab initio}. Analytical calculations become possible only if the problem is substantially simplified.
During the past decade it has been shown that important aspects of spatial clustering of particles in turbulence can be understood in terms of statistical models.
The problem becomes tractable when the turbulent velocity fluctuations are replaced by those of a random velocity field with appropriate correlation functions.
One of the most frequently studied statistical models of this kind is the Kraichnan model \citep{Kra68}.
It exhibits a range of spatial scales representing the inertial range of scales in turbulence.
Other models are kinematic turbulence models \citep{Bag09,Ijz10},
so-called random renovating flows \cite{Gil92,Elp02}, and telegraph models \cite{Fal07}.
In this review we employ spatially smooth Gaussian models
\citep{Wil03,Dun05,Meh05,Wil05,Wil07,Bec08,Gus11a}
with a single correlation time and a single correlation length
that represents the small spatial scales of three-dimensional homogeneous isotropic turbulence.
 We describe the questions, methods, and conclusions
 that have led to the  present understanding of small-scale spatial clustering in turbulent aerosols,
and explain its relation to preferential sampling.

The remainder of this article is organised as follows.
In Section~\ref{sec:Model} we introduce the statistical model for the dynamics of heavy, small spherical  particles in turbulence
and briefly mention other models that have been analysed
in the literature. Section~\ref{sec:measures} summarises
common ways of quantifying spatial clustering in turbulent aerosols
on different scales. In Section~\ref{sec:mechanisms} we explain the mechanisms
of preferential concentration and multiplicative amplification and arrive at a synthesis
that consistently accounts for spatial clustering in turbulent aerosols.
We review different limits of the statistical-model analysis that connect to results obtained by other means. Section \ref{sec:comparison} contains a comparison between the statistical-model results and those of numerical simulations of particles in turbulence.
Finally, in Section \ref{sec:Method} we review the  methods that made it possible
to obtain the results described in this review. The first five Sections of this review can be read without reference to Section \ref{sec:Method}. But we believe that it is nevertheless important
to describe the mathematical methods. This allows the reader to judge the limitations of the different methods, and to understand how they may be generalised for use in other contexts. Last but not least it becomes clear
how the questions discussed in this review are connected to other
problems in condensed-matter theory, statistical physics, and dynamical-systems theory.
Section \ref{sec:conc} contains our conclusions.

\section{Statistical model}
\label{sec:Model}
In this Section we describe the statistical model for the inertial dynamics of small heavy spherical particles in fully developed homogeneous isotropic turbulence, at large Reynolds numbers \cite{Fri97,Pope2000}. When the particles are very small, only the dissipative range of the turbulence matters where the fluid-velocity field is spatially smooth. This range extends up to length scales of about $10\,\eta_{\rm K}$ where $\eta_{\rm K}$ is the Kolmogorov length of the turbulent flow, of the order of
the size of the smallest turbulent eddies \cite{Saw08,Ish08}.
\subsection{Equation of motion}
\label{sec:model}
The general problem of determining the dynamics of small particles in turbulence is a very difficult one.
If interactions between the particles are important for their dynamics we are faced with a many-body problem. If there are no long-range interactions and if the suspension is dilute
then {direct interactions between the particles can be neglected.} But {even} to determine the dynamics of a single particle in a flow is a formidable task. In general, it requires solving the non-linear coupled particle and Navier-Stokes equations, taking into account the time-dependent boundary conditions of the fluid problem due to the moving particle. In certain limiting cases the problem simplifies, however.
The centre-of-mass dynamics of a heavy, small spherical particle suspended in a fluid is commonly approximated by the equation of motion
\begin{eqnarray}
\dot{\ve x}&={\ve v}\,,\quad \dot{\ve v}=\ve g+\gamma (\ve u-\ve v)\,.
\label{eq:stokes}
\end{eqnarray}
Dots over variables denote time derivatives, $\ve x\equiv \ve x(t)$ is the particle position {at time $t$}, $\ve u\equiv\ve u(\ve x(t),t)$
is the fluid velocity evaluated at the particle position at time $t$, $\ve g$ is the
gravitational acceleration of magnitude $g$, and $\gamma \equiv 9\rho_{\rm f}\nu/(2a^2\rho_{\rm p})$ is the Stokes damping constant that characterises the damping of differences in particle and fluid velocities. Here $\rho_{\rm f}$ is the fluid-mass density, $\rho_{\rm p}$ is the particle-mass
density,
$a$ is the particle radius, and $\nu$ is the kinematic viscosity.
Eq.~(\ref{eq:stokes})
takes into account the effect of particle inertia (it is an equation for the particle acceleration $\dot{\ve v}$),
but neglects the effect of fluid inertia, both convective and unsteady, as well as buoyancy. Under which circumstances is this a good approximation? First, buoyancy can be disregarded if the particle density is much larger than
that of the fluid, $\rho_{\rm p}\gg \rho_{\rm f}$.
Second, for convective effects to be negligible it is necessary that the particle Reynolds number ${\rm Re}_p = a v_{\rm s}/\nu$ is small ($v_{\rm s} = |\ve u- {\ve v}|$ is the magnitude of the slip velocity).  This condition is satisfied for a sufficiently small particle. For larger particle Reynolds numbers heuristic non-linear drag laws are used in numerical simulation studies \cite{Sommerfeld}. In this review we only consider the linear drag law (\ref{eq:stokes}). Third, the effect of unsteady fluid inertia can be neglected provided that the correlation time of the flow is larger than the viscous time $a^2/\nu$. In turbulence this is the case if the particle is sufficiently small: the condition
$a^2/\nu\ll \tau_{\rm K}$ corresponds to $a\ll \eta_{\rm K}$ where $\tau_{\rm K}$ and $\eta_{\rm K}$ are the Kolmogorov time and length scales of the turbulent flow.
For heavy small particles the effects of particle inertia may nevertheless be very important
because we may have $\gamma \tau_{\rm K} \ll 1$ if $\rho_{\rm p}\gg \rho_{\rm f}$.

For the most part of this review we disregard the effect of gravitational settling.
This is justified when the turbulence is intense and leads
to the simplified equation of motion:
\begin{eqnarray}
\dot{\ve x}&={\ve v}\,,\quad \dot{\ve v}=\gamma (\ve u-\ve v)\,.
\label{eq:eom}
\end{eqnarray}
Section \ref{sec:gs} explains under which circumstances settling makes a difference to spatial clustering,
and how the effect of settling can be computed.
Eq.~(\ref{eq:eom}) gives rise to complex dynamics characterised by large fluctuations in the distribution of particle positions, caused by the non-linearity of the turbulent driving: the fluid-velocity field depends non-linearly on the particle position.

An important question is how Eqs.~(\ref{eq:stokes},\ref{eq:eom}) must be modified to describe the motion of neutrally buoyant particles in turbulence \cite{Cal12,Oli14}, of bubbles \cite{Pra12},
or of larger particles \cite{Gua08,Tra15}. It has been suggested \cite{Dai15} to use an
approximate equation of motion \citep{Max83,Aut88,Gat83} that is often referred to as the \lq Maxey-Riley equation\rq{}. This equation describes the
centre-of-mass motion of a small sphere in an unsteady non-uniform flow, generalising earlier work of Tchen, Basset, and Boussinesq. The approximation rests
on solutions of the time-dependent Stokes equation. The effect of convective fluid inertia is neglected. In many cases this is an excellent approximation.
But in turbulence the effects of convective fluid inertia are likely to be
important \cite{Candelier2016}. It may thus be difficult to justify using the Maxey-Riley equation in turbulence, and it remains an open problem to formulate a general equation of motion for the
dynamics of particles in turbulence when fluid inertia matters. In some cases it may be sufficient to account for finite particle size by so-called \lq Fax\'e{}n corrections\rq{} \cite{Aut88}
but it is not known under which circumstances this is a good approximation. An example where these questions definitely matter is the dynamics of air bubbles in water \cite{Mat15}.

The above discussion refers to particles that are small, yet much larger than the mean free path of the fluid. But in the astrophysical example mentioned in the Introduction the turbulent gas is very dilute, so that micron-sized dust grains are much smaller than the mean free path. In this limit (the so-called \lq Epstein limit\rq{} \cite{Epstein24})
the drag law is still of the form (\ref{eq:eom}), but there are two important modifications. First the constant $\gamma$ depends differently on particle size. The mean free path $\ell$ in the expression for the viscosity $\nu \propto c_{\rm s} \ell$ must be replaced
by the smallest length scale of the problem, the particle size $a$. Here $c_{\rm s}$ is the r.m.s molecular speed of the gas. Second, molecular diffusion gives rise to an additional stochastic term in the equation of motion.
In the remainder of this review, however, we stick to Eqs.~(\ref{eq:stokes}) and (\ref{eq:eom}).

\subsection{Statistical model for the velocity field}
\label{sec:rvf}
\begin{table}[b]
\caption{\label{tab:1} Conversion table for different parameterisations of the compressibility of $d$-dimensional random velocity fields. The parameter $\beta$ is defined in the main text, Eq.~(\ref{eq:velocity_field}).
An alternative parameterisation is in terms of
the ratio $\Gamma$ between radial and transversal diffusion coefficients for particle separations~\cite{Wil07}, or in terms of the ratio  $\wp\equiv\langle(\ve\nabla\cdot\ve u)^2\rangle_\infty/\langle\tr(\ve\nabla\ve u^{\sf T})^2\rangle_\infty$ \cite{Fal01,Bec04}.  The entries
of this Table show how  $\Gamma$ and $\wp$ are related to $\beta$.}
\mbox{}\\[-5mm]

\begin{tabular}{cccc}
\hline\hline\\
& $\beta^2$ & $\Gamma$  & $\wp$ \\[1mm]
\hline\\[-2mm]
$\beta^2$ & -                                    & $\frac{d+1+\Gamma(1-d)}{3\Gamma-1}$    & $\frac{\wp(1-d)}{\wp-1}$ \cr
$\Gamma$ & $\frac{d+1+\beta^2}{d-1+3\beta^2}$   & -   & $\frac{d+1-2\wp}{(d-1)(1+2\wp)}$    \cr
$\wp$ & $\frac{\beta^2}{d-1+\beta^2}$ & $\frac{d+1+\Gamma(1-d)}{2(1+\Gamma(d-1))}$ &  -  \cr\\
\hline\hline
\end{tabular}
\end{table}
Different models of the turbulent velocity fluctuations have been employed in the literature.
Most of the results summarised in this review were obtained for a synthetic-flow model, a Gaussian random velocity field $\ve u(\ve x,t)$
with a given correlation length and correlation time.
To define the model it is convenient to decompose the fluid-velocity field into its compressible and incompressible parts.
In one-, two- and three spatial dimensions we write
\begin{subequations}
\label{eq:velocity_field}
\begin{equation}
\label{eq:velocity_field_1d}
\mbox{}\hspace*{-34mm} u(x,t){\equiv}{\cal N}_1\beta\nabla{\phi}(x,t)\,,\\[-0.7cm]
\end{equation}
\begin{equation}
\label{eq:velocity_field_2d}
\ve u(\ve x,t){\equiv}{\cal N}_2\Big[
\lp
\begin{array}{c}
\partial_y{\psi}(\ve x,t)\cr
-\partial_x\psi(\ve x,t)
\end{array}
\rp
+\beta\ve\nabla\phi(\ve x,t)\Big]\,,\\[-0.7cm]
\end{equation}
\begin{equation}
\mbox{}\hspace*{-5mm}\ve u(\ve x,t)\equiv{\cal N}_3\left[\ve\nabla\times\ve A(\ve x,t)+\beta\ve\nabla\phi(\ve x,t)\right]\,,
\label{eq:velocity_field_3d}
\end{equation}
with normalisation constants
\begin{equation}
\mbox{}\hspace*{-1.9cm}{\cal N}_d\equiv[d(d-1+\beta^2)]^{-1/2}\,.
\label{eq:normalisation}
\end{equation}
\end{subequations}
Here $d$ is the spatial dimension.
The parameter $\beta$ determines the degree of compressibility of the two-
and three-dimensional velocity fields (since ${\cal N}_1 = \beta^{-1}$ we see that the one-dimensional velocity field $u(x,t)$ in Eq.~(\ref{eq:velocity_field_1d}) is independent of $\beta$).
The limit $\beta \rightarrow \infty$ corresponds
to potential, entirely compressible velocity fields. For $\beta=0$,
by contrast, the velocity field is solenoidal (incompressible).
Other authors have parameterised the degree of compressibility in other ways.
In Ref.~\cite{Wil07} the ratio $\Gamma$ between radial and transversal diffusion coefficients for particle separations is used. Falkovich {\em et al.} \cite{Fal01} define the parameter  
$\wp\equiv\langle(\ve\nabla\cdot\ve u)^2\rangle_\infty/\langle\tr(\ve\nabla\ve u^{\sf T})^2\rangle_\infty$. Table~\ref{tab:1} translates between the different conventions.

Turbulent aerosols are usually suspensions of particles in incompressible flows,
this makes the limit $\beta=0$ most relevant. It is nevertheless interesting
to consider compressible flows. Particles floating on a turbulent flow, for example, experience
a compressible surface flow \cite{Som93,Bof04,Cre04,Bof06,Lov13}. More generally particle dynamics in partly compressible flows has been studied \cite{Gaw00,Gus13b,Vincenzi}, see Ref.~\cite{Fal01} for a review.

The functions $\phi$ and $\psi$ as well as the components of the three-dimensional vector field $\ve A$ are taken to be independent Gaussian random functions.
They are assumed to have the same steady-state correlation functions,
\begin{eqnarray}
\label{eq:velocity_field_C} C(\ve x_1,t_1;\ve x_2, t_2)&\equiv& u_0^2\eta^2\exp\Big(-\frac{| \ve x_2-\ve x_1|^2}{2\eta^2}-\frac{|t_2-t_1|}{\tau}\Big)\,.
\end{eqnarray}
Here $\tau$, $\eta$ and $u_0$ are characteristic time-, length- and speed scales of the flow.
The normalisation constants ${\cal N}_d$ [Eq.~(\ref{eq:normalisation})]
are chosen so that $\langle|\ve u|^2\rangle=u_0^2$.
The important property of (\ref{eq:velocity_field_C}) is that this equation describes fields
that are smooth in space and time, modeling the dissipative range of turbulence. The analytical methods summarised later in this review do not rely upon the particular functional form (\ref{eq:velocity_field_C}) of the correlation function.

The statistical model is intended to describe small-scale  spatial clustering in
the dissipative range of homogeneous isotropic turbulence, at separations smaller than the Kolmogorov length
where the turbulent velocity field is smooth. Unlike the Kraichnan model \cite{Kra68} or kinematic turbulence models \cite{Bag09,Ijz10} the statistical model
does not have  an inertial range. The inertial ranges of two- and three-dimensional turbulence are quite
different, but both exhibit spatially smooth dissipative ranges.
Therefore small-scale spatial patterns are expected to be essentially similar in two and three spatial dimensions.

The statistical model as formulated above avoids a problem that occurs in kinematic models with an inertial range.
In turbulence small vortices are swept along by large-scale velocities. There is no \lq sweeping\rq{} in kinematic models, and this
affects the particle dynamics \cite{Che06,Vos15}. Particles are removed from small-scale turbulent structures due to large-scale turbulent velocities
much like particles falling through small vortices as they settle under gravity \cite{Gus14e}.

It is convenient to construct the fields $\phi$, $\psi$, and $\ve A$  using Ornstein-Uhlenbeck processes
with the steady-state correlation function (\ref{eq:velocity_field_C}).
The Gaussian spatial decay in (\ref{eq:velocity_field_C}) is obtained by representing
the fields in terms of Fourier series of the form
\begin{equation}
\label{eq:phixt}
\phi(\ve x,t) \equiv (2\pi)^{d/4}\frac{u_0\eta^{d/2+1}}{L^{d/2}}\sum_{\ves k} a_{\ves k}(t) {\rm e}^{{\rm i} \ves k\cdot \ves x-k^2\eta^2/4}
\end{equation}
with random time-dependent Fourier coefficients $a_{\ves k}(t)$ that have zero means and unit variances.
The allowed values of the wave vector $\ve k$ are determined by imposing periodic spatial boundary conditions
in a finite box of side length $L$. The side length is taken
to be much larger than the correlation length $\eta$ of the fluid-velocity field.  The random Fourier coefficients $a_{\ves k}(t)$ are determined by independent Ornstein-Uhlenbeck processes:
\begin{equation}
\label{eq:OUak}
\delta a_{\ves k} = -a_{\ves k}{{\delta t}/{\tau}} + {\delta w_{\ves k}}\,.
\end{equation}
The Gaussian random increments ${\delta w_{\ves k}}$ are complex, they satisfy
$ \delta w_{-\ves k} = \delta w_{\ves k}^\ast$ (the asterisk denotes complex conjugation). The increments have
vanishing means, and covariances $\langle {\delta w_{{\ves k}_1}\delta w^*_{{\ves k}_2}}\rangle = {2\delta_{{\ves k}_1,{\ves k}_2}\delta t/\tau}$.
It follows that the coefficients $a_{\ves k}$ average to zero, $\langle a_{\ves k}(t)\rangle_\infty =0$.
Their steady-state covariances are given by:
\begin{eqnarray}
\label{eq:akt}
&
\langle a_{{\ves k}_1}(t_1) a^\ast_{{\ves k}_2}(t_2)\rangle_\infty
= \delta_{{\ves k}_1{\ves k}_2}{{\rm e}^{-|t_1-t_2|/\tau}}\,,\,\,
 \langle a_{{\ves k}_1}(t_1) a_{{\ves k}_2}(t_2)\rangle_\infty
 =\langle a_{{\ves k}_1}^*(t_1) a_{{\ves k}_2}^*(t_2)\rangle_\infty =0\,.
\end{eqnarray}
Steady-state ensemble averages of the process (\ref{eq:OUak}) are denoted by $\langle \cdots \rangle_\infty$.

The statistical model describes spatially and temporally fluctuating
fluid-velocity fields that are stationary, and spatially isotropic and homogeneous. The model represents
a highly idealised caricature of stationary, isotropic, and homogeneous turbulence.
But as we show in this review this simple model does a good job at explaining
the mechanisms that cause spatial patterns of particles in turbulent flows.

\subsection{Dimensionless numbers and variables}
\label{sec:dimlessvar}
\subsubsection{Dimensionless numbers}
We must consider two sets of dimensionless numbers: those that determine whether Eq.~(\ref{eq:stokes}) is valid, and those that distinguish between different dynamical behaviours of this equation.

First, the suspension must be dilute so that direct interactions between the particles are negligble.
In deriving Eq.~(\ref{eq:stokes}) it is assumed that the particle Reynolds number ${\rm Re}_p$ is small, that the particle is much smaller than the Kolmogorov length, $a/\eta_{\rm K} \ll 1$,
and that the particle density is much larger than that of the fluid, $\varrho_{\rm p}/\varrho_{\rm f}\gg 1$.

Second, what are the dimensionless numbers that determine the
statistical-model dynamics?
It is assumed throughout that the correlation length $\eta$ of the fluid velocity is much
smaller than the side length $L$ of the simulation box. This leaves
five dimensional parameters: $u_0$, $\tau$, $\eta$, $\gamma$, and $g$.
Out of these parameters three independent dimensionless numbers can be formed:
\begin{equation}
\st \equiv 1/(\gamma \tau)\,,\quad \ku \equiv {u_0 \tau}/{\eta}\,,\quad {\rm F} \equiv {g\tau}/{u_0}\,.
\label{eq:StKu}
\end{equation}
The Stokes number ${\rm St}$ is a dimensionless measure of the importance of inertial effects.
The Kubo number $\ku$ is a dimensionless measure of how rapidly the fluid velocity fluctuates.
The number ${\rm F}$ characterises the importance of settling. In the literature this
parameter (sometimes its inverse) is referred to as the Froude number \cite{Sha03,Dev12}.

\subsubsection{Dimensionless variables}
When the turbulence is intense then ${\rm F}\ll 1$ and gravitational settling can be neglected.  In this
case the equation of motion is Eq.~(\ref{eq:eom}). It can be de-dimensionalised in different ways. Which scheme
is most convenient? This depends on the limit of the problem considered. At finite Kubo and Stokes numbers
we use the dimensionless variables
\begin{equation}
\label{eq:dim1}
 t'\equiv t/\tau\,,\quad \ve x'\equiv \ve x/\eta\,, \quad\ve v'\equiv \ve v/u_0\,, \quad
\ve u'\equiv \ve u/u_0\,.\end{equation}
In these variables Eq.~(\ref{eq:eom})  reads
\begin{eqnarray}
\dot{\ve x}'&=\ku\,{\ve v}'\,, \quad \dot{\ve v}'= (\ve u'-\ve v')/\st\,.
 \label{eq:langevin_dimless}
\end{eqnarray}
The correlation function (\ref{eq:velocity_field_C}) becomes
 $C(\ve x_1',t_1';\ve x_2', t_2') = {\rm e}^{-{| \ves x_2'-\ves x_1'|^2}/{2}-{|t_2'-t_1'|}}$.

The second de-dimensionalisation scheme
\begin{equation}
\label{eq:dim2}
\tilde t \equiv t\gamma\,,\quad
\tilde{\ve x} \equiv \ve x/\eta\,,\quad
\tilde{\ve v} \equiv \ve v/(\eta\gamma)\,,\quad
\tilde{\ve u} \equiv \ve u/(\eta\gamma)
\end{equation}
is useful in two limiting cases of the model. These limiting cases are discussed below.
In the variables defined by Eq.~(\ref{eq:dim2}) the equation of motion (\ref{eq:eom}) takes the form
\begin{eqnarray}
\dot{\tilde{\ve x}}&=\tilde {\ve v}\,,\quad \dot{\tilde {\ve v}}= \tilde{\ve u}-\tilde{\ve v}\,.
 \label{eq:langevin_dimless_2}
\end{eqnarray}
The correlation function becomes
\begin{align}
\label{eq:CorrWNUnits}
C(\tilde{\ve x}_1,\tilde t_1;\tilde{\ve x}_2,\tilde t_2)= \ku^2\,\st^2 {\rm e}^{-{|\tilde{\ves x}_2-\tilde{\ves x}_1|^2}/{2}-|\tilde t_2-\tilde t_1|\st}\,.
\end{align}
\subsubsection{Two limiting cases}
\label{sec:limits}
In the dimensionless variables (\ref{eq:dim2}) the equation of motion (\ref{eq:langevin_dimless_2})
is independent of $\st$ and $\ku$. The dynamics depends on these dimensionless numbers only through
the correlation function (\ref{eq:CorrWNUnits}). There are two limiting cases of the statistical model
in which this correlation function takes particularly simple forms,
the \lq white-noise limit\rq{} and the \lq persistent-flow limit\rq{}.
In these limits the dynamics does not depend on $\st$ and $\ku$ separately
but only on combinations of these numbers.

The  white-noise limit corresponds to $\tau\rightarrow 0$, or $\st\!\rightarrow\!\infty$.
In this limit the fluid velocities fluctuate very rapidly in time, so that diffusion approximations can be used.
The correlation function (\ref{eq:CorrWNUnits}) tends to
\begin{equation}
\label{eq:corr2}
C(\tilde{\ve x}_1,\tilde t_1;\tilde{\ve x}_2,\tilde t_2)=2\, \ku^2\,\st \,{\rm e}^{-{|\tilde{\ves x}_2-\tilde{\ves x}_1|^2}/{2}}\,\delta(\tilde t_2-\tilde t_1)\,.
\end{equation}
As $\st \rightarrow \infty$ the parameter $\ku$ must be taken to zero so that
$\ku^2\st$ remains constant. In this limit it is the product $\ku^2\st$ that determines the particle dynamics. We define the dimensionless parameter
\begin{equation}
\label{eq:defeps}
\varepsilon^2\equiv\frac{d-1+3\beta^2}{d(d-1+\beta^2)} \ku^2\st\,.
\end{equation}
The prefactor in (\ref{eq:defeps}) is chosen so as to make $\varepsilon^2$ equal to the radial diffusion coefficient for particle separations \cite{Wil07}. In summary, the white-noise limit is determined by
\begin{equation}
\label{eq:wnlimit}
\ku\to 0 \quad \mbox{and}\quad \st\to\infty\,,\quad\mbox{so that}\quad\mbox{$\varepsilon^2 =$ constant.}
\end{equation}
In incompressible flows a second limiting case is important,
the persistent-flow limit. This is the case where the correlation time $\tau$ is larger than all other 
time scales in the system, so that $\st\ll 1$ and $\ku\gg 1$.
As $\tau$ becomes much larger than $\eta/u_0$  the decay of fluid-velocity correlations as seen by the particles is
governed by the spatial part of the correlation function:
\begin{equation}
\label{eq:corr3}
C(\tilde{\ve x}_1,\tilde t_1;\tilde{\ve x}_2,\tilde t_1)= \ku^2\,\st^2 \,{\rm e}^{-{|\tilde{\ves x}_2-\tilde{\ves x}_1|^2}/{2}}\,.
\end{equation}
We shall see later that it is the time scale $\eta/u_0$ that corresponds to the Kolmogorov time $\tau_{\rm K}$ of fully developed turbulent flows. Comparisons between statistical-model results and those of numerical simulations of turbulent aerosols must refer to this Lagrangian time scale.
Eq.~(\ref{eq:corr3}) shows that the dynamics is described by
a single dimensionless parameter, the product $\ku\,\st$.  We define
\begin{align}
\label{eq:kappa}
\kappa\equiv\sqrt{d+2}\,\ku\,\st\,.
\end{align}
The prefactor is chosen so that
$\kappa = \gamma^{-1} \langle \tr(\ma A \ma A^{\sf T})\rangle_\infty^{1/2}$,
evaluated at $\st=0$ (\Secref{sec:comparison}).
In summary the persistent-flow limit is given by
\begin{equation}
\label{eq:klimit}
\ku\gg 1 \quad \mbox{and}\quad \st\ll 1\,,\quad\mbox{with parameter}\quad\mbox{$\kappa \propto \ku\,\st = u_0/(\gamma\eta)$.}
\end{equation}
We conclude by remarking that compressible flows may trap particles in certain regions. In this case the temporal part of the correlation function may still matter at  large values of $\tau$.

\subsection{Other models with finite time correlations}
\label{sec:othermodels}
The perhaps simplest models used to analyse heavy-particle dynamics are time-independent incompressible  cellular fluid-velocity fields \cite{Sto10,McL88},
corresponding to $\ku\rightarrow \infty$. These models have been refined in several ways. First, Maxey and Corrsin \cite{MC86} have constructed
homogeneous and isotropic ensembles of steady fluid-velocity configurations by randomly
shifting and rotating the unit cell.
Second, several authors have introduced a time-periodic modulation of the fluid-velocity field, in order to take into account the fact
that the fluid velocity changes as the particles move through the flow (see Ref. \cite{Ang07} and references cited therein).
Third, an alternative way to construct a time-dependent flow is to keep the fluid velocity field constant for Poisson-distributed
times, and to change the field randomly after each time interval. Such models are commonly referred to as random \lq renovating\rq{} or \lq renewing\rq{}
flows \cite{Gil92,Elp02}.
A number of authors have studied spatial pattern formation due to clustering of particles in closely related  models
by means of numerical simulations~\cite{Ijz10,Per12,Vincenzi}.

\label{page:definition_of_A}
Falkovich {\em et al.} \cite{Fal07} have analysed a one-dimensional model for the dynamics of the separation between two close-by particles,
 assuming that the fluid-velocity gradient $A \equiv \partial_x u$ follows a telegraph process.
In this model $A$ can only assume two values: $-A_0$ and $A_0$. The variable switches from $-A_0$ to $A_0$ at rate $\nu_+$. The reverse process occurs at rate $\nu_-$.
The amplitude $A_0$ and the rates $\nu_\pm$ are not independent, they are
constrained by the requirement that this \lq telegraph model\rq{} admits a steady state~\cite{Fal07}.
Since $A(t)^2=A_0^2$ is constant, gradient statistics and other observables
can be calculated exactly in this model. For the average fluid-velocity gradient one finds
$\langle A \rangle_\infty = -3\ku$ to lowest order in the Kubo number.
Here $\ku$ is defined as in Eq.~(\ref{eq:StKu}), slightly different from the convention employed in Ref.~\cite{Fal07}.
The Stokes number is defined in the usual way,
$\st\equiv{1}/{\gamma\tau}$, where $\tau\equiv(\nu_++\nu_-)^{-1}$
is the correlation time of the telegraph process.
The fact that $\langle A\rangle_\infty < 0$ is a consequence of preferential sampling: particles sample $-A_0$ more often than $A_0$. We shall see that the
statistical model gives a result equivalent to $\langle A \rangle_\infty = -3{\ku}$ in one spatial dimension, to first order in $\ku$.

It has been suggested to neglect preferential sampling but to keep the finite time correlations of the fluid-velocity field and its derivatives. We refer to this approximation as the \lq coloured-noise approximation\rq{}. The resulting equations can be analysed by perturbation theory and asymptotic methods \cite{Kan89,Wil10,Gus13a}, but we shall see that preferential sampling has an important effect upon small-scale spatial clustering
except in the white-noise limit (\ref{eq:wnlimit}) where, however, the time correlations are negligible.

\section{Spatial clustering}
\label{sec:measures}
\begin{figure}[t]
\begin{overpic}[width=\textwidth,clip]{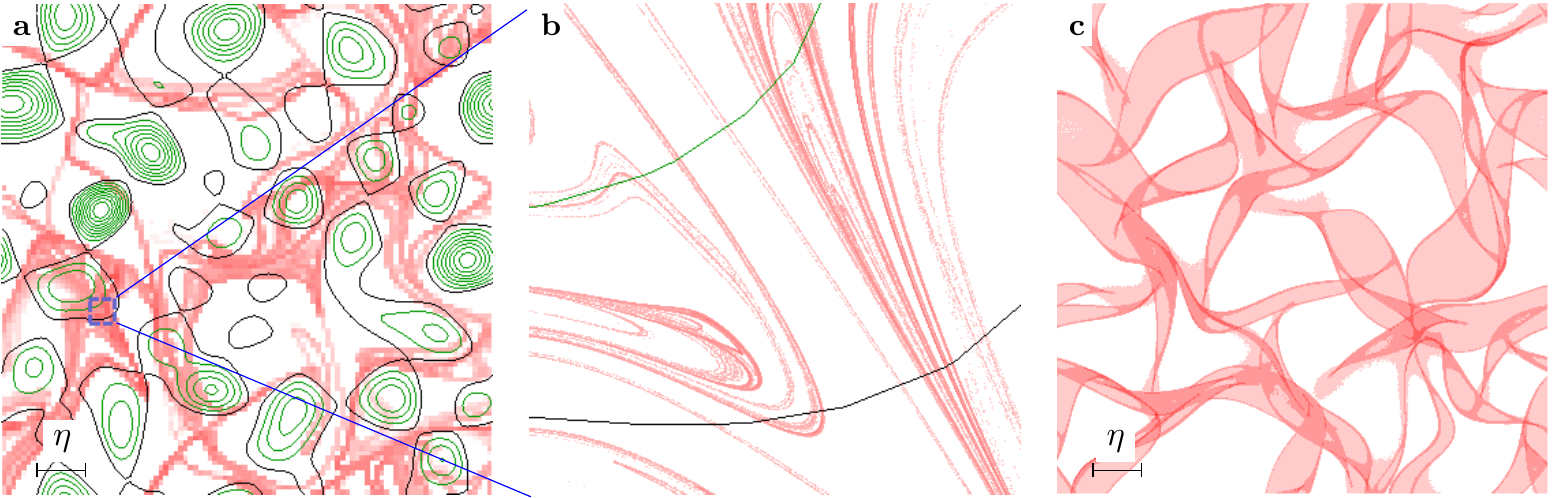}
\end{overpic}
\caption{\label{fig:mechs}
Illustration of spatial clustering mechanisms in incompressible flow.
 {\bf a} Preferential sampling. Particle-number density colour-coded on a logarithmic scale
from low concentration (white) to high concentration (red).
Black lines separate the flow into regions where either vorticity or strain dominates.
In regions of high vorticity green lines show
contours of squared vorticity minus squared strain.
The scale bar shows the correlation length $\eta$ of the random velocity field.
 Parameters: $\ku=10$, $\st=0.025$, and ${\rm F}=0$.
{\bf b} Small-scale clustering.  Magnification of the small blue square of size $0.5\,\eta\times 0.5\,\eta$ shown in panel {\bf a}.
 {\bf c} Short-time clustering due to caustics. Particle-number density obtained
after a short time starting from a uniform scatter of particles.
  Regions surrounded by caustic singularities exhibit higher particle-number density.
  Parameters: $\ku=1$, $\st=10$, ${\rm F}=0$. }
\end{figure}

Spatial patterns in turbulent aerosols occur on a range of time and length scales, illustrated by Fig.~\ref{fig:mechs}. Panel {\bf a}
shows a snapshot of the particle-number density obtained from steady-state simulations of the statistical model.
Also shown is the vorticity of the
underlying velocity field at the same time. Spatial variations on the scale of the correlation length $\eta$ of the flow are clearly visible.
It appears that the particles tend to avoid vortical flow regions.

Resolving the spatial distribution of the particles at scales much smaller than the correlation length (panel {\bf b}) shows that there is \lq small-scale clustering\rq{}. One observes filamentary structures on scales much smaller than $\eta$.

Panel {\bf c} shows transient spatial patterns obtained by following an initially uniform distribution of particles for a short time.
The distinct inhomogeneities visible in panel {\bf c} are
caused by singularities in the particle dynamics (so-called \lq caustics\rq{}).

\subsection{Small-scale clustering}
\label{sec:fractal}
\subsubsection{Lyapunov exponents}
Sommerer and Ott \cite{Som93} described spatial small-scale clustering of particles floating on the surface of a turbulent flow,
on spatial scales much smaller than its correlation length. They characterised
this small-scale clustering in terms of spatial Lyapunov exponents of the particle
dynamics.
In $d$ spatial dimensions there are $d$ such exponents \cite{Ott02},
describing the long-term dynamics of spatial separations between a test particle and particles
in its vicinity. In three spatial dimensions one writes:
\begin{subequations}
\label{eq: 2.4}
\begin{equation}
\label{eq:l1_def}
\mbox{}\hspace*{18mm}
\lambda_1\equiv\lim_{t\to \infty} t^{ -1}\log({\RRR_t/\RRR_0})\,, \\[-8mm]
\end{equation}
\begin{equation}
\mbox{}\hspace*{10mm} \lambda_1+\lambda_2\equiv\lim_{t \to \infty}  t^{ -1}\log({\AAA_t/\AAA_0})\,,
\\[-8mm]
\end{equation}
\begin{equation}
\label{eq:l123}
\mbox{}\hspace*{0mm}\lambda_1+\lambda_2+\lambda_3\equiv\lim_{t \to \infty} t^{ -1}
\log({\VVV_t/\VVV_0}) \,.
\end{equation}
\end{subequations}
Here $t$ denotes time, {$\RRR_t$} is an initially small distance between two close-by particles, $\AAA_t$ is the
initially infinitesimal area of the parallelogram spanned by the two separation vectors between three particles that are close together, and $\VVV_t$ is the small volume of the parallelepiped spanned by the three separation vectors between four close-by particles. The Lyapunov exponents determine how an initially infinitesimally small cloud of particles contracts or expands in the long run.

It follows from
Eq.~(\ref{eq: 2.4}) that the Lyapunov exponents are ordered, $\lambda_1 \geq \lambda_2 \geq \dotsc \geq \lambda_d$.
For heavy particles in a three-dimensional incompressible turbulent flow at Stokes numbers of order unity  one finds that $\lambda_1>0$, $\lambda_1+\lambda_2 > 0$,
but $\lambda_1+\lambda_2+\lambda_3 < 0$. Small separations and areas expand in the long run, but
volumes contract. This implies that the particles in a small volume are swept onto
a set with dimension smaller than three, resulting in spatial clustering. In the special case where $\lambda_1+\lambda_2=0$ the attracting set is two dimensional. But when $\lambda_1+\lambda_2 >0$ the situation is  more complicated: in this case the particles form a fractal set with a fractal dimension that lies between two and three. Such attractors may arise in systems with a positive maximal Lyapunov exponent
\cite{Ott02}.

\subsubsection{Fractal dimensions}
\begin{figure}[t]
\begin{overpic}[width=\textwidth,clip]{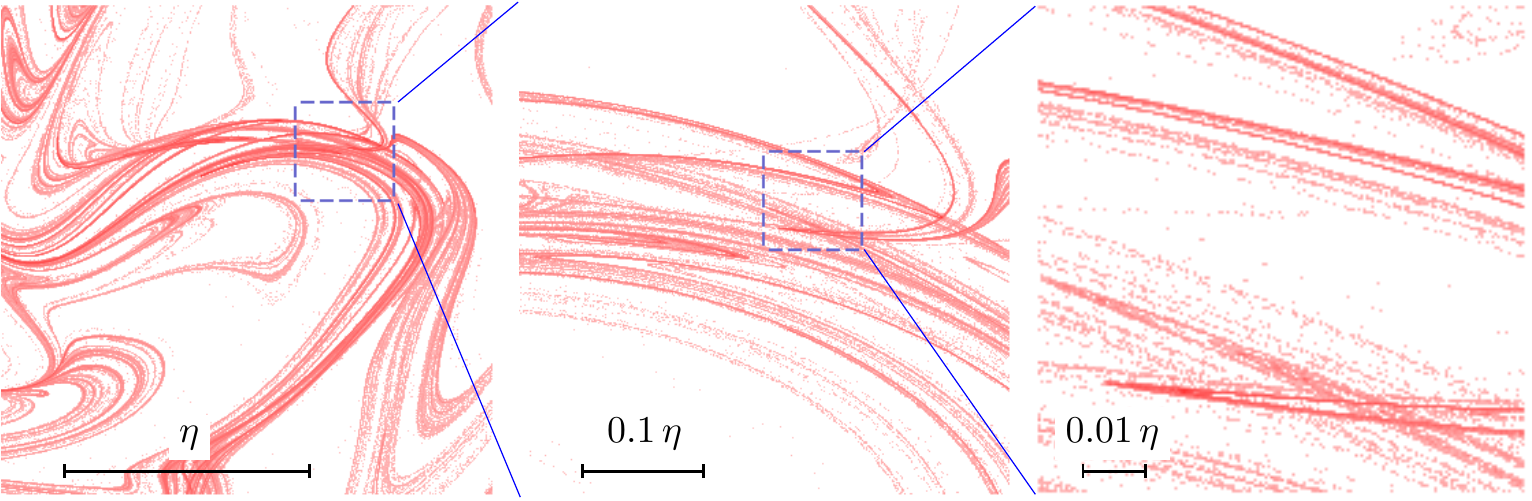}\\[-4mm]
\end{overpic}
\caption{\label{fig:fractal}
Illustration of fractal clustering in
the statistical model (Section \ref{sec:Model}), for a two-dimensional incompressible fluid velocity field. Parameters: $\ku = 0.1$,
$\st = 10$. For these parameters the Lyapunov dimension (\ref{eq:dL1}) takes the value $\dL \approx 1.6$.
Left: small square of side length $2\eta$ containing $1.2\times 10^6$ particles, cut from the simulation domain of side length $L=10\eta$. Centre panel: a small part of the left panel (dashed square) enlarged by a factor five ($2\times 10^5$ particles).
Right panel: magnification of a part of the centre panel, again
by a factor of five ($3.2\times 10^4$ particles). In all cases a coarse-grained particle-number density is shown, colour coded from white (no particles) to red (highest density).
Scale bars are given in units of the flow-correlation length $\eta$. }
\end{figure}

Fig.~\ref{fig:fractal} shows an example of a fractal distribution of particles. It was obtained by
following a large number of particles using the two-dimensional version of the statistical model described in Section \ref{sec:Model}.
The simulation was run long enough to ensure that Fig.~\ref{fig:fractal} represents the steady
state of the system. Two magnifications of the particle patterns are shown, to illustrate the fact that
the patterns look similar on smaller scales when magnified. This statistical self similarity is an important
property of fractals. For particles in turbulence the statistical self similarity is broken
at very small scales determined by the particle size, or by the length scale associated with molecular diffusion.

In the simplest case the fractal geometry is described by a so-called \lq fractal dimension\rq{}, a number that is not  necessarily an integer \cite{Mandelbrodt,Ott02}. The fractal dimension determines to which degree the particles fill out space.

Spatial clustering in turbulent aerosols is \lq multifractal\rq{} \cite{Bec05,Bec05a}.
The degree of spatial clustering is quantified by so-called \lq generalised fractal dimensions\rq{} $d_q$ \citep{Gra83a}. Different dimensions (corresponding to different values of $q$) assume different values. They characterise different geometrical aspects of the particle distribution by weighing fluctuations of the particle-number density on the fractal attractor in different ways.
The dimension $d_0$ is called the \lq box-counting dimension\rq{}, it is a measure how space-filling the fractal is. The larger $q$ the more weight
is given to regions with high particle-number densities.

Often the \lq{}correlation dimension' $d_2$ is used to characterise small-scale spatial clustering: in experiments \citep{Saw08,Sal08} and in
numerical simulations \cite{Rea00a,Chu05,Bec07,Bec07b,Fal04}. 
The correlation dimension is defined as follows.
Consider the number ${\cal N}$ of particles inside a ball of radius $\RRR$.
For uniformly distributed particles ${\cal N} \sim \RRR^d$ for small values of $\RRR$, where $d$ is the
spatial dimension. If the particles cluster onto a fractal set one writes ${\cal N} \sim \RRR^{d_2}$ where $d_2$ defines the correlation dimension.
The correlation dimension is of physical importance
because it characterises the form of the pair correlation function $g(\RRR)$ at small separations: $g(\RRR) \equiv \RRR^{-(d-1)}{\rm d}{\cal N}/{\rm d}\RRR \sim \RRR^{d_2-d}$.

The correlation dimension is difficult to compute analytically. It can be expressed in terms of the large-deviation statistics of finite-time Lyapunov exponents \cite{Bec04,Gra84,Ott02}, but the correlation dimension has only been computed in certain limiting cases: for small values of $\st$~\cite{Fal04,Chu05}, and in the white-noise limit~\cite{Wil10b} (see also \cite{Bec08}).
Ref.~\cite{Bra15a} analyses a \lq drift velocity of particle separations\rq{} \cite{Zai07,Zaichik} that gives rise to small-scale clustering, and in principle allows to compute $g(\RRR)$.
\begin{figure}
\includegraphics[width=7cm,clip]{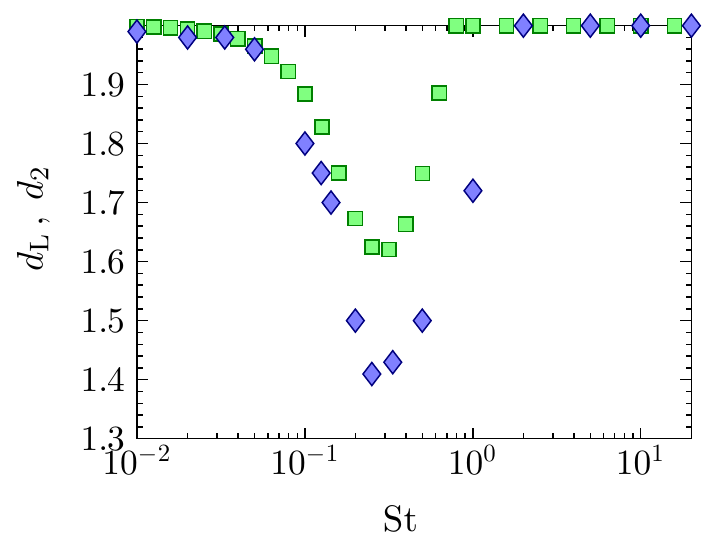}
\caption{\label{fig:fd}
Fractal Lyapunov dimension $\dL$ (green filled $\Box$) and correlation dimension $d_2$ (blue filled $\Diamond$)
as functions of $\st$. Shown are numerical results of statistical-model simulations of an incompressible flow in $d=2$ dimensions. Parameters: $\ku=1$, and $F=0$. The data for $d_{\rm L}$ is taken from 
Ref.~\cite{Gus11a}.}
\end{figure}
In general, all the fractal dimensions $d_q$ are difficult to calculate, numerically and analytically.

Kaplan and Yorke  \citep{Kap79} proposed a dimension, $\dL$, that is easier to analyse. It is
based on the spatial Lyapunov exponents. This \lq Lyapunov dimension\rq{} interpolates between integer dimensions $d=1,2,\ldots$ using the values of the Lyapunov exponents as a guideline:
\begin{equation}
\label{eq:dL1}
\dL \equiv \kappa + \frac{\sum_{\mu = 1}^\kappa \lambda_\mu}{|\lambda_{\kappa+1}|}\,.
\end{equation}
Here $\kappa$ is the largest integer for which the sum of Lyapunov exponents up to and including $\kappa$ is positive.
A uniform scatter of particles corresponds to $\dL =d$ where
$d$ is the spatial dimension. Fractal clustering occurs when $\dL < d$.
The so-called \lq dimension deficit\rq{} is defined as $\DeltaL \equiv d-\dL$.
The fractal Lyapunov dimension of the patterns in Fig.~\ref{fig:fractal} lies between
unity and two, $\dL \approx 1.6$, reflecting the fact that the system has two spatial
dimensions, and that $\lambda_1>0$ but $\lambda_1+\lambda_2 < 0$.

In order to understand which mechanisms cause fractal clustering in turbulent aerosols many authors
have analysed the Lyapunov exponents and the Lyapunov dimension numerically \citep{Som93,Bec03,Bec06,Bec07}.
Fig.~\ref{fig:fd} shows numerical results for the Lyapunov dimension for the statistical
model as a function of the Stokes number. Also shown is the correlation dimension $d_2$.
We see that both dimensions show qualitatively similar behaviour, but they are not equal. This is expected because the particles cluster on a multifractal, as mentioned above.

\subsection{Preferential sampling}
\label{sec:LSC}
Heavy particles may detach from the flow, and this may result
in spatial clustering.  Maxey \cite{Max87} suggested that heavy particles gather
in straining regions of the flow because they are centrifuged out of vortical regions.
 This is an example of \lq preferential sampling\rq{}.
Particles prefer to sample certain regions in the flow (straining regions),
and are biased to avoid other regions (vortices).
Maxey reached this conclusion using a small-$\st$ expansion. The centrifuge mechanism is commonly invoked to explain clustering in turbulent aerosols, observed in direct numerical simulations \citep{Bec07}, in experiments \citep{Cal08,Saw08,Gib12}, in astrophysics \citep{Hod98,Bra99},
atmospheric physics \citep{Sha03,Mar94,Pin97}, and in biology~\citep{Rei03}.

Fully-developed turbulence exhibits a hierarchy of eddies of different sizes (the \lq inertial range\rq{}), ranging from the smallest eddies
(their size is of the order of the Kolmogorov length $\eta_{\rm K}$) to much larger eddies with sizes of the order
of the macroscopic driving scale~\cite{Pope2000}. Preferential sampling of different scales in the inertial range of turbulence has been observed in simulations of turbulent-aerosol dynamics (Fig.~2 in Ref.~\cite{Bec07} and Fig.~3a in Ref.~\cite{Col09}).
Falkovich {\em et al.} discuss inertial-range clustering in terms of a scale-dependent Stokes number \cite{Fal03b}.
An alternative description of inertial-range clustering is suggested in Ref.~\cite{Bra15b}.
Vassilicos and collaborators~\obs{\cite{Vas08,Che06,Got06}} analysed \obs{inertial-range} preferential sampling \obs{of a different property of the fluid flow}:
they found that the particles stick to regions of low fluid acceleration as these are swept through the flow \obs{(Fig.~8 in Ref.~\cite{Got06}).}

How is preferential sampling usually measured? Many studies rely on the observation that snapshots of particle positions and fluid vorticity show a certain degree of instantaneous anti-correlation.  Fig.~\ref{fig:mechs}{\bf a} is an example. It shows the spatial particle distribution (red) and the vorticity pattern of the flow (green \obs{contour lines}).  We observe spatial variations in both patterns of the order of the correlation length of the flow, and it is plausible that the particles tend to avoid vortical regions.
To quantify these correlations Bec {\em et al.}~\cite{Bec06d}
computed how the probability of particles sampling straining regions
depends on the Stokes number and on the Reynolds number of the turbulent flow.

Fessler {\em et al.} \cite{Squ91,Fes94} measured spatial patterns of heavy particles in wind-tunnel experiments. Fig.~5 in Ref.~\citep{Fes94} shows intricate spatial patterns of $28\mu$m-sized  Lycopodium particles with a Stokes number of order unity. The inhomogeneities in the particle-number density occur  at scales much larger than the Kolmogorov length, in the inertial range. The authors of Ref.~\cite{Fes94} analysed the patterns by computing the distribution of the local particle-number density, evaluated  at different scales
ranging from $10\eta_{\rm K}$ to $100 \eta_{\rm K}$. The degree of clustering is determined by comparing the standard deviations of these distributions to that of a Poisson distribution.
We remark that the particle size in the experiment is of the same order as the Kolmogorov length. In this case corrections to Eq. (\ref{eq:eom}) become important.

Monchaux {\em et al.} \cite{Mon10} studied spatial clustering of water droplets in a wind tunnel.
Fig.~3(b) in this reference shows the spatial droplet distribution in a thin slice exhibiting voids of the size of $\sim 10\eta_{\rm K}$. The authors analysed the patterns by a Voronoi construction. Nilsen {\em et al.} \cite{Nil13} used Voronoi analysis
to characterise preferential sampling observed in numerical simulations of heavy particles in a turbulent channel flow.

\subsection{Clustering by caustics}
\label{sec:clustering_caustics}
The patterns in Fig.~\ref{fig:fd}{\bf c} are caused by singularities in the particle dynamics.
Consider the dynamics of the small parallelepiped spanned by the three separation vectors between four close-by particles. When two of the separation vectors become collinear the volume \obs{$\mathscr{V}_t$} of the parallelepiped    
collapses to zero for an instant of time.
It is intuitively clear that $\mathscr{V}_t\rightarrow 0$ corresponds to {spatial} clustering: all particles initially contained in $\mathscr{V}_0$ are brought close together at this moment of time.
The resulting density patterns (Fig.~\ref{fig:mechs}{\bf c})  are analogous to light patterns seen at the bottom of a swimming pool on a sunny day~\cite{Wil05}. The phenomena of random focusing of \obs{sunlight} due to fluctuating optical-path lengths and \obs{of spatial clustering of heavy particles in turbulence
as $\mathscr{V}_t\rightarrow 0$} are mathematically closely related. The
singularities in the heavy-particle dynamics correspond to \lq caustic\rq{} patterns in \obs{ray dynamics of light}. We therefore refer to the singularities in the \obs{heavy-}particle dynamics as caustics~\cite{Wil03,Wil05,Cri92}. The patterns observed by Martin and Meiburg \cite{Mar94} in simulations of particle dynamics in a steady vortex are caustic patterns.

Caustics are in effect singularities of the inertial phase-space dynamics
\obs{where an initially single-valued  phase-space manifold folds over and becomes multivalued \cite{Fal02,Wil05}.} 
Caustics can be located in different ways. One possibility is to refer to the deformations of small volume elements as explained above.
An alternative is to follow the dynamics of the particle-velocity gradients $\partial v_i(t)/\partial x_j(t)$ where $x_j(t)$ and $v_i(t)$ are components of the vectors of particle position and velocity at time $t$. At a caustic singularity the trace of the matrix of particle-velocity gradients diverges \cite{Fal02,Wil05,Gus12}. Bewley {\em et al.} \cite{Bew13} measured particle-velocity gradients for microscopic water droplets in air turbulence for different droplet sizes (corresponding to different Stokes numbers). In this way the authors of Ref.~\cite{Bew13} could indirectly demonstrate the occurrence of caustics.

A related phenomenon occurs in \obs{the} Burgers equation. Its solutions form shocks that correspond to caustic singularities in the inviscid limit \cite{Fal11}.

\section{Clustering mechanisms}
\label{sec:mechanisms}

\subsection{Small-scale clustering}
\label{sec:mulamp}
What is the process that causes small-scale clustering in the steady state? Consider an infinitesimally small cloud of particles. Its volume $\mathscr{V}_t$
randomly contracts and expands.  Depending upon whether the random product of expansion and contraction factors decreases or increases in the limit of large times, fractal clustering may or may not occur in the steady state. This mechanism is referred to as \lq multiplicative amplification\rq{} \cite{Gus11a}.
According to this picture the degree of spatial clustering is determined by the history of fluid-velocity gradients the particles have experienced in the past.

The small volume element $\mathscr{V}_t$
is related to its initial value $\mathscr{V}_0$ by a random product  of possibly correlated expansion and contraction factors. Following this random multiplicative
process for a sufficiently long time $t$ results in a
log-normal distribution of the volume $\mathscr{V}_t$ \cite{Fal01}.
Whether $\log \mathscr{V}_t$ is negative or positive at long times determines whether small-scale clustering occurs or not.
This logarithm increases, on average, linearly with time.
Eq.~(\ref{eq: 2.4}) says that its rate of increase is given by the sum of the Lyapunov exponents.
More precisely, consider the dynamics of a small signed volume element ${\cal V}_t$
spanned by the $d$ infinitesimal separation vectors $\ve X_\mu(t)$ between
a test particle and $d$ nearby particles,
${\cal V}_t  =  \det\big (\ve X_1(t),\ldots,\ve X_d(t)\big)$, so that ${\VVV_t} \equiv |{\cal V}_t|$.
While the sign of ${\cal V}_t$ may change,$\VVV_t$ remains positive.
When all particles are initially close to each other and remain closely together on the same branch of a smooth phase-space manifold, 
their infinitesimal separation vectors change according to
\begin{equation}
\label{eq:VZ}
\ve X_\mu(t+\delta t) =   \big(\ma I+\ma Z(t)\,\delta t\big)\, \ve X_\mu (t)
\end{equation}
in a small time step $\delta t$. Here and in the following Greek indices are used to label particles, and Roman indices label spatial vector- and matrix-components. The matrix $\ma Z$ is the same for all values of $\mu$.  Its elements are the particle-velocity gradients
\begin{equation}
\label{eq:defZ}
Z_{ij}(t) \equiv \frac{\partial v_i(t)}{\partial x_j(t)}\,.
\end{equation}
At $\st=0$ the particles follow the flow. In this case $\ma Z(t)$
equals the matrix $\ma A$ of fluid-velocity gradients
\begin{equation}
A_{ij}\big(\ve x(t), t\big) \equiv \frac{\partial u_i(\ve x(t),t)}{\partial x_j(t)}
\end{equation}
evaluated at the particle position $\ve x(t)$, at time $t$.
When $\st>0$ we  define $\ma Z$ through its equation of motion
\begin{equation}
\label{eq:langevin_Zmatrix_0}
\dot{ \ma Z} =\gamma (\ma A-\ma Z)-{\ma Z}^2\,,
\end{equation}
rather than evaluating $\ma Z$ by following a cloud of particles that remains infinitesimally small. Equation (\ref{eq:defZ}) is obtained by differentiating Eq.~(\ref{eq:eom}) with respect to $\ve x$.
Since $\ma A$ depends on the particle path $\ve x(t)$, it follows that $\ma Z(t)$ is a functional of $\ve x(t)$. In other words,
Eq.~(\ref{eq:langevin_Zmatrix_0}) must be integrated along with the particle dynamics. The solution describes infinitesimal velocity differences between particles that remain infinitesimally close at all times. In this sense the matrix $\ma Z$ is defined locally.

Returning to Eq.~(\ref{eq:VZ}) we infer that:
\begin{equation}
{\cal V}_{t+\delta t}=\det\big(\ma I+\ma Z(t)\,\delta t\big){\cal V}_{t}\,.
\label{eq:calVt}
\end{equation}
Differentiating  with respect to $\delta t$ and taking the limit $\delta t\to 0$ gives
\begin{equation}
\label{eq:Vdot}
\frac{{\rm d} {\cal V}_{t}}{{\rm d}t}=\mbox{tr}\ma Z(t)\,{\cal V}_{t}
\end{equation}
where $\tr\, \ma Z \equiv \ve \nabla \cdot \ve v$ denotes the trace of $\ma Z$.
Eq.~(\ref{eq:Vdot}) shows that the sum of the spatial Lyapunov exponents, Eq.~(\ref{eq:l123}),
is given by the time average of $\tr\, \ma Z$.
In the limit $t\to\infty$ this average is evaluated as an ensemble average, $\langle \tr\, \ma Z\rangle_\infty$, of $\tr\, \ma Z$ along particle paths:
\begin{equation}
\label{eq:suml}
\lambda_1+\ldots+\lambda_d = \langle \tr\,\ma Z\rangle_\infty\,.
\end{equation}
The condition $\langle \tr\, \ma Z\rangle_\infty < 0$ implies spatial clustering.
Eq.~(\ref{eq:suml})  describes the rate at which the volume
of a small cloud of particles contracts. To estimate
the fractal dimension (\ref{eq:dL1}) we require individual Lyapunov exponents.
These can also be computed from $\ma Z$ (Section \ref{sec:Method}).
Eqs.~(\ref{eq:langevin_Zmatrix_0}) and (\ref{eq:suml}) demonstrate that small-scale clustering is determined by the history of fluid-velocity gradients experienced by the particles.
 \begin{figure}[t]
\begin{overpic}[height=5cm,clip]{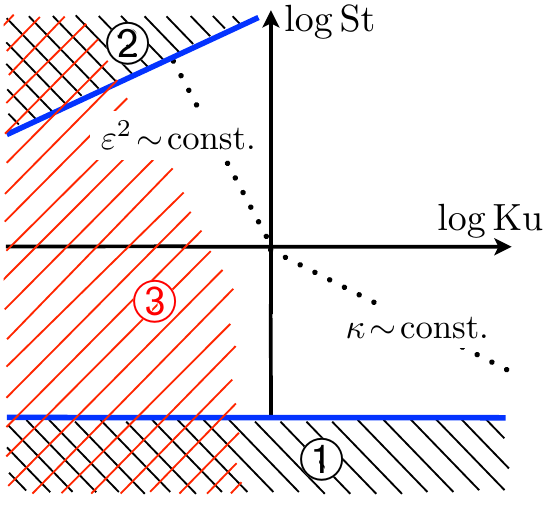}
\end{overpic}\mbox{}\hspace*{10mm}
\caption{\label{fig:pd} Schematic phase diagram in the $\ku$-$\st$-plane
for the statistical model of a turbulent aerosol.
The diagram distinguishes different asymptotic parameter regions in the model.
Region $1$ corresponds to small Stokes numbers. In this limit fractal clustering is weak (Section \ref{sec:expadv}).
Region $2$ corresponds to the white-noise limit of the model, Eq.~(\ref{eq:wnlimit}), Section \ref{sec:WNL}. In this limit the parameter $\varepsilon^2\propto\ku^2\st$ [Eq.~(\ref{eq:defeps})] determines how strongly the particles cluster. In the limit
of large Kubo and small Stokes numbers by contrast spatial clustering is determined by the parameter
$\kappa \propto \ku\,\st$  [Eq.~(\ref{eq:kappa})].
The trajectory expansion described in Section \ref{sec:Method} results
in an expansion in $\ku$, valid not only at small and large Stokes numbers,
but also at Stokes numbers of order unity. The expansion works well in region $3$  (Section \ref{sec:finiteStKu}).}
\end{figure}

Fig.~\ref{fig:pd} shows the parameter plane of the statistical model for ${\rm F}=0$,
 the $\ku$-$\st$-plane. In different regions of this diagram, different approximations are employed to evaluate Eq.~(\ref{eq:suml}), and to compute individual Lyapunov exponents.
At small Stokes numbers (region 1) perturbation theory in $\st$ is appropriate. In the white-noise limit (\ref{eq:wnlimit}), region 2, diffusion approximations can be used.
In region $3$ a perturbation expansion around deterministic particle paths allows to evaluate Eq.~(\ref{eq:suml}), individual Lyapunov exponents, and the fractal Lyapunov dimension for small and large Stokes numbers.  We begin by discussing the simplest case, the limit of particles
advected at $\st=0$. When the velocity field is incompressible there is no clustering in this limit. But a compressible component causes fractal clustering by multiplicative amplification.

\subsubsection{Clustering of particles advected in compressible velocity fields.}
\label{sec:adv}
Particles advected in a smooth incompressible random velocity field at $\st=0$
must become and remain uniformly distributed.
Yet in a smooth compressible flow
an initially uniform scatter of particles does not remain uniform.
An example of this effect is  described by
Sommerer and Ott \cite{Som93} who performed experiments observing fluorescent tracer particles floating
on the surface of an unsteady flow. The floating particles experience local up- and
down-welling regions as sources and sinks, corresponding to a compressible surface flow.
Sommerer and Ott demonstrate that the particles form fractal patterns, and
characterise these patterns in terms of their Lyapunov fractal dimensions.
Refs.~\cite{Bof04,Cre04} report on  direct numerical simulations of tracer particles floating on the surface
of a turbulent flow, and quantify the clustering of the particles in terms of the fractal correlation dimension
(capillary effects may give rise to different clustering patterns on water waves~\cite{Fal05}).

At $\st=0$ the particles are constrained to follow the flow, $\ve v = \ve u$  and $\ma Z=\ma A$, as mentioned directly below Eq.~(\ref{eq:defZ}).
If the flow has a compressible component then $\langle \tr\,\ma Z\rangle_\infty=\langle \tr\,\ma A\rangle_\infty = \langle \ve \nabla\cdot \ve u\rangle_\infty<0$, and this implies spatial clustering.
Many authors have studied spatial clustering in this limit by numerically computing Lyapunov exponents of particles advected in turbulent, random, or chaotic velocity fields.

Analytical calculations are difficult for velocity fields that possess finite time correlations.
But in the limit where the fluid-velocity field varies rapidly in time ($\ku\ll 1$)
more is known. In this limit the Lyapunov exponents can be calculated using diffusion approximations \citep{Jan85}.
For particles advected in Kraichnan velocity fields at $\st=0$
the Lyapunov exponents were obtained in Ref.~\cite{Che98},
the results are reviewed in Section~II.C.3 of Ref.~\citep{Fal01}.
The Lyapunov exponents describe fluctuations of
small separations between particles, much smaller than the smallest length
scale in the Kraichnan model.
Inertial-range fluctuations are thus not relevant in the limit of small Kubo numbers. Consequently, the results for smooth random velocity fields of the form (\ref{eq:velocity_field})
and for the Kraichnan model are equivalent in this limit.
In the notation employed in this review, the Lyapunov exponents
are given by
\begin{equation}
\label{eq:ladv}
\lambda_\mu\tau  = \ku^2\, \frac{d(d+1-2\mu)+\beta^2(d-4\mu)}{d(d-1+\beta^2)}
\end{equation}
for $\mu = 1,\ldots, d$, and $\beta$ is the compressibility parameter introduced in Section~\ref{sec:Model}.
It follows from Eq.~(\ref{eq:ladv})  that the maximal Lyapunov exponent $\lambda_1$
is positive for $\beta < {\beta_{\rm c}}$ and negative for $\beta > {\beta_{\rm c}}$ where
$\beta_{\rm c}={\sqrt{d(d-1)/(4-d)}}$.
In other words, a \lq path-coalescence transition\rq{} occurs
in two and three spatial dimensions.

When $\lambda_1>0$ particle pairs tend to diverge and the distribution of particle positions approaches a fractal
for velocity fields with a compressible component.  From Eq.~(\ref{eq:ladv}) one finds
\begin{equation}
\label{eq:dL2}
\dL = d-\frac{d+2}{d-1} \beta^2 + \ldots
\end{equation}
for the Kaplan-Yorke dimension (\ref{eq:dL1}) for $d>1$,
and to second order in $\beta$.
Eq.~(\ref{eq:dL2}) shows that the fractal dimension tends to the spatial dimension $d$ as $\beta\rightarrow 0$.
This is expected since particles advected in incompressible velocity fields at $\st=0$ cannot cluster.
In one spatial dimension
Eq.~(\ref{eq:dL2}) does not apply. The one-dimensional velocity field is potential, it is independent of $\beta$ according
to Eqs.~(\ref{eq:velocity_field}a,d), and $\dL=0$.

\subsubsection{Fractal clustering at small Stokes numbers}
\label{sec:expadv}
When the Stokes number is not zero then inertia allows the particles to detach from the flow, and this may result
in clustering, $\langle \tr\,\ma Z\rangle_\infty < 0$.
For small Stokes numbers one can compute $\langle \tr\,\ma Z\rangle_\infty$ as a systematic expansion in $\st$.
This expansion is most conveniently expressed in the dimensionless variables (\ref{eq:dim1}). To lowest order we find:
\begin{equation}
\label{eq:nablav0}
\langle\tr\,\ma Z'\rangle_\infty \approx -\ku\,\st^2\frac{\partial}{\partial\st}\langle {\tr\,{\ma A'}^2}\rangle_\infty\Big|_{\st=0}\,.
\end{equation}
Using the trajectory expansion for the statistical model (Section \ref{sec:Method})
we find to order $\ku^2$:
\begin{equation}
\label{eq:trA2_av}
\langle {\tr\,{\ma A'}^2}\rangle_\infty=
\frac{(d+1)(d+2)\ku^2\,\st}{d(d-1)(1+\st)^2(1+2\st)}\,.
\end{equation}
According to Eq.~(\ref{eq:nablav0}) $\tr\, \ma Z$  is negative on average if the derivative of $\langle \tr\,{\ma A}^2\rangle_\infty$ with respect to the Stokes number is positive.
It follows from (\ref{eq:trA2_av}) that $\langle \tr\,{\ma A}^2\rangle_\infty$ vanishes at $\st=0$ and is positive at finite values of the Stokes number.
Negative values of $\langle \tr \,\ma Z\rangle_\infty$ thus correspond to $\langle \tr\,\ma A^2\rangle_\infty>0$.
This means that particles with small $\st$ are on average attracted to regions where $\tr\,{\ma A}^2>0$.
So where in the flow do the particles go?
This question was answered by Maxey \cite{Max87} by writing $\tr\, \ma A^2$ in terms of the symmetric and anti-symmetric parts
of $\ma A$, namely the strain-rate matrix $\ma S \equiv (\ma A+\ma A^{\sf T})/2$ and the rotational part $\ma O \equiv (\ma A-\ma A^{\sf T})/2$:
\begin{equation}
\label{eq:SO}
\tr\, \ma A^2  =  \tr\, \ma S^{\sf T} \ma S-\tr\, \ma O^{\sf T} \ma O\,.
\end{equation}
Since the terms $\tr\, \ma S^{\sf T} \ma S$ and ${\tr\, \ma O^{\sf T} \ma O}$ are both positive it follows
that particles tend to be expelled from vortical regions and preferentially sample regions of large strain rate. Preferential sampling is
discussed in more detail in Section \ref{sec:prstr}. We note here that
this appealing picture is valid for small Stokes numbers only.
Higher-order corrections to (\ref{eq:nablav0}) result in more complicated relations between $\langle \tr\,\ma Z\rangle_\infty$ and the fluid-velocity statistics.
Terms corresponding to higher orders in the Stokes number depend on time derivatives of the fluid-velocity gradients.
This has a significant consequence. Clustering can no longer be explained solely in terms of instantaneous fluid-velocity gradients, the history of the fluid-velocity gradients becomes important, too \cite{Gus11a}.

Returning to the limit of small Stokes numbers, consider now the fractal Lyapunov dimension.
Using the trajectory expansion for the statistical model (Section \ref{sec:Method}) one finds to lowest order in $\st$ and $\ku$ for $d>1$:
\begin{equation}
\label{eq:DeltaL0}
\dL =  d-\frac{(d+1)(d+2)}{d(d-1)} \ku^2\st^2\,.
\end{equation}
We infer that the fractal dimension deficit $\Delta_{\rm L} = d-\dL$ vanishes as $\sim\st^2$ in the limit $\st\rightarrow 0$.
This prediction has been verified
by numerical model calculations (inset of Fig.~4 in Ref. \cite{Bec05a}, see also Ref.~\cite{Wil07}).

Comparing Eqs.~(\ref{eq:DeltaL0}) and (\ref{eq:dL2}) suggests the following approximation.
For small values of $\st$ and $\ku$ one approximates the inertial dynamics by
advection in an \lq effective\rq{} Gaussian random velocity field $\ve v_\beta$ of the form (\ref{eq:velocity_field_2d},\ref{eq:velocity_field_3d}). The compressibility of this vector field,
\begin{equation}
\label{eq:beta2}
 \beta^2 = \frac{d+1}{d}\ku^2\st^2\,,
\end{equation}
is induced by weak particle inertia \cite{Elp02,Elp96,Bal01,Fal02,Fal04,Wil07}. In this model the particle-number density $n(\ve x,t)$ evolves according to the continuity equation
\begin{equation}
\label{eq:hd}
\partial_t n + \ve \nabla (\ve v_\beta n) = 0\,.
\end{equation}
Eq.~(\ref{eq:hd}) relates fluctuations of the particle-number density $n(\ve x,t)$ to those of a
random  velocity field $\ve v_\beta(\ve x,t)$.
Eq.~(\ref{eq:hd}) allows to compute small-scale clustering using the methods mentioned in Section \ref{sec:adv},
leading to a prediction for the $\st$-dependence of the correlation-dimension deficit $\Delta_2\equiv d-d_2$ that is analogous to (\ref{eq:DeltaL0}),
namely $\Delta_2 \sim \st^2$ \cite{Bal01,Fal02,Fal04,Gus11a}.
The same law was obtained
by Chun {\em et al.}~\cite{Chu05},
and also by Zaichik and Alipchenkov~\cite{Zai03}. All theories predict $\Delta_2 \sim \st^2$
at small Stokes numbers, but the prefactors differ \cite{Fal04}.

The origin of these differences is not entirely clear. There are several possible reasons. First,
the numerical factors in Eqs.~(\ref{eq:dL2}), (\ref{eq:DeltaL0}), and (\ref{eq:beta2}) are not exactly known in turbulence, Refs.~\cite{Chu05} and \cite{Zai03} use different approximation schemes. Second, the mapping of the inertial problem to the compressible-flow problem (\ref{eq:hd}) cannot be exact.  The trajectory expansion (Section \ref{sec:Method}) shows that the distributions of the strain and rotation matrices evaluated along the particle paths are not in general Gaussian \cite{Gus13b}, not even in the statistical 
model. This implies that the fluctuations cannot be represented by an \lq effective\rq{} Gaussian model. We conclude that approximations based on the assumption that the fluctuations of $\tr\,\ma A^2$ along particle paths can be described by advection in a compressible Gaussian random velocity field $\ve v_\beta$ cannot be entirely correct. Third,  the arguments outlined above refer to
small Stokes numbers,  but in deriving (\ref{eq:beta2}) diffusion approximations are employed that require
the limit $\tau\to 0$ to be taken. To ensure that $\st$ remains
small, the limit $\gamma\to\infty$ must be taken at the same time. It is not immediately obvious how to do this in a consistent way.
These remarks show why it is important to develop systematic
methods to calculate the fractal dimension, that do not rely on a mapping to compressible-flow models. One such method
(the trajectory expansion) is described in Section \ref{sec:Method}.

\subsubsection{Fractal clustering in the white-noise limit}
\label{sec:WNL}
\begin{figure}
\begin{overpic}[height=5cm,clip]{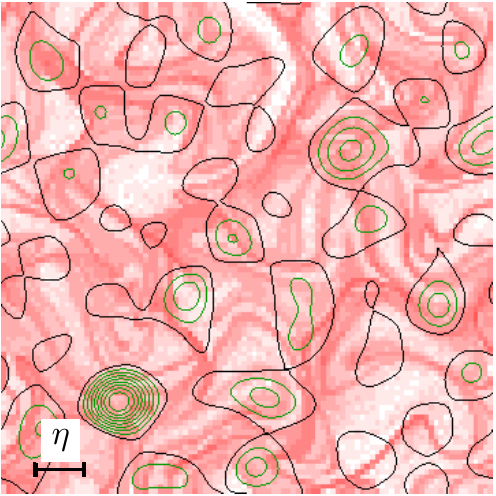}
\end{overpic}
 \caption{\label{fig:wnl}  Spatial clustering in the white-noise limit. Results obtained
from numerical simulations of the statistical model for a two-dimensional incompressible random velocity
field. Parameters: $\ku=0.1$, $\st=10$, $F=0$.
Data is plotted as in Fig.~\ref{fig:mechs}{\bf a}.
There a no apparent correlations between straining regions and particle positions.
}
\end{figure}
The full inertial dynamics (\ref{eq:langevin_dimless_2}) in incompressible flows is difficult to analyse in general. But in the white-noise limit (\ref{eq:wnlimit}) the problem simplifies substantially. In this limit
the fluid velocity $\ve u(\ve x,t)$ fluctuates so rapidly that vortical regions
do not persist long enough to efficiently push out particles.  Nevertheless strong clustering is observed in this limit (Fig.~\ref{fig:wnl}). It cannot be due to the centrifuge effect.
In this limit fractal clustering is explained by  \lq multiplicative amplification\rq{}: it follows from Eq.~(\ref{eq:calVt}) that small-scale clustering is the result of the sequence of random expansions and contractions of volume elements, determined by a random product of expansion and contraction factors
$\det[\ma I+\ma Z\big(\ve x(t),t)\delta t]$. Eq.~(\ref{eq:langevin_Zmatrix_0}) shows that the dynamics of the matrix $\ma Z$ of particle-velocity gradients is driven 
by the fluid-velocity gradients $\ma A$ evaluated at the particle position. This means
that $\ma Z$ depends upon the history of fluid-velocity gradients encountered by the particles.
But in the white-noise limit the distribution of $\ma A$ is unbiased and the elements of $\ma A$ are uncorrelated in time, so that diffusion approximations can be used to compute the time evolution of $\ma Z$ from Eq.~(\ref{eq:langevin_Zmatrix_0}). In one spatial dimension one finds for the Lyapunov exponent in the white-noise limit \citep{Wil03,Gus13a}
\begin{equation}
\frac{\lambda_1}{\gamma} = -\frac{1}{2}{\mathcal Re}\left[\frac{ \frac{\rm d}{{\rm d}y}{\rm Ai}(y)}{\sqrt{y} {\rm Ai}(y) }+1\right]\Bigg|_{y=[-1/(8\varepsilon^2)]^{2/3}}\,,
\label{eq:lyapunov_1d_WN}
\end{equation}
where ${\mathcal Re}$ the denotes real part, ${\rm Ai}$ is the Airy function and $\varepsilon^2 = 3 \ku^2 \st$.
Expanding Eq.~(\ref{eq:lyapunov_1d_WN}) for small $\varepsilon$ yields $\lambda_1/\gamma  \sim -\varepsilon^2$.
When the Lyapunov exponent is negative, separations between two neighbouring particles must eventually contract.
This small-$\varepsilon$ regime is therefore a \lq path-coalescence regime\rq{} \cite{Wil03}.
A path-coalescence transition occurs where $\lambda_1$ changes sign,
at $\varepsilon_{\rm c}^2 \approx 1.77$.
Eq.~(\ref{eq:lyapunov_1d_WN}) was also obtained by Piterbarg~\cite{Pit01} who  surmised that
the expression (\ref{eq:lyapunov_1d_WN}) equals the maximal Lyapunov exponent
in two spatial dimensions in the white-noise limit.
However it is shown in Ref.~\cite{Meh04} that this is not the case.

In higher spatial dimensions exact expressions for the Lyapunov exponents have not yet been found,
not even in the white-noise limit. We are forced to seek approximate solutions of the corresponding diffusion equation. This equation is equivalent to the quantum-mechanical problem of $d^2$ interacting harmonic oscillators. Algebraic perturbation theory
in the parameter $\varepsilon^2$ allows to compute series expansions for the Lyapunov exponents
(Section \ref{sec:Method}).
In this way Lyapunov exponents have been calculated in different spatial dimensions for different degrees of compressibility \citep{Wil03,Meh04,Meh05,Wil05,Dun05,Wil07}.
To order $\varepsilon^4$ these results can be summarised as follows:
\begin{eqnarray}
\label{eq:lwn}
\frac{\lambda_\mu}{\gamma} &=&
\frac{d(d+1-2\mu)+\beta^2(d-4\mu)}{d-1+3\beta^2}\,\varepsilon^2\\
     &-&\Big[{\frac { ( 3\,{d}^{2}\!-\!(6\,\mu\!-\!2)\,d+(3\,{\mu}\!-\!2)\,\mu )
 ( d\!+\!1\!+\!{\beta}^{2} ) ^{2}}{ ( d-1+3\,{\beta}^{2}) ^{2}}}\nn\\
      &&+{\frac { ( -(2+8\,\mu)\,d+(6\,\mu+4)\,{\mu}
 )  ( d\!+\!1\!+\!{\beta}^{2} ) }{d-1+3\,{\beta}^{2}}}+2\,\mu +3\,{\mu}^{2}\Big] \,\varepsilon^4\,.\nn
\end{eqnarray}
The lowest order in $\varepsilon$ in Eq.~(\ref{eq:lwn}) agrees with Eq.~(\ref{eq:ladv}).
It is worth noting though that Eq.~(\ref{eq:lwn})
is valid at large Stokes numbers (region $2$ in Fig.~\ref{fig:pd}),
whereas Eq.~(\ref{eq:ladv}) was derived
in region 1 in Fig. \ref{fig:pd}. Eq. (\ref{eq:lwn})
corresponds to the underdamped limit of the problem, while Eq. (\ref{eq:ladv})
was derived in the overdamped limit. Ref.~\cite{Wil07} explains why the Lypapunov exponents are the same in these two limits, to lowest order in $\ku$. Higher orders differ.

From the series expansions for the Lyapunov exponent we can calculate the Lyapunov dimension using the Kaplan-Yorke formula (\ref{eq:dL1}).
Extending  Eq.~(\ref{eq:lwn}) to order $\varepsilon^6$ for $d>1$ and $\beta=0$
one finds:
\begin{equation}
\label{eq:dLd}
 \dL = d- \frac{(d+2)(d+1)}{d-1} \,\varepsilon^2
- \frac{(d+2)(d+1)(d^2+3d-20)}{(d-1)^2}\,\varepsilon^4+\cdots\,.
\end{equation}
Eq.~(\ref{eq:dLd}) describes fractal clustering in the white-noise limit. Since $\varepsilon^2\propto \ku^2 \st$ the dimension deficit of the Lyapunov dimension is proportional to $\st$
to lowest order:
\begin{equation}
\label{eq:DeltaL_WN}
\DeltaL \equiv d-\dL \propto \ku^2 \st\,.
\end{equation}
In region 1 in Fig. \ref{fig:pd}, by contrast, the dimension deficit scales as $\st^2$, Eq.~(\ref{eq:DeltaL0}).  We conclude that the fractal dimension differs in regions 1 and 2.

The algebraic perturbation theory leading to (\ref{eq:lwn}) can be iterated to high orders:
\begin{equation}
\label{eq:ptGamma}
\lambda_\mu/\gamma = \sum_{k} c_k^{(d,\mu)}\!(\beta) \,\varepsilon^{2k}\,.
\end{equation}
The coefficients for $k=1$ and $2$ are given in Eq.~(\ref{eq:lwn}).
High-order coefficients $c_k^{(d,\mu)}\!(\beta)$ in two and three spatial dimensions were computed in Refs.~\cite{Meh04,Meh05,Dun05,Wil07}.
The results  indicate that the series are asymptotically divergent,
a common property of perturbation expansions.
Under certain conditions asymptotically divergent series can
be resummed to yield highly accurate results even for large values of $\varepsilon$.
In Section \ref{sec:Method} we explain
briefly how the series can be resummed,
and list the open questions that remain to be answered concerning perturbation expansions in the white-noise limit.

\subsubsection{Clustering at finite Stokes and Kubo numbers}
\label{sec:finiteStKu}
Spatial clustering was explained in Sections \ref{sec:expadv} and \ref{sec:WNL} in two
different limits, at small and large Stokes numbers.
The question is: how can one understand and quantify clustering in turbulent aerosols
at Stokes numbers of order unity where the flow and particle relaxation
times are of the same order, allowing the particles to react strongly to the flow?

In order to pin down the clustering mechanism at finite Stokes numbers one may try to expand around
the advective limit by computing higher-order corrections to Eq.~(\ref{eq:nablav0}).
This is possible (Section \ref{sec:Method}). But such higher-order corrections consist of more and more complicated combinations of gradients and time-derivatives
of the fluid-velocity field. This shows that clustering at finite Stokes numbers
is not simply determined by $\tr\, \ma A^2$, as in Eq.~(\ref{eq:nablav0}), and it raises the question which fluid-velocity
configurations are preferentially sampled at finite Stokes numbers.

But there is a second, more significant difficulty. The fact that time derivatives of the fluid-velocity field arise implies that the clustering mechanism is not instantaneous.
By contrast, small-scale clustering is in general determined by the
history of fluid-velocity gradients $\ma A(\ve x(t),t)$ evaluated along particle paths $\ve x(t)$,
as Eqs.~(\ref{eq:langevin_Zmatrix_0}) and (\ref{eq:calVt}) show.
\begin{figure}[tp]
\begin{overpic}[height=5cm,clip]{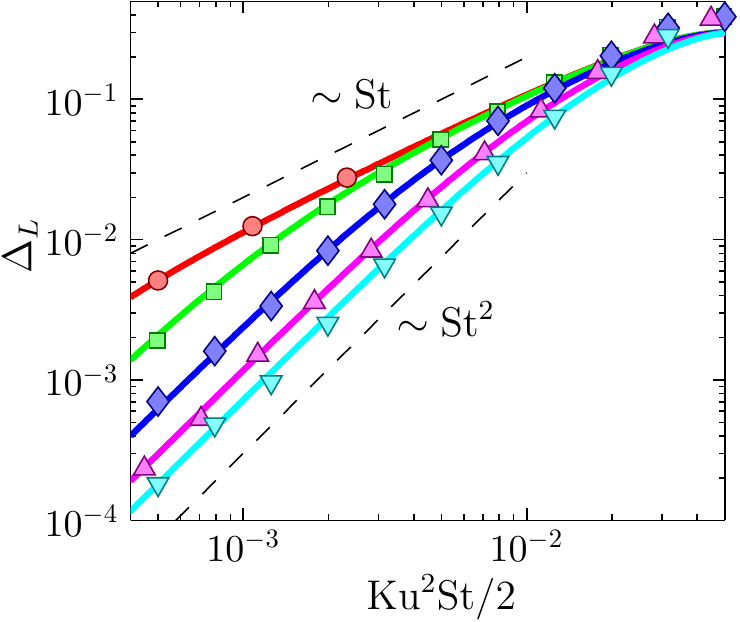}
\end{overpic}
\caption{\label{fig:Lyapunov_dimension}
Fractal dimension deficit $\DeltaL$ for the statistical model (two-dimensional incompressible random velocity field)
described in \Secref{sec:Model} as a function of $\varepsilon^2=\ku^2\st/2$.
Numerical results are shown as symbols. The theoretical results according to \eqnref{eq:DeltaLd} are
 plotted as solid lines (to order $\ku^4$).
Parameters: $\ku=0.02$ (red filled $\circ$), $\ku=0.05$ (green filled $\Box$), $\ku=0.1$ (blue filled $\Diamond$), $\ku=0.15$ (magenta filled $\vartriangle$), and $\ku=0.2$ (cyan filled $\triangledown$).  This Figure is similar to Fig.~2{\bf a} in Ref.~\cite{Gus11a}.}
\end{figure}
At finite Kubo and Stokes numbers, the distribution of $\ma A$ is biased
by preferential sampling, and its elements are correlated in time. As a consequence, to determine the time evolution of $\ma Z$, 
Eq.~(\ref{eq:langevin_Zmatrix_0}) must be integrated along with the particle dynamics. This is a difficult task.

The trajectory expansion explained in Section \ref{sec:Method} makes it possible to solve  this problem approximately at finite $\ku$ and $\st$. The method allows to compute $\langle\tr\, \ma Z\rangle_\infty$ as well as individual Lyapunov exponents as expansions in $\ku$,
at finite values of $\st$. For incompressible flows one finds to order $\ku^4$ for the Lyapunov dimension:
\begin{eqnarray}
\label{eq:DeltaLd}
\DeltaL&\equiv&d-\dL= \frac{(d+1)(d+2)\ku^2\st^2}{(d-1)d(1+\st)^3}(1+3\st+\st^2)
 \\&
 +&\frac{(d+1)(d+2)\ku^4\st^2\big(f(\st)+ d\, g(\st)+ d^2\, {h}(\st)\big)}{3d^2(d-1)^2(1+\st)^6(2+\st)^2(1+2\st)^2}\,.
 \nn
\end{eqnarray}
Here $f$, $g$, and $h$ are polynomials in $\st$ of order $10$ with integer coefficients.
One finds $f \sim -24 \,\st$, $g \sim 24+288 \,\st$, and  $h\sim 20+196\,\st$ for small values of $\st$
and $f \sim -240 \,\st^{10}$, $g \sim 36 \,\st^{10}$, and $h\sim 12\,\st^{10}$ for large values of $\st$.
A corresponding expression in two spatial dimensions ($d=2$) was first derived in Ref.~\cite{Gus11a}.
Eq.~(\ref{eq:DeltaLd}) is valid for small but finite Kubo numbers,
in region 3 in Fig.~\ref{fig:pd}. Region $3$ connects regions $1$ and $2$ in this Figure. The theoretical prediction \eqnref{eq:DeltaLd} is compared to data from numerical simulations of the statistical model in Fig.~\ref{fig:Lyapunov_dimension}. We observe excellent agreement between theory and simulations for the statistical model for the values of $\ku$ and $\st$ shown (the parameters lie in region $3$ in Fig.~\ref{fig:pd}).

Let us discuss several important features of Eq.~(\ref{eq:DeltaLd}).
First,
$\DeltaL$ vanishes upon setting  $\st=0$. This simply reflects the fact that particles suspended in an incompressible flow
do not cluster. Second, in the white-noise limit Eq.~(\ref{eq:DeltaLd}) reduces
to (\ref{eq:dLd}).
In this region the dimension deficit scales as $\varepsilon^2 \sim \ku^2 \st$,
linearly in Stokes number just as in Eq.~(\ref{eq:dLd}).
In this limit clustering is determined by the history of fluid-velocity gradients. Instantaneous flow configurations that
could give rise to rapid convergence of particles are too short-lived to have any effect at all.   Third, for small Stokes numbers one finds to order $\ku^4$:
\begin{equation}
\label{eq:dLMaxey}
\DeltaL \sim \frac{(d+1)(d+2)}{d(d-1)}\ku^2\st^2
+\frac{(d+1)(d+2)(5d+6)}{3d(d-1)^2}\ku^4\st^2\,.
\end{equation}
 This equation shows that the dimension deficit is $\propto \st^2$ for small Stokes numbers,  while we saw above that
 it is $\propto \st$ for large Stokes numbers, in the white-noise limit (Fig.~\ref{fig:Lyapunov_dimension}).
 This confirms the picture described in Sections \ref{sec:expadv} and \ref{sec:WNL}.

\begin{figure}[t]
\begin{overpic}[height=5.5cm,clip]{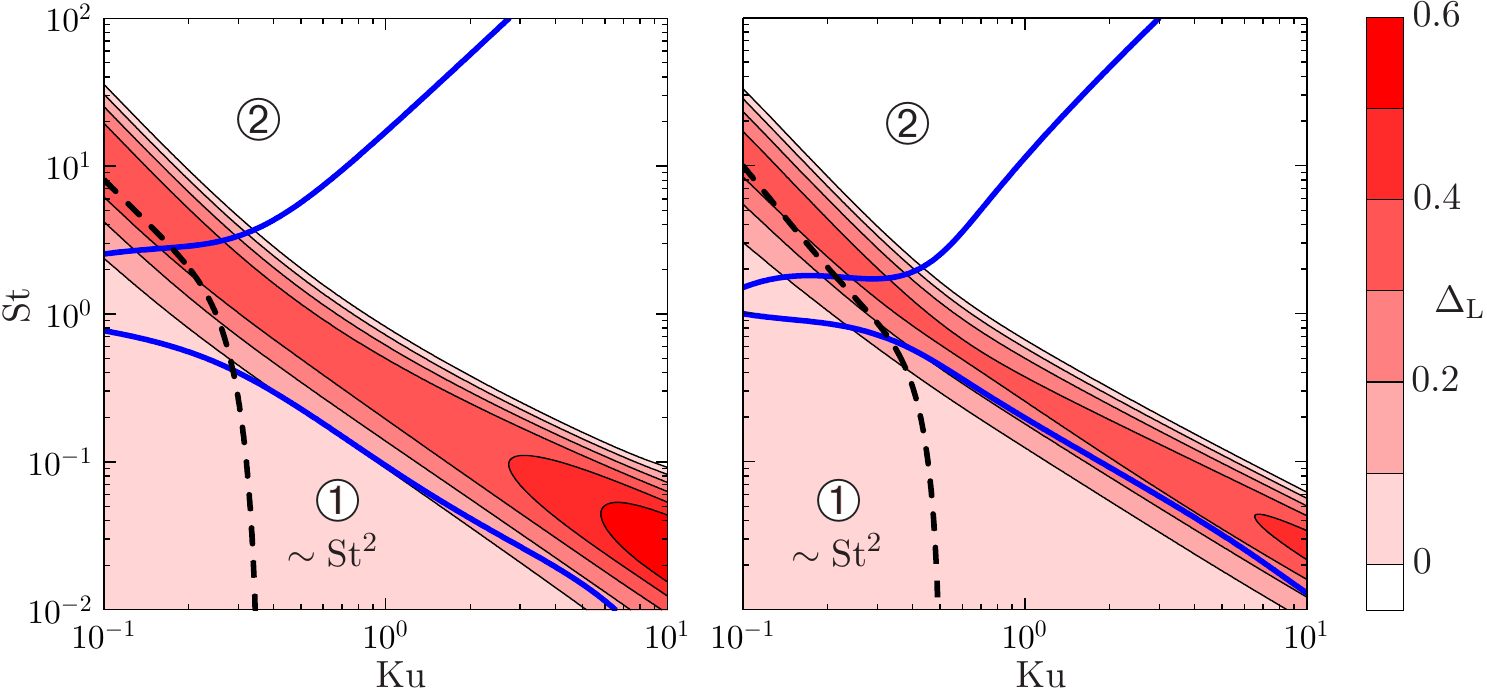}
\put(9,41.5){\bf a}
\put(52,41.5){\bf b}
\end{overpic}
\caption{\label{fig:phase_diagram}
Contour plot of Lyapunov dimension deficit $\Delta_{\rm L}$ in the statistical model, for an incompressible random velocity field in two ({\bf a}) and three ({\bf b}) spatial dimensions.
Regions 1 and 2 correspond to those in the schematic phase diagram in Fig.~\ref{fig:pd}.
Region 1 is bounded by a blue line below which the scaling $\Delta_{\rm L}\sim\st^2$ holds.
Region 2 is bounded by a blue line above which {$1+\lambda_1/\lambda_2$} ($d=2$) and {$1+(\lambda_1+\lambda_2)/\lambda_3$} ($d=3$)
are well approximated by white-noise results.
The above expressions equal $\Delta_{\rm L}$ when positive, in two and
three dimensions, respectively.
In region $2$ we have that $\Delta_{\rm L}\sim\st$ for small values of $\epsilon$.
To the left of the black-dashed lines the trajectory expansion (Section~\ref{sec:Method}) predicts $\Delta_{\rm L}$ with good accuracy.
}
\end{figure}

Fig.~\ref{fig:phase_diagram} shows contour plots
 of $\Delta_{\rm L}$ at finite Stokes and Kubo numbers,
 obtained from statistical-model simulations. Regions corresponding to strong small-scale clustering are dark red.
 The boundaries of regions $1$ and $2$ where the small-$\st$ and white-noise approximations work well (Fig.~\ref{fig:pd}) are shown as solid blue lines. They were determined by analysing where the numerical results exhibit the scaling $\Delta_{\rm L}\sim\st^2$, or by comparing with results of Langevin (white-noise) simulations.

Within the region bounded by the dashed line Eq.~(\ref{eq:DeltaLd}) works well
(region $3$ in Fig.~\ref{fig:pd}). When $\st$ is small the theory leading to Eq.~(\ref{eq:DeltaLd}) fails for values of $\ku$ larger than $\approx 0.2$.  For large values of $\st$ the theory fails when caustics become frequent.  This may be due to the fact that the theory treats the $\ma Z^2$-term in  \Eqnref{eq:langevin_Zmatrix_0} perturbatively
(Section \ref{sec:Method}).

Fig.~\ref{fig:phase_diagram} also illustrates the limiting cases mentioned in Section \ref{sec:Model}. In the white-noise limit (\ref{eq:wnlimit}) the fractal dimension deficit depends on $\ku$ and $\st$ only through the parameter $\varepsilon$, Eq.~(\ref{eq:defeps}). In the persistent-flow limit (\ref{eq:klimit}) $\Delta_{\rm L}$ depends on $\ku$ and $\st$ through the parameter $\kappa$, Eq.~(\ref{eq:kappa}).
In these two limits the contour lines approach curves of constant $\epsilon$ and $\kappa$ (\Figref{fig:pd}).

To conclude this Section we remark that the small-$\st$ result (\ref{eq:dLMaxey}) was derived by averaging over the multiplicative process
that determines the dynamics of the particle-velocity gradients $\ma Z$.
Eq.~(\ref{eq:dLMaxey}) agrees with Eq.~(\ref{eq:DeltaL0})
to lowest order in $\ku$. Since Eq.~(\ref{eq:DeltaL0}) was obtained by averaging $\tr\,\ma A^2$ at the particle positions we infer that fractal clustering becomes a local effect in the limit of $\st \to 0$. In this limit
the clustering mechanism relies on instantaneous correlations between particle positions and fluid-velocity structures. As the Stokes number tends to zero the fluid-velocity history experienced by the particles becomes less and less important.
Fractal clustering at small Stokes numbers is directly determined by preferential sampling. This is discussed in more detail in the next Section. But in general small-scale clustering is only indirectly affected by preferential sampling, it biases the process of multiplicative amplification.

\subsection{Preferential sampling}
\label{sec:prstr}
\subsubsection{Maxey's centrifuge mechanism}
\label{sec:mcm}
Maxey \cite{Max87} suggested an intuitive picture of preferential sampling at small Stokes numbers.
He approximated  the dynamics (\ref{eq:eom}) by advection in a \lq synthetic\rq{} $\st$-dependent
velocity field $\ve v_{\rm s}(\ve x,t)$. This field is obtained by expanding (\ref{eq:eom})
to first order in $\st$:
\begin{equation}
\label{eq:maxey}
\dot {\ve x} = \ve v_{\rm s}\,, \quad \ve v_{\rm s} = \ve u -\frac{1}{\gamma}\, \big( \partial_t \ve u + (\ve u\cdot \ve \nabla) \ve u\big)\,.
\end{equation}
Eq.~(\ref{eq:maxey}) shows that the divergence of $\ve v_{\rm s}$ need not vanish even though $\ve \nabla \cdot \ve u =0$.
Consequently, particles may aggregate in the sinks of the
synthetic velocity field $\ve v_{\rm s}$.
The divergence of this velocity field is given by
 $\ve \nabla \cdot \ve v_{\rm s} = -\ve \nabla\cdot \big[(\ve u\cdot \ve \nabla) \ve u\big]/\gamma$, so that
\begin{equation}
\label{eq:decompose}
\ve \nabla \cdot \ve v_{\rm s} = -\frac{1}{\gamma}\,\tr\, \ma A^2=
  \frac{1}{\gamma}(\tr\, \ma O^{\sf T} \ma O-\tr\, \ma S^{\sf T} \ma S)\,,
\end{equation}
where Eq.~(\ref{eq:SO}) was used.  Eq.~(\ref{eq:decompose}) implies that regions of large strain rate are sinks of the synthetic velocity field $\ve v_{\rm s}$.
This means that the particles preferentially sample straining regions of the flow
and avoid vortices (the corresponding sources).
This mechanism is an example of preferential sampling because it invokes correlations
between the local particle-number density
and the flow configuration at the same position and time.

To quantify the effect of this preferential sampling, Eq.~(\ref{eq:decompose})  must be averaged. It is important that $\tr\,\ma A^2$ must be evaluated along particle trajectories. Averaging $\tr\, \ma A^2$ at a fixed position $\ve x_0$ gives zero for a homogeneous and isotropic flow since the averages
of $\tr\,\ma S^{\sf T} \ma S$ and $\tr\, \ma O^{\sf T} \ma O$ are equal when evaluated
at a fixed position. Pinsky and Khain have evaluated the average of $\ve \nabla \cdot \ve v_{\rm s}$
along particle paths numerically in a statistical model for isotropic and homogeneous
turbulence not unlike the model described in Section \ref{sec:Model}, see Fig.~7 in Ref.~\cite{Pin97}.

In the statistical model the average along particle paths can be explicitly performed. At small Stokes numbers the result is given by Eqs.~(\ref{eq:nablav0}) and (\ref{eq:trA2_av}).
Let us compare \Eqnref{eq:nablav0} with (\ref{eq:decompose}).
Since Eq.~(\ref{eq:decompose}) averages to zero at $\st=0$ it follows that clustering is a second-order effect in $\st$.
This means that $\st^2$-corrections to (\ref{eq:maxey}) could contribute.
It turns out that this is not the case for homogeneous flows, consistent with \Eqnref{eq:nablav0}. But it must be emphasised that small-$\st$ clustering in inhomogeneous flows could be different.

\subsubsection{Probability of sampling straining regions}
\begin{figure}[tp]
\begin{center}
\includegraphics[height=5.5cm]{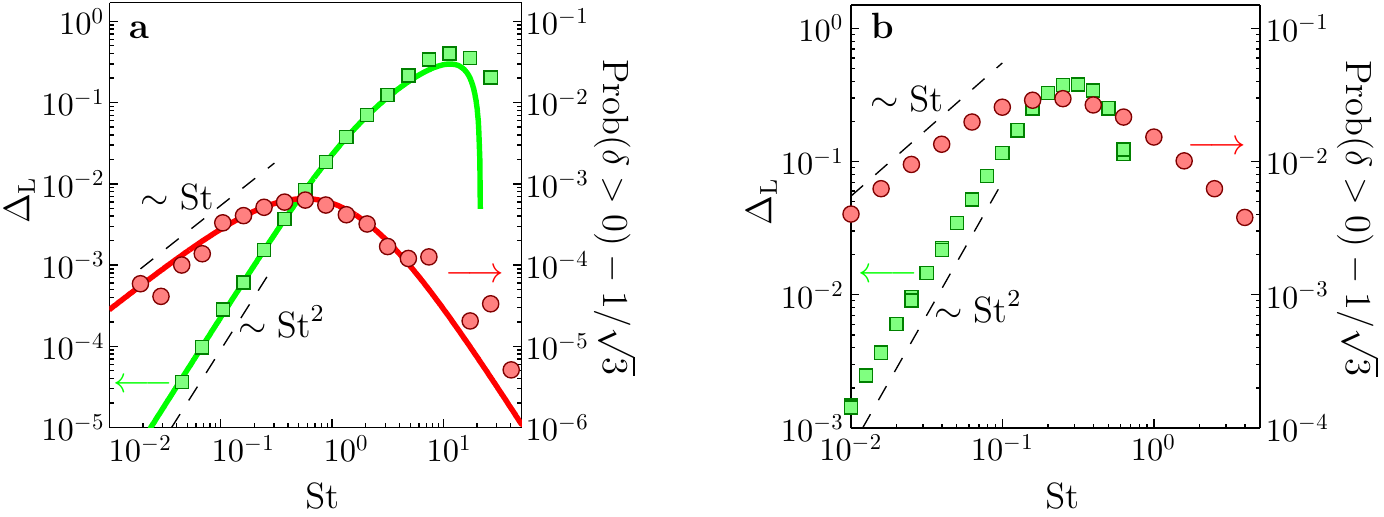}
\caption{\label{fig:Prob_Pos_Delta} Lyapunov dimension deficit $\DeltaL$ (green filled $\Box$) and
probability of sampling straining regions ${\rm Prob}(\delta>0)$ minus the probability for a straining region to occur in the incompressible fluid (red filled $\circ$) plotted against $\st$ for {\bf a} ${\rm F}=0$, $\ku=0.1$ and {\bf b} ${\rm F}=0$, $\ku=1$ in two spatial dimensions.
Symbols show results of numerical simulations of the statistical model.
The theory $\DeltaL$ is given by \eqnref{eq:DeltaLd} (to order $\ku^4$) and
the theory for ${\rm Prob}(\delta>0)$ is given by Eq.~\eqnref{eq:P_Delta_gt_0}.
The probability for a straining region to occur in the fluid is given by \Eqnref{eq:P_Delta_gt_0} with $\st=0$.
Coloured arrows indicate which axis corresponds to which data.}
\end{center}
\end{figure}
The trajectory expansion (Section~\ref{sec:Method}) allows to compute the probability of sampling straining
regions in the statistical-model flow. The calculation is simpler in two spatial dimensions, therefore we discuss this case here. The corresponding calculations in three dimensions have not yet been performed but we expect analogous results.
In two spatial dimensions the two eigenvalues $\sigma_\pm$ of the matrix $\ma A$ of fluid-velocity gradients
 are given by $\sigma_{\pm}=(\tr\,\ma A\pm\sqrt{2\tr\,\ma A^2-(\tr\,\ma A)^2})/2$.
The discriminant $\delta\equiv [2\tr\,\ma A^2-(\tr\,\ma A)^2] (\eta/u_0)^2$ reflects the local nature of the underlying flow: for $\delta >0$ the flow is straining ($\ma A$ has real eigenvalues), but for $\delta<0$ the flow is vortical (complex eigenvalues) corresponding to vortical regions in incompressible flows.

The probability ${\rm Prob}(\delta >0)$ for the particles to preferentially sample straining regions can be computed as follows.
One first uses Eqs.~(\ref{eq:velocity_field_2d}), (\ref{eq:normalisation}),  and (\ref{eq:velocity_field_C}) to find the distribution of $\delta$ to order $\ku^0$:
\begin{equation}
P_0(\delta)=\frac{1}{4}\sqrt{\frac{1+\beta^2}{3+\beta^2}}{\rm e}^{-\delta/4}\left[1-\Theta(-\delta){\rm erf}\left(\sqrt{-(3+\beta^2)\frac{\delta}{8}}\right)\right]\,.
\end{equation}
Here $\Theta$ is the Heaviside function.
In a second step an ansatz of the form
(\ref{eq:distz_1d_ansatz}) is used to determine finite-$\ku$ corrections, as described in Section \ref{sec:Method}.
To order $\ku^2$ one finds:
\begin{equation}
\label{eq:Pf}
P(\delta)=P_0(\delta)\left[1+\ku^2\frac{\st}{4(1+\st)^2(1+2\st)}\left(\delta - \frac{4\beta^2}{1+\beta^2}\right)\right]\,.
\end{equation}
By integrating $P(\delta)$ one obtains the probability to be in straining regions:
\begin{equation}
{\rm Prob}(\delta>0)=\sqrt{\frac{1+\beta^2}{3+\beta^2}}+\frac{\ku^2\st}{\sqrt{(1+\beta^2)(3+\beta^2)}(1+\st)^2(1+2\st)}\,.
\label{eq:P_Delta_gt_0}
\end{equation}
Fig.~\ref{fig:Prob_Pos_Delta}{\bf a} shows this probability as a function of $\st$ for an incompressible flow ($\beta=0$)
at $\ku=0.1$. Excellent agreement is observed with results of numerical simulations of the statistical model.
Also shown is the Lyapunov dimension deficit $\DeltaL$ indicating the strength of clustering.
The two curves have quite different functional forms. At small Stokes numbers
${\rm Prob}(\delta >0)$ is linear in $\st$, whereas $\DeltaL \propto \st^2$.
There is substantial clustering (strongest at $\st=10$), but the bias
of preferentially sampling straining regions
is very small for all Stokes numbers.
Panel {\bf b} shows the same quantities at $\ku=1$. The results were obtained
by numerical simulations of the statistical model. The picture is qualitatively similar to the case of $\ku=0.1$,
but the bias of preferentially sampling    straining regions is larger.

\subsection{Clustering by caustics}
\label{sec:caustics}
Consider the signed volume element ${\cal V}_t \equiv \det(\ve X_1,\ldots,\ve X_d)$ spanned by the $d$
separation vectors $\ve X_\mu$ ($\mu=1,\ldots,d$) between a test particle and $d$ nearby particles.
The determinant ${\cal V}_t$ fluctuates randomly as a function of time, a consequence of the impulses experienced by the particles as they move through the fluid. From time to time ${\cal V}_t$ may collapse
to zero for an instant of time.
These singularities occur
when two separation vectors become collinear. At this point the determinant of the deformation matrix $\ma J$ vanishes. This matrix
has elements
\begin{equation}
\label{eq:defJ}
J_{ij}(t) \equiv \frac{\partial x_i(t)}{\partial x_j(0)}\,,
\end{equation}
where $\ve x(t)$ is the position of a particle at time $t$ and $\ve x(0)$ is its initial position.
The determinant of $\ma J$ determines how ${\cal V}_t$ depends upon time:
\begin{equation}
\label{eq:VJ}
{\cal V}_{t}=\det\big(\ma J(t)\big)\, {\cal V}_{0}\,.
\end{equation}
Eq.~(\ref{eq:VJ}) is equivalent to Eq.~(\ref{eq:Vdot}) that describes the time evolution of ${\cal V}_t$ in terms of the trace of the matrix $\ma Z$ of particle-velocity gradients. The matrices $\ma Z$ and $\ma J$ are related by
\begin{equation}
\label{eq:ZJ}
\tr\,\ma Z=\frac{{\rm d}}{{\rm d}t} \log\det\ma J\,.
\end{equation}
Substituting this relation into \obs{(\ref{eq:Vdot})} yields \obs{Eq.~(\ref{eq:VJ})}.

\begin{figure}
\includegraphics[height=5cm,clip]{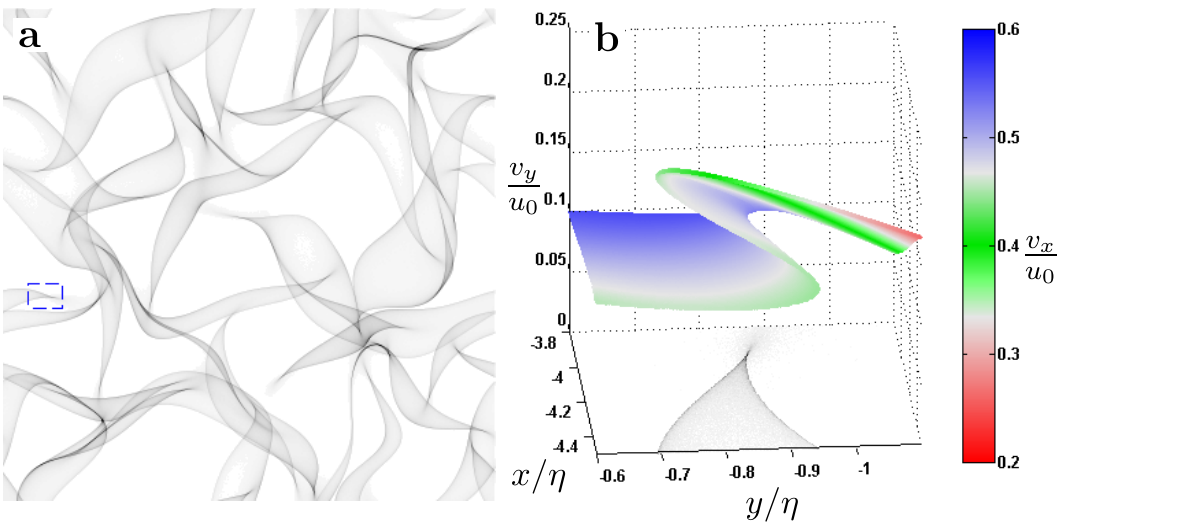}
\caption{\label{fig:caustic}{\bf a}
Particle-number density in the $x$-$y$-plane for particles in a two-dimensional incompressible 
velocity field (gray scale).  {\bf b} Shows folds of the phase-space manifold $v_y$ as a function of $x$ and $y$, corresponding to a small part
of the $x$-$y$-plane shown in panel {\bf a} (indicated by a dashed rectangle). Parameters: $\ku=1$, $\st=10$, $F=0$.}
\end{figure}

Fig.~\ref{fig:caustic}{\bf a} shows the particle-number density in a two-dimensional
incompressible random velocity field. The particles were initially at rest and uniformly distributed in space. Fig.~\ref{fig:caustic}{\bf a}
was obtained after integrating
the dynamics for eight correlation times.
The Figure demonstrates strong clustering caused by caustic singularities.

From Eq.~(\ref{eq:ZJ}) it follows
that the singularities $\det \ma J \rightarrow 0$ correspond to instances where $\tr\,\ma Z\rightarrow -\infty$~\cite{Gus12}.
In other words the phase-space manifold describing the spatial dependence of the particle velocities
folds over at caustic singularities, giving rise to a singularity in $\tr\,\ma Z$.
The fold singularity is a \lq catastrophe\rq{} in mathematical terms \cite{Arn92}.
An example of such a fold is shown in Fig.~\ref{fig:caustic}{\bf b}. Such folds have two consequences.

First, between caustic singularities the particle-velocity field becomes multi-valued, allowing for nearby particles to move at large relative velocities.
Large relative velocities between nearby particles were first observed
in numerical simulations of particle dynamics in turbulence in Ref.~\cite{Sun97}. Caustics can substantially increase the rate at which the particles approach~\cite{Fal02,Wil06,Meh07,Gus11b,Gus12,Gus13a,Gus13c}. Ref. \cite{Bew13} gives experimental evidence for the formation of caustics.

Second, at a given fold singularity the spatial particle-number density diverges, giving
rise to an algebraic tail in the distribution of particle-number densities.
That the particle-number density is enhanced is clearly visible in Fig.~\ref{fig:caustic}.
The rate of caustic formation $\mathscr{J}$ can be computed by mapping the problem onto a Kramers
escape problem \cite{Kra40} (\Secref{sec:Method}).  In one dimension the steady-state rate of caustic formation is obtained exactly in the white-noise limit \cite{Gus13a}
\begin{equation}
\label{eq:Jc}
\frac{\mathscr{J}}{\gamma} = \frac{1}{2\pi}  {\mathcal Im}\left[\frac{ \frac{{\rm d}}{{\rm d}y} {\rm Ai}(y) }{\sqrt{y} {\rm Ai}(y) }+1\right]\Bigg|_{y=(-1/(8\varepsilon^2))^{2/3}} \sim \frac{1}{{2\pi}} {\rm e}^{-1/(6\varepsilon^2)}\,,
\end{equation}
with $\varepsilon^2 = 3\ku^2\st$,
and where ${\mathcal Im}$ denotes the imaginary part.  At finite Kubo numbers $\mathscr{J}$ depends sensitively upon $\st$, too, but the
functional form is slightly different. The rate of caustic formation for the one-dimensional model at finite $\ku$ was computed in Ref.~\cite{Gus13a}.
In two and three spatial dimensions a similarly sensitive dependence upon $\st$ is observed \cite{Wil05,Dun05}.
In the white-noise limit and at small values of $\varepsilon$ the functional form of $\mathscr{J}$ is given by the asymptote (\ref{eq:Jc}), but the prefactors of $\exp[-1/(6\varepsilon^2)]$ differ by factors of order unity.

The caustic formation rate can be estimated from numerical simulations by integrating Eq.~(\ref{eq:langevin_Zmatrix_0}) along with the particle dynamics, to sample the steady-state frequency at which $\tr\ma Z$ passes $-\infty$.
When caustics occur frequently it is important to accurately evaluate the $\ma Z$-dynamics when elements of $\ma Z$ become large.  If caustics are rare, however, this is not as important. 
In this case the elements of $\ma Z$ can be set to zero (corresponding to the stable fixed point of the noiseless dynamics) whenever $\tr\ma Z$ passes a large negative threshold.

We remark that the close relation between the one-dimensional results for $\lambda_1$, Eq. (\ref{eq:lyapunov_1d_WN}), and for $\mathscr{J}$, Eq. (\ref{eq:Jc}), is no coincidence. A corresponding relation is known in the theory of one-dimensional disordered quantum systems \cite{Thouless,Hal65} (Section \ref{sec:Method}).

What are the consequences of caustic formation for the long-time evolution of the particle-number density?
Since caustics contract as a consequence of the dissipative nature of the phase-space dynamics their effect on particle clustering in the steady state is small
when the average particle-number density is small: caustics are rarely resolved.
Could caustics make an additional contribution to small-scale clustering besides that described by the Lyapunov exponents and the Lyapunov dimension (\ref{eq:dL1})?
In the white-noise limit this is not the case \cite{Wil07}. But at finite Kubo numbers and large Stokes numbers this remains to be proven.

\subsection{Gravitational settling}
\label{sec:gs}
\begin{figure}[t]
\begin{overpic}{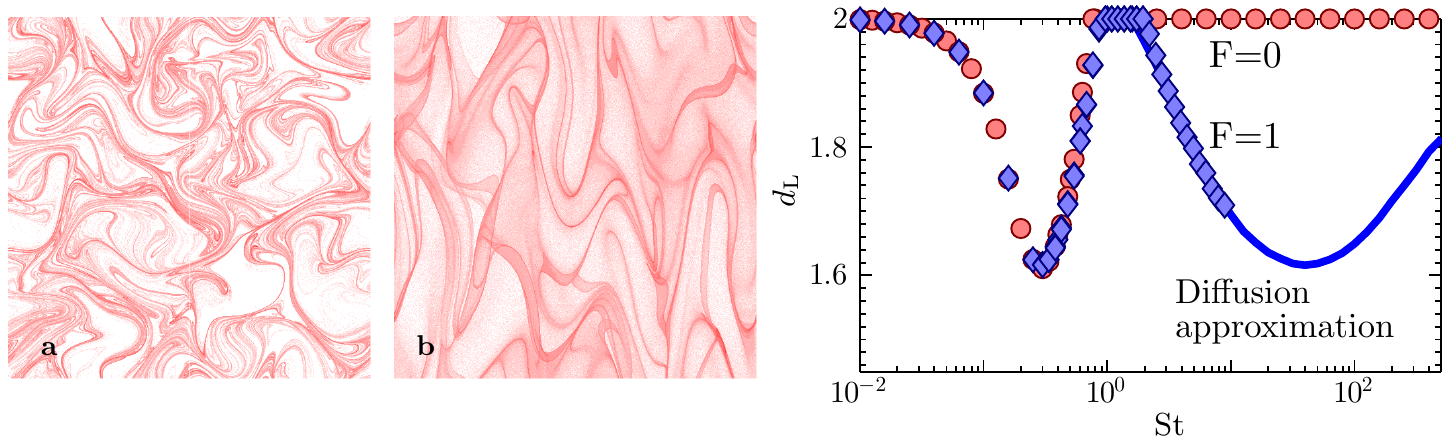}
\put(61,6){{\bf \small c}}
\end{overpic}
\caption{\label{fig:gs}
Spatial clustering of particles falling in a turbulent aerosol. Red regions show high particle-number densities.  {\bf a} slow settling, {\bf b} fast settling.
Computer simulation of the statistical model described in \Secref{sec:Model} in two spatial dimensions.
Parameters:  $\ku=1$, ${\rm F}=1$, $\st=0.2$ ({\bf a}) and $\st=10$ ({\bf b}).
Panel {\bf c} shows the fractal dimension for ${\rm F}=0$ (red filled $\circ$) and ${\rm F}=1$ (blue \obs{filled $\Diamond$)} as a function
of Stokes number. Adapted from Ref.~\cite{Gus14e}.
}
\end{figure}
When the turbulence is strong enough to suspend heavy particles,
then the parameter ${\rm F}$ [Eq.~(\ref{eq:StKu})] is small. Otherwise the particles settle through the turbulent aerosol.
How does settling affect spatial clustering? One might argue that the gravitational term in Eq.~(\ref{eq:stokes}) is just a constant,  and that spatial clustering is not affected by this term because it disappears upon taking the divergence of this equation, leaving Eq.~(\ref{eq:langevin_Zmatrix_0}) unchanged. It is correct that the gravity term does not affect
the form of (\ref{eq:langevin_Zmatrix_0}). But settling changes the paths the particles take through the flow. This changes the history of fluid-velocity gradients experienced by the particles. Gravitational settling may thus affect spatial clustering indirectly. Fig.~\ref{fig:gs} demonstrates the effect of gravitational settling on spatial clustering at ${\rm F}=1$. Panel ({\bf a}) shows spatial patterns formed by identical particles that settle slowly while the particles in panel ({\bf b}) settle faster.  We see that fast settling breaks the isotropy of the patterns.

Does settling decrease or increase the strength of fractal clustering? It has been argued \cite{Wan06,Fra07,Aya08a} that falling particles cluster less because Maxey's centrifuge is less efficient. This argument fails to take into account that clustering is in general determined by the history of the fluid-velocity gradients, and we note that Woittiez {\em et al.}~\cite{Woi09} have reported that settling may increase clustering.

To address this question it is necessary to compute how the time-series of fluid-velocity gradients changes due to settling. The trajectory expansion (Section \ref{sec:Method}) allows to achieve just that for the
statistical model \cite{Gus14e}. The resulting fractal dimension is shown in
Fig.~\ref{fig:gs}{\bf c} as a function of $\st$. At small Stokes numbers $d_{\rm L}$ is almost the same for ${\rm F}=0$ and  ${\rm F}=1$. At much larger values of the settling parameter ${\rm F}$, small-scale clustering at small Stokes numbers is reduced more. The results of Ref.~\cite{Gus14e} show that this is due to the fact that settling weakens preferential sampling.

A more striking effect occurs at larger Stokes numbers. The ${\rm F}=0$-particles remain homogeneously distributed at large Stokes numbers, their motion resembles that of kinetic-gas particles because caustics are abundant. But there is strong clustering when ${\rm F}=1$. In Ref.~\cite{Gus14e}
this surprising observation was explained as follows. Rapid settling at large Stokes numbers reduces the effective correlation time of the flow as seen by the particles. In this limit the fluid-velocity gradients encountered by the particles appear as a white-noise signal, and strong small-scale clustering may result from multiplicative amplification. Diffusion approximations can be used to compute the dynamics in this limit, with an effective parameter
$\varepsilon_{\rm eff}^2 = \varepsilon^2/(\ku\,\st\, {\rm F})^{3/2}$ instead of (\ref{eq:defeps}).
The solid line in Fig.~\ref{fig:gs}{\bf c}  is obtained by numerically integrating the corresponding diffusion equation.
But we remark that it is not enough to simply replace $\varepsilon\to\varepsilon_{\rm eff}$ in the results in Section~\ref{sec:WNL} to find the clustering in the limit considered here.
The reason is that the anisotropy induced by gravity changes the properties of the diffusion process~\cite{Gus14e}.

In summary gravitational settling may decrease fractal clustering somewhat
at small Stokes numbers, but it can substantially increase fractal clustering at large Stokes numbers, by multiplicative amplification.

\subsection{Summary}
\label{sec:sum}
Small-scale clustering is determined by the random (but correlated) product of expansion and contraction factors $\det[\ma I+\ma Z\big(\ve x(t),t)\delta t]$ that describe how an initially infinitesimally small cloud of particles to expand or contract in the long run. This process depends on the history of fluid-velocity gradients that the particles experienced in the past. This mechanism (multiplicative amplification) is most easily analysed in the white-noise limit where the fluid-velocity gradients seen by the particle can be approximated by a white-noise signal.

But the white-noise results do not directly apply in general, since the particle paths are biased by preferential effects at finite Kubo numbers. This can be taken into account perturbatively using the trajectory expansion described in Section \ref{sec:Method}. This method allows to compute how  the
fluid-velocity gradients sampled by the particles reflect preferential effects, in other words
how small-scale clustering is  indirectly affected by preferential sampling.

Preferential sampling also has a direct effect: it determines large-scale inhomogeneities in the particle-number density, on the scale of the turbulent eddies.  In general, these two effects - direct and indirect \ - are distinct. The first one is determined by instantaneous correlations between particle positions and fluid-velocity configurations, while the second one is determined by the history of fluid velocities seen by the particle. But as $\st$ tends to zero spatial clustering becomes an instantaneous effect, the flow history becomes less important. In this limit clustering is weak, yet the mechanisms causing spatial patterns on small and large scales become more and more similar.

When the turbulence is intense then the settling parameter ${\rm F}$ is small and small-scale clustering is only weakly affected by settling (Fig.~\ref{fig:gs}{\bf c}). At larger values of ${\rm F}$ settling may reduce preferential sampling at small Stokes numbers. At large Stokes numbers rapid settling can significantly increase clustering.

\section{Comparison with results of direct numerical simulations}
\label{sec:comparison}
In the previous Section we explained the mechanisms that cause steady-state spatial clustering
in the statistical model.  In this Section we show by comparison with numerical simulations of turbulent aerosols that
the statistical model yields, in some cases, a quantitative description of the numerical-simulation results for Navier-Stokes turbulence.

\subsection{Small-scale clustering}
\label{sec:compare}
Where in the phase diagram Fig.~\ref{fig:pd} should we place numerical results that are based on direct numerical simulations (DNS) of turbulence? This requires some discussion.

First, turbulence has finite time correlations. This means that we must evaluate the statistical model at finite Kubo numbers.

Second, DNS of incompressible fully developed turbulence refer to the \lq Kolomogorov time\rq{}. From DNS
results this time scale can be inferred in different ways. A common choice is
\begin{equation}
\label{eq:tauk}
\tau_{\rm K}\equiv \langle{\tr(\ma A}{\ma A}^{\sf T})\rangle^{-1/2}_\infty
\end{equation}
where the average is over Lagrangian fluid trajectories. Lagrangian correlation functions of turbulent fluid-velocity gradients are determined by this time scale, for example $\langle\tr[\ma S(t)\ma S(0)]\rangle_\infty \sim \exp[-t/(2\tau_{\rm K})]$ and $\langle\tr[\ma O(t)\ma O(0)]\rangle_\infty \sim \exp[-t/(7\tau_{\rm K})]$~\cite{Gir90}, provided that $t$ is at least of the order of $\tau_{\rm K}$. Therefore it is natural to express the Stokes number in terms of this  Lagrangian time scale \cite{Bec06}.
The relevant limit of the statistical model is therefore (\ref{eq:klimit}). In this limit the statistical-model dynamics is determined by the
time-scale $\eta/u_0$, and the correlations of the fluid-velocity gradients approach
$\langle\tr[\ma S(t)\ma S(0)]\rangle_\infty \sim \exp[-t/(2\tau_{\rm K})]$ and $\langle\tr[\ma O(t)\ma O(0)]\rangle_\infty \sim \exp[-t/(4.3\tau_{\rm K})]$.

Third,  in the statistical model the time-scale (\ref{eq:tauk}) evaluates to $\tau_{\rm K}=\eta/(u_0\sqrt{d+2})$. This shows
that the Stokes number used in numerical studies of particle dynamics in Navier-Stokes turbulence, $(\gamma\tau_{\rm K})^{-1}$, corresponds to the parameter $\kappa$ defined in Eq.~(\ref{eq:kappa}).

\begin{figure}[t]
\begin{overpic}[height=5.5cm,clip]{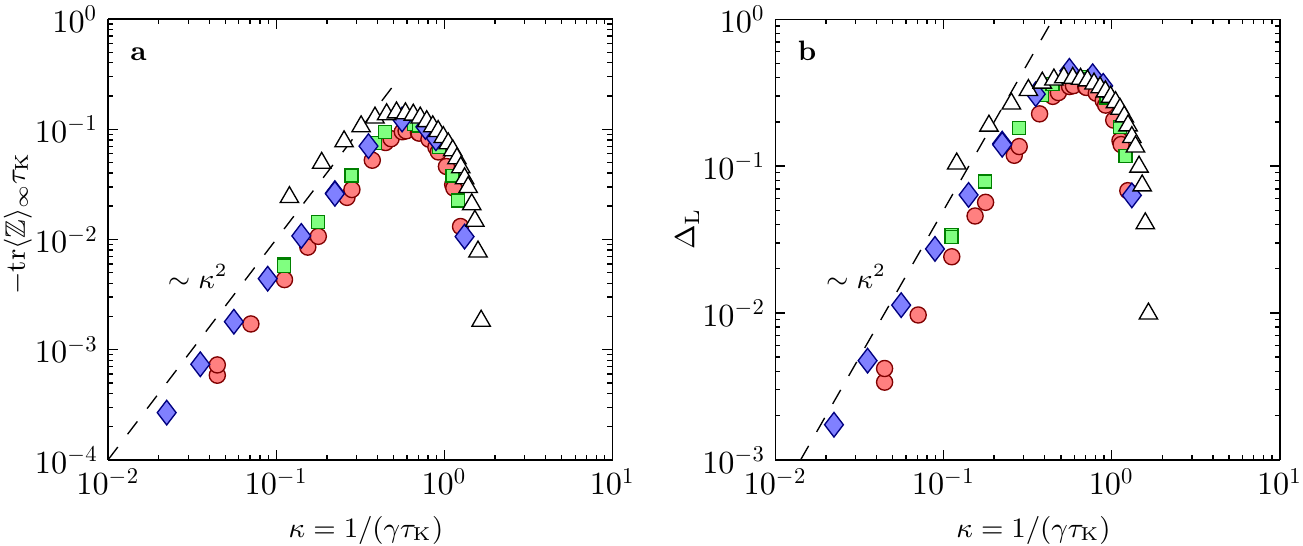}
\end{overpic}
\caption{\label{fig:compare}
Small-scale clustering,  $-\langle\tr\,\ma Z\rangle_\infty$ ({\bf a}) and the Lyapunov dimension deficit
$\Delta_{\rm L}$ ({\bf b}). Shown are results from statistical-model simulations (coloured symbols), and
from numerical simulations of heavy particles in Navier-Stokes turbulence.
Parameters: $\ku=2$ (red filled $\circ$), $\ku=5$ (green filled $\Box$), and $\ku=10$ (blue filled $\Diamond$).
Data from numerical simulations with ${\rm Re}_\lambda=185$ (empty $\vartriangle$) were taken from Ref.~\cite{Bec06}. Dashed lines show the asymptote $\sim\kappa^2$.
}
\end{figure}

\Figref{fig:compare} shows the comparison between small-scale clustering obtained through
simulations of the statistical model, and from results relying on DNS of turbulence~\cite{Bec06}. Shown are the average of $\langle \tr \ma Z\rangle_\infty$ and the fractal dimension deficit $\Delta_{\rm L}$ as functions of the Stokes number $\kappa$. The statistical-model results were obtained for $\ku=\{2,5,10\}$ and are roughly independent of $\ku$, as expected in this limit. We observe that the statistical-model predictions are very close to the DNS results. There is no fitting parameter. The deviations are largest at small values of $\kappa$, at small Stokes numbers. In this limit preferential sampling of straining regions becomes important. We shall see in the following Section that preferential sampling is qualitatively similar in the statistical model and in DNS of turbulence, but that there are quantitative differences.

\subsection{Preferential sampling}
\label{sec:pv}
At small Stokes numbers, numerical results based on DNS of turbulence show that  there is a strong anti-correlation between the particle locations  and vortical regions (Fig.~2(left) in Ref.~\cite{Bec06d}, at $\kappa = 0.16$).  At larger Stokes numbers this anti-correlation persists.
But one also observes that large parts of the straining regions have very small particle concentrations (Fig.~2(right) in Ref.~\cite{Bec06d}, $\kappa = 2.03$). This indicates that preferential sampling alone does not explain small-scale clustering, and supports the conclusions drawn from the statistical-model analysis: small-scale clustering is determined by the history of the fluid-velocity gradients encountered by the particles. It is this mechanism that gives rise to the small-scale filamentary structures in Fig.~\ref{fig:mechs}{\bf b} in a region where the strain is essentially constant.

\begin{figure}[t]
\begin{overpic}[height=5.5cm,clip]{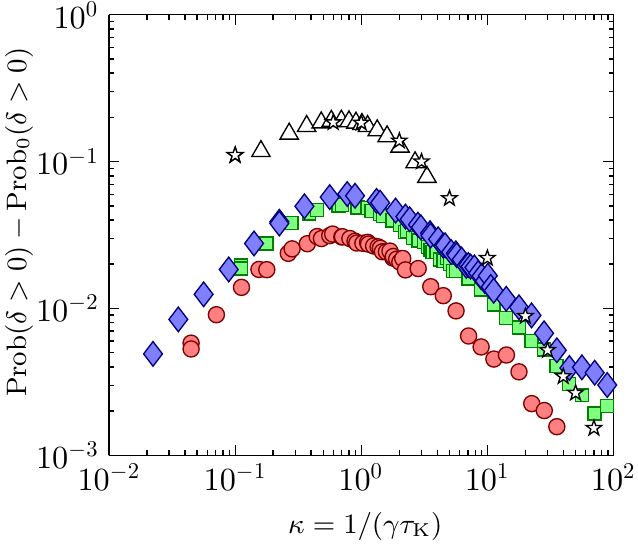}
\end{overpic}
\caption{\label{fig:compare2}Preferential sampling, probability ${\rm Prob}(\delta >0)$ of sampling straining regions minus the probability for a straining region in the fluid, ${\rm Prob}_{0}(\delta >0)$.
Comparison between simulation results of statistical model (coloured symbols) and from direct numerical simulations
of Navier-Stokes equations (empty symbols). Data with ${\rm Re}_\lambda=185$ taken from Ref.~\cite{Bec07} (empty triangles), complemented with data with ${\rm Re}_\lambda=400$ obtained from the DNS of Ref.~\cite{Bec10} (empty stars).  Parameters: $\ku=2$ (red filled $\circ$), $\ku=5$ (green filled $\Box$), and $\ku=10$ (blue filled $\Diamond$).
}
\end{figure}

Fig.~\ref{fig:Prob_Pos_Delta} quantifies the strength of preferential sampling of straining regions observed in the statistical model. The Figure shows how the probability of sampling straining regions in the flow depends on the Stokes number and on the Kubo number.
Numerical simulations of heavy-particle dynamics in turbulence
yield results that are qualitatively similar to the large-$\ku$ results
for the statistical model. This is demonstrated in Fig.~\ref{fig:compare2}.
We see that the statistical-model curves and those obtained
from numerical simulations of Navier-Stokes turbulence
show similar shapes. But the magnitude of the effect is smaller by a factor of $\approx 3$ in the statistical model.

There are at least two possible reasons for these deviations.
First, the correlation time of turbulent vortices is slightly larger than the corresponding
time in the statistical model at large Kubo numbers, by a factor of $\approx 1.6$ according to the numbers quoted in \Secref{sec:compare}.
This difference is probably caused by long-lived regions of high vorticity in turbulence (so-called \lq vortex tubes\rq{} \cite{She90}).
Vortex tubes are created by stretching of vortical structures by the turbulent shear, the vorticity in such regions can be substantially larger than its mean value, and these tubes can persist for many Kolmogorov times. It is plausible that the centrifuge effect is therefore somewhat
stronger in  Navier-Stokes turbulence than in the statistical model.
Second, the probability for a straining region to occur in the flow, 
${\rm Prob}_{0}(\delta >0)$, is smaller in the statistical model than in the DNS data.
In the statistical model we have ${\rm Prob}_{0}(\delta >0)=\sqrt{1/3}$, 
this follows from the first term on the r.h.s. of Eq.~(\ref{eq:P_Delta_gt_0}) for $\beta=0$.
The DNS data gives ${\rm Prob}_{0}(\delta >0)\approx 0.40$. Particles are therefore less likely to hit straining regions after being expelled from vortices or straining regions due to inertia.
Can the statistical model be improved to more quantitatively describe the preferential sampling of straining regions seen in Fig.~\ref{fig:compare2}?
This would require to more precisely account for the universal dissipative-scale fluctuations of turbulence~\cite{Sch14}.

We conclude this Section by discussing preferential sampling on length scales in the inertial range \cite{Bec07,Col09}.
The statistical model has only one length scale, the correlation length $\eta$. This model cannot answer the question
which flow properties are sampled on length scales in the inertial range.
Different suggestions have been put forward in the literature. Under which circumstances the different mechanisms apply depends on the Stokes number,
on the Reynolds number of the turbulence, and on the dimensionality. Vassilicos and collaborators~\cite{Che06,Vas08,Got06}
observe that heavy particles preferentially sample regions of
low acceleration in three-dimensional Navier-Stokes turbulence.  At not too large values of $\st$ this follows from \Eqnref{eq:maxey}. This equation can be written as
\begin{equation}
\label{eq:af}
\ve v_{\rm s} = \ve u -\gamma^{-1}\ve a_{\rm f}
\end{equation}
where $\ve a_{\rm f}$ is the fluid-acceleration field defined by $\ve a_{\rm f}\equiv\partial_t \ve u + (\ve u\cdot \ve \nabla) \ve u$.
Consider a particle in a region of small fluid acceleration,  $|\ve a_{\rm f}|\ll 1$ or $\ve v_{\rm s} \approx \ve u$. Provided that the Stokes number
is small enough one expects the particle to stick to this region as it is swept through the flow. This indicates that particles may preferentially sample regions of low fluid acceleration,
if the Stokes number is not too large.
For large Stokes numbers results of numerical simulations suggest
that preferential sampling of low fluid acceleration is negligible~\cite{Bra15b}.
The authors of Ref.~\cite{Bra15b} argue  instead that the particles sample regions of positive \lq coarse-grained\rq{} $\tr\ma A^2$, at Stokes numbers of order unity and larger. They show that at small Stokes numbers coarse graining is not important. In this case the mechanism reduces to that described in Section \ref{sec:expadv}, i.e.,
small-scale clustering because $\tr \ma Z$ is negative (corresponding to positive values of $\tr\ma A^2>0$).

The above discussion illustrates that particles in turbulence may preferentially sample different flow properties (for example $\ve a_{\rm f}\!=\!0$-regions or regions where $\ve\nabla\cdot\ve a_{\rm f}=\tr\ma A^2>0$.  The statistical-model calculations indicate that preferential sampling
may or  may not affect small-scale clustering, depending on the value of $\st$ and of other dimensionless parameters. At small Stokes numbers the bias due to preferential sampling is more important. This is consistent with Figs.~\ref{fig:compare} and \ref{fig:compare2}. Fig.~\ref{fig:compare2} shows that preferential sampling is stronger in DNS of turbulence compared with the statistical model. This affects small-scale clustering more at small Stokes numbers, less at larger Stokes numbers (Fig.~\ref{fig:compare}).

\subsection{Lyapunov spectrum in turbulent flows}
\label{sec:lambda_2}

Section \ref{sec:pv} briefly describes how intense vortex tubes in fully developed turbulence may affect the preferential sampling of particles.
Such intense, long-lived structures are absent in the statistical model. A further difference between the statistical model
and fully developed turbulence is the following: the statistical model is time-reversal invariant while turbulence is not~\cite{Pum14}.
The broken time-reversal invariance in turbulence implies that the second Lyapunov exponent $\lambda_2$ must approach a finite (positive) value as $\st\rightarrow 0$.
This can be seen in Fig.~1 in Ref.~\cite{Bec06}.  In the three-dimensional statistical model, by contrast, time-reversal invariance implies
that the second Lyapunov exponent approaches zero as the Stokes number tends to zero (Section \ref{sec:Method}).
But the second exponent is generally much smaller than $\lambda_1$ and $\lambda_3$.  Therefore breaking of time-reversal invariance affects small-scale clustering only weakly.

\subsection{Gravitational settling}
\label{sec:gravs}
The computer simulations of turbulent aerosols discussed in
Sections \ref{sec:compare}, \ref{sec:pv}, and \ref{sec:lambda_2} are based
on Eq.~(\ref{eq:eom}) in combination with direct numerical simulations of turbulence. Settling due to gravity is not considered (${\rm F}=0$).
Those numerical studies that take into account settling tend to show that
settling weakens {preferential sampling and small-scale }spatial clustering~\cite{Pum04,Wan06,Fra07,Aya08a,Woi09}. This effect is commonly discussed in terms
of Maxey's centrifuge. It is argued that clustering is reduced
because vortices do not have enough time to efficiently spin out the particles.
The statistical-model analysis shows that this mechanism applies at small Stokes numbers, and how it depends on the value of the parameter ${\rm F}$.
Details are given in Ref.~\cite{Gus14e}.

Woittiez {\em et al.}~\citep{Woi09} reported that settling may increase clustering when $\st$ is larger than order unity.
The statistical model provides a qualitative explanation of
this surprising effect. Small-scale clustering at large Stokes numbers is caused by multiplicative amplification. When the particles fall rapidly then the turbulent velocity gradients seen by the particle can be approximated as a white-noise signal. Diffusion theory shows that clustering in this limit is determined by the parameter $\kappa F$. This parameter can be of order unity at large Stokes numbers, causing small-scale clustering.

We conclude by commenting on the persistent-flow limit, \Eqnref{eq:klimit}. When gravitational settling
cannot be neglected then the dynamics is described by a second dimensionless parameter,
$v_{\rm T}=\kappa {\rm F}\st/\sqrt{d+2}$, in addition to $\kappa$. The new parameter $v_{\rm T}$ is the dimensionless terminal settling velocity in the absence of turbulent fluctuations ($\ve u=0$).
If this parameter is large, then the decay of fluid-velocity correlation functions is dominated by spatial displacements due to settling.
It was explained in \Secref{sec:compare} that the Stokes number used in numerical simulations of particles in turbulence, $(\gamma\tau_{\rm K})^{-1}$, corresponds to $\kappa$.
Similarly, the gravity parameter used in numerical simulations of particles in turbulence, $g\tau_{\rm K}/u_{\rm K}$, corresponds to $v_{\rm T}\eta/(\kappa^2\eta_{\rm K})$.

The Lyapunov dimension has not yet been calculated numerically for particles settling in Navier-Stokes turbulence. But the correlation dimension was computed in Ref.~\cite{Bec14}. The results are qualitatively
similar to those obtained in the statistical model, showing increased small-scale clustering at large values of $\st$ when ${\rm F}>0$.

Settling not only affects spatial clustering but also caustic formation.
Statistical-model calculations~\cite{Gus14e} show that the rate of caustic formation is reduced for large values of $\st$ when ${\rm F}>0$.
This is consistent with the simulation results discussed in Refs.~\cite{Jin15,Ire15}.  It would be of interest to compare statistical-model predictions in the persistent-flow regime with the recent DNS results in Refs.~\cite{Bec14,Jin15,Ire15}.

\section{Methods}
\label{sec:Method}
In this Section we describe the methods that allow to compute the results
quoted in Sections \ref{sec:measures}, \ref{sec:mechanisms}, and \ref{sec:comparison}. To keep the formulae simple, we concentrate on one spatial dimension
where the Stokes equation (\ref{eq:eom}) takes the form
\begin{eqnarray}
\dot x &=& v\,,\quad
\dot v = \gamma\,(u-v)\,.
\label{eq:langevin_1d_dimless}
\end{eqnarray}
To compute the Lyapunov exponent $\lambda_1$, 
we need to follow the dynamics of two initially nearby particles. Equations for the
distance $X\equiv x_2-x_1$ and the velocity difference $V\equiv v_2-v_1$ between the two particles are obtained by linearising Eq.~(\ref{eq:langevin_1d_dimless}) in $X$:
\begin{equation}
\label{eq:dxdv}
\dot X = V\,,\quad \dot V=   \gamma(A X - V)\,.
\end{equation}
Here $A$ is the fluid-velocity gradient at the particle position.
A change of variables to $z = V/X$ results in
\begin{subequations}
\begin{equation}
\label{eq:1d}
\hspace*{-10mm}
 \frac{{\rm d}}{{\rm d} t} \log |X| = z\,,\\[-.7cm]
\end{equation}
\begin{equation}
\label{eq:z1d}
\frac{{\rm d}z}{{\rm d}t}  = \gamma(A -z)- z^2\,.
\end{equation}
\end{subequations}
Eq.~(\ref{eq:z1d}) is the one-dimensional version of (\ref{eq:langevin_Zmatrix_0}).
It follows from Eqs.~\eqnref{eq: 2.4} and (\ref{eq:1d}) that the Lyapunov exponent
can be computed as
\begin{eqnarray}
\lambda_1&=&\lim_{t\to\infty}\frac{1}{t}\int_0^t{\rm d}t_1\,z(t_1)=\langle z \rangle_\infty\,.
\label{eq:lambda_1d_def}
\end{eqnarray}
Here $\langle \cdots \rangle_\infty$ is the ensemble average of $z(t)$ determined along particle paths,  in the limit of $t\to\infty$.

In the following we briefly describe the methods that allow to calculate this average, Lyapunov exponents, and the fractal Lyapunov dimension in $d$ dimensions.  In the white-noise limit diffusion approximations can be used that neglect preferential sampling, but at finite Kubo and Stokes numbers it is important to take into account how the particles move through the flow.

\subsection{White-noise approximations}
\label{sec:mwna}
In this Section we describe methods that apply to the white-noise limit
of the problem, $\ku\rightarrow 0$ and $\st \rightarrow \infty$
so that $\varepsilon^2 = 3\ku^2 \st$ remains constant.  In this limit we use the dimensionless variables (\ref{eq:dim2}). All equations in this Section are expressed in the variables (\ref{eq:dim2}). For notational convenience we drop the tilde.
\subsubsection{Mapping to disordered quantum system}
\label{sec:liap1d}
In the dimensionless variables (\ref{eq:dim2}) Eq.~(\ref{eq:dxdv}) takes the form (dropping the tildes):
\begin{equation}
\label{eq:XV}
 \frac{ {\rm d}{X}}{{\rm d} t} =  V\,,\quad  \frac{{\rm d} V}{{\rm d} t} =  A X- V\,.
\end{equation}
In the white-noise limit the fluctuations of the fluid-velocity gradient $ A$ are determined by the correlation function
(\ref{eq:corr2}):
\begin{equation}
\langle  A( t_1)  A(t_2) \rangle_\infty = 2\varepsilon^2\, \delta(t_1- t_2)\,.
\end{equation}
The parameter $\varepsilon$ is defined in Eq. (\ref{eq:defeps}), $\varepsilon^2 = 3\ku^2 \st$. The $\delta$-function is given an infinitesimal width
in order to make the problem well-defined.  The resulting problem is closely related
to the quantum-mechanical problem of calculating the density of states and the localisation length of a particle in a one-dimensional
random potential \cite{Wil03}. To make the connection explicit consider the change of variables
\cite{Der07} $X(t) \equiv \exp(-t/2) \varphi(t)$  and $t=x$.
Eq.~(\ref{eq:XV}) transforms into:
\begin{equation}
\label{eq:s}
-\frac{{\rm d}^2\varphi}{{\rm d}x^2}  +A \varphi = -{1\over 4} \varphi\,.
\end{equation}
This is a Schr\"odinger equation for the wave function  $\varphi(x)$ of a particle in the potential $A$
with dimensionless \lq disorder strength\rq{} $\varepsilon^2$,
reduced Planck constant $\hbar=1$,  mass $m=1/2$, and energy $E=-1/4$. Eq.~(\ref{eq:s}) corresponds to Eq.~(1.1) in Ref.~\cite{Hal65}.
The rate (\ref{eq:Jc}) of caustic formation corresponds
to the integrated density of states $N(E)$ of the quantum model, Eq.~(1.62) in Ref.~\cite{Hal65}, converted to dimensional variables by inserting the \lq energy scale\rq{}
$(D^2 m\hbar^{-2})^{1/3}$ and length scale $(D m^2 \hbar^{-4})^{-1/3}$ where $D$ is the dimensional disorder strength. The localisation length in the quantum problem is related
to the density of states by a dispersion relation
(\lq Thouless formula\rq{}), Eq.~(16) in Ref.~\cite{Thouless}:
\begin{equation}
\label{eq:dr}
\ell^{-1}(E) = -\pi N_0(E) + \int_{-\infty}^\infty {\rm d}E'\, [\rho(E')-\rho_0(E')] \log |E-E'|\,.
\end{equation}
Here $\rho(E) = {\rm d}N(E)/{\rm d}E$ and $N_0(E)$ is the density of states of the system in the absence of disorder.
The close relation between the Lyapunov exponent (\ref{eq:lyapunov_1d_WN}) and the rate of caustic formation (\ref{eq:Jc}) in our problem
follows from the dispersion relation (\ref{eq:dr}). Thouless~\cite{Thouless} explains the dispersion relation in terms of a complex wave vector,
its real part gives the localisation length whereas its imaginary part determines the number of nodes of the wave function and
thus the integrated density of states. In our problem the Airy-function expression in (\ref{eq:lyapunov_1d_WN}) and (\ref{eq:Jc})
plays the role of this wave vector. The sensitive dependence  of the rate of caustic formation  (\ref{eq:Jc}) upon $\varepsilon$
corresponds to \lq Lifshits tails\rq{} of the quantum density of states outside the energy band of the clean system.

\subsubsection{Mapping to Kramers problem}
\label{sec:kramers}
In the white-noise limit the problem of calculating the Lyapunov exponents can be mapped onto the problem of computing the escape
of a random variable from a fixed point \cite{Wil03,Meh04,Hal65}.
This problem is similar to the stochastic escape problem of a \lq reaction coordinate\rq{} over a barrier formulated by Kramers to explain
Arrhenius' law, the sensitive dependence of chemical reaction rates upon temperature \cite{Kra40}.
To simplify the notation we just discuss the one-dimensional case, following Ref. \cite{Wil03}.
The starting point is Eq.~(\ref{eq:z1d}) that describes the stochastic dynamics of the variable $z = V/X$.
Formulating the problem in terms of the variable $z$
is convenient \cite{Wil03,Meh05,Wil07} because it approaches a steady state, as opposed to the variables $X$ and $V$ {in an unbounded system}. Furthermore Eq.~(\ref{eq:lambda_1d_def}) shows
that the Lyapunov exponent is given by the steady-state average of $z$. In the white-noise limit
the dynamics of $z$ decouples from the particle dynamics as $X\rightarrow 0$.  This reduces the task of calculating the Lyapunov exponent from the
two-dimensional problem (\ref{eq:dxdv})
to the one-dimensional problem of determining the steady-state distribution of $z$.
Finally  the $z$-dynamics also determines the rate of caustic formation. This rate
is given by the rate at which $z$ escapes to $-\infty$ as explained below.
\begin{figure}[tp]
\includegraphics[width=6cm]{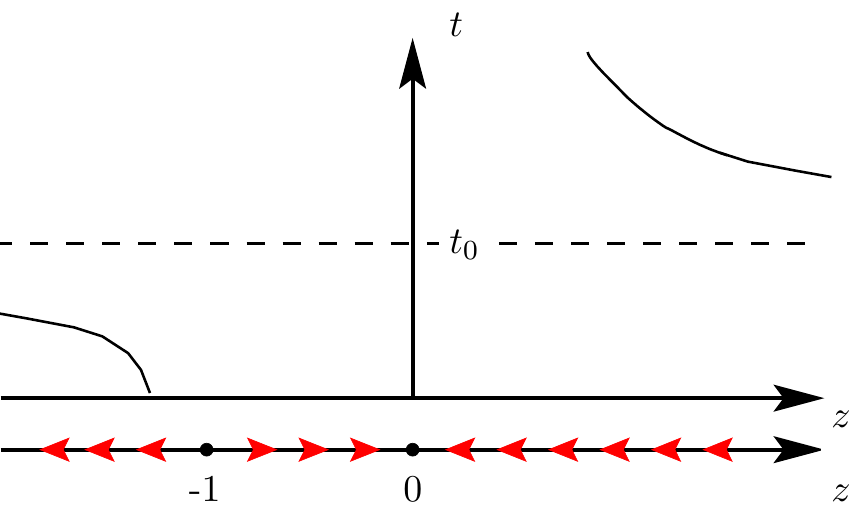}
\caption{\label{fig:1d} Illustrates the drift part of the dynamics (\ref{eq:Langevinz}). There
are two fixed points: $z=0$ is stable and $z=-1$ is unstable. Starting at $z_0 < -1$ the variable
$z$ reaches $-\infty$ in the finite time $t_0 = \log[z_0/(z_0+1)]$ and re-appears at $+\infty$.
Eventually it converges to the stable fixed point.}
\end{figure}

In the white-noise limit (\ref{eq:wnlimit}) it follows
from (\ref{eq:z1d}) that the dynamics of $z$ is determined by
the Langevin equation
\begin{equation}
\label{eq:Langevinz}
\delta z = (-z-z^2 )\delta t+ \delta w
\end{equation}
in the dimensionless variables (\ref{eq:dim2}).
Here $\delta w$ is a Gaussian random increment with zero mean and variance $\langle \delta w^2 \rangle = 2\varepsilon^2 \delta t$.
The drift part of this equation [obtained by setting $\delta w=0$ in Eq. (\ref{eq:Langevinz})] has two fixed points, a stable one at $z=0$ and an unstable one at $z=-1$, see Fig.~\ref{fig:1d}.
The deterministic solution $z(t) = z_0/[{\rm e}^t(z_0+1)-z_0]$ has a finite-time
singularity. Starting at $z_0 < -1$ one
reaches $-\infty$ in the finite time $t_0 = \log[z_0/(z_0+1)]$. This singularity is a caustic singularity where the phase-space manifold
describing the $x$-dependence of the particle velocity folds over, and thus $\partial_x v \rightarrow -\infty$. This correspondence
shows that the appropriate boundary condition for (\ref{eq:Langevinz}) is the following: when $z$ tends to $-\infty$ it re-appears at $+\infty$ with
the same rate of change.
Eventually the variable $z$ reaches its stable fixed point $z=0$. 

Now consider the effect of the noise $\delta w$. It may
allow the variable $z$ to escape from the vicinity of $z=0$ to $-\infty$,  via the unstable fixed point $z=-1$. In this case a caustic occurs, and $z$ returns 
via $+\infty$ to the stable fixed point $z=0$.
The rate of caustic formation is thus given by the rate at which $z$ escapes from its stable fixed point. The solution of the problem is analogous
to Kramers' solution \cite{Kra40} in terms of a Fokker-Planck equation for the variable $z$. This equation determines
the distribution $P(z,t)$ of $z$ at time $t$, subject to boundary conditions that $P(z,t)$ must be
normalised and that the $z$-dynamics must respect \lq particle exchange symmetry\rq{} $P(z,t)= P(-z,t)$ as $z\to\infty$.
In the dimensionless variables (\ref{eq:dim2}) $P(z,t)$ satisfies \cite{Wil03}:
\begin{equation}
\label{eq:FP_1d}
\partial_t P = \partial_z\big(z + z^2 + \varepsilon^2\, \partial_z\big) P \,.
\end{equation}
The exact steady-state solution of (\ref{eq:FP_1d}) is readily obtained~\cite{Wil03}.
Substituting the steady-state solution of Eq. (\ref{eq:FP_1d}) into
\begin{equation}
\label{eq:lambdaz}
\frac{\lambda_1}{\gamma} = \langle z\rangle_\infty =\int_{-\infty}^\infty\!\!\!\!{\rm d}z \, z\, P(z)
\end{equation}
results in the exact white-noise expression (\ref{eq:lyapunov_1d_WN}).
We note that (\ref{eq:lambdaz}) differs from (\ref{eq:lambda_1d_def}) by a factor of $\gamma$ because
Eq.~(\ref{eq:lambdaz})  is written in the dimensionless variables (\ref{eq:dim2}).

The rate of caustic formation (\ref{eq:Jc}) is determined  by the
finite steady-state probability current of Eq. (\ref{eq:FP_1d}).
The corresponding excursions of $z$ cause algebraic tails
of the steady-state distribution: in the limit of large values of $|z|$ only the second term on the r.h.s. of
Eq.~(\ref{eq:FP_1d}) survives. In this case the steady-state condition becomes $z^2 P(z) = {\rm const}$. This implies
$P(z) \sim |z|^{-2}$.
Particle-exchange symmetry ensures that the left and right tails are symmetric. As a consequence
the Lyapunov exponent (\ref{eq:lambdaz})  is well defined. But higher steady-state moments of $z$ diverge.

Corresponding escape problems were formulated in two and three spatial dimensions in the white-noise limit \cite{Meh04,Meh05,Dun05,Wil07}.
But unlike the one-dimensional case these diffusion problems have not yet been exactly solved. The difficulty is that the multi-dimensional
deterministic drift part of the resulting set of Langevin equations is neither potential nor solenoidal. In the next Section we illustrate
how the diffusion problem can be solved by algebraic perturbation theory.

\subsubsection{Algebraic perturbation theory for the Lyapunov exponent}
\label{sec:pt}
In this Section we describe the perturbative method employed in Refs.~\cite{Meh04,Meh05,Dun05,Wil07} to compute the Lyapunov spectrum in the white-noise limit.
This perturbation theory is closely related to the perturbative treatment of  quantum non-linear oscillators.

\begin{figure}
\begin{overpic}[width=10cm,clip]{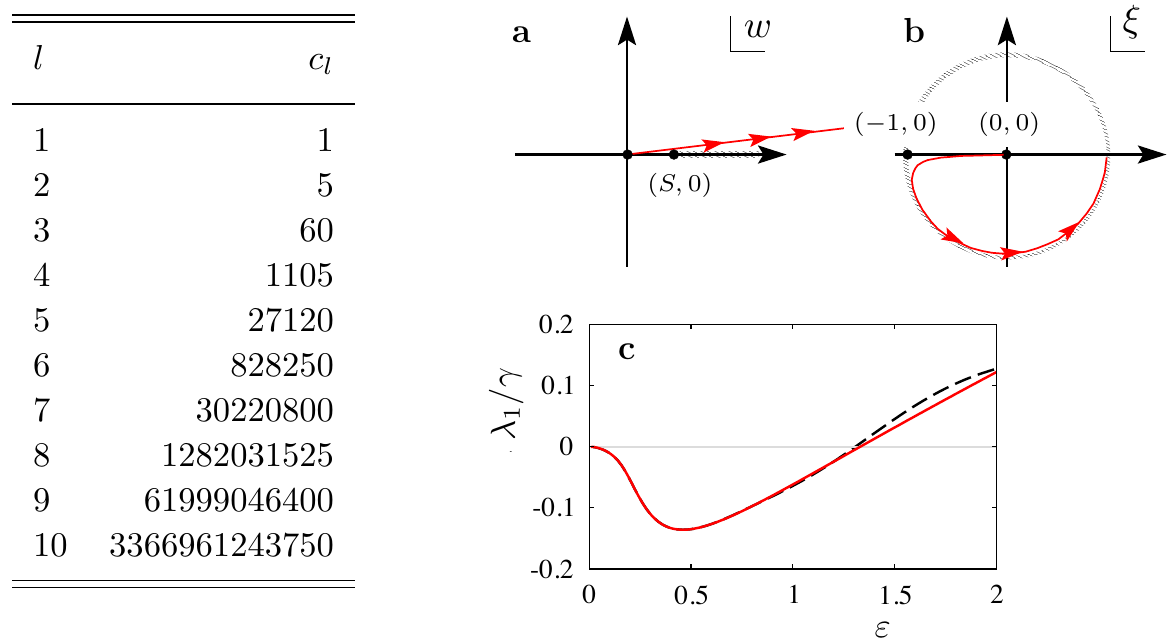}
\end{overpic}
\caption{\label{fig:con} Table: first 10 coefficients of the perturbation
 expansion for the maximal Lyapunov exponent. Panel {\bf a} shows integration
contour $\cal C$ (red) for the principal-value integral (\ref{eq:borelint}) in the $w$-plane. The Borel sum $B(w)$ has a pole at $S=1/6$ and may have further
poles in the interval $[S,\infty)$.  {\bf b} Corresponding integration contour in the $\xi$-plane. The interval $[S,\infty)$ is mapped to the unit circle.
{\bf c} One-dimensional Lyapunov exponent in the white-noise limit.
Shown is the exact result (\ref{eq:lyapunov_1d_WN}), solid red line, and
the result from conformal Borel summation to order $l_{\rm max} = 24$, black dashed line.}
\end{figure}

To keep the notation simple we illustrate this calculation in one spatial dimension (where the exact solution is known, as explained above).
To begin with we change variables to {$y = z/\varepsilon$}:
\begin{equation}
\label{eq:FPy}
\partial_t P =  \partial_y \big(y + \varepsilon\,y^2 + \partial_y\big) P .
\end{equation}
Using Dirac notation \cite{Dir30} we write the steady-state solution of this equation as $P(y)\equiv\langle y|P\rangle$.
We denote the differential operator on the r.h.s. of Eq.~(\ref{eq:FPy}) by $\hat F$. Its action is defined by
 $\langle y |\hat F|P\rangle \equiv \partial_y (y + \varepsilon\,y^2 + \partial_y) P(y)$.
The steady-state solution of Eq.~(\ref{eq:FPy}) is determined by the  condition:
\begin{equation}
\label{eq:FP}
\hat F |P\rangle = 0\,,\quad \mbox{with}\quad
\hat F \equiv \hat F_0 + \varepsilon \hat F_1\,,\quad
\hat F_0  \equiv  \partial_y y + \partial_y^2\,,\quad \mbox{and} \quad
\hat F_1  \equiv  \partial_y y^2\,,
\end{equation}
and the Lyapunov exponent is computed as
\begin{equation}
\label{eq:l1}
\frac{\lambda_1}{\gamma}= \varepsilon \frac{\langle 0|\hat y|P\rangle}{\langle 0|P\rangle}\,.
\end{equation}
In the limit of $\varepsilon\rightarrow 0$ the symmetric positive solution
of Eq.~(\ref{eq:FP}) is simply a Gaussian, ${\rm e}^{-y^2/2}$, corresponding to the ground state $|0\rangle$
of a quantum-mechanical harmonic oscillator (so that $\langle y|0\rangle = {\rm e}^{-y^2/2}$).
An algebraic perturbation expansion around this solution is constructed as follows.
The perturbation is $\varepsilon\hat F_1$.
Expanding $|P\rangle$ in powers of $\varepsilon$,
$|P\rangle = |0\rangle + \varepsilon |P_1\rangle +\varepsilon^2 |P_2\rangle+\cdots$ we find the recursion
\begin{equation}
\label{eq:recursionQ}
\hat F_0|P_{k+1}\rangle  = - \hat F_1 |P_k\rangle\,.
\end{equation}
This recursion is evaluated using
raising and lowering operators just as for the quantum harmonic oscillator \cite{Dir30}. As a result we obtain
the maximal Lyapunov exponent as an expansion in $\varepsilon$:
\begin{equation}
\label{eq:pts1d}
\frac{\lambda_1}{\gamma} = -\sum_{l=1}^\infty c_l \varepsilon^{2l}\,.
\end{equation}
The first few coefficients $c_l$ in this expansion are listed in Fig.~\ref{fig:con}.
The coefficients in (\ref{eq:pts1d}) can be computed to very high orders $l$ and turn out to be the
same as those obtained
by a series expansion of the exact white-noise result (\ref{eq:lyapunov_1d_WN}).
These coefficients satisfy the recursion \citep{Pra03}
\begin{equation}
c_{l+1} = (6l-2)c_{l}+\sum_{j=1}^{l} c_j c_{l+1-j}
\label{eq:cl_recursion}
\end{equation}
with initial condition $c_1=1$.
The same recursion also determines the
moments of the so-called \lq Wiener index\rq{}  for a certain
class of random graphs \cite{Jan03}. The Wiener index of a connected graph is
the sum over distances (number of edges) between all pairs
of vertices of the graph. It is known
that the moments of the Wiener index are related to moments of so-called
Brownian excursions \cite{Ald97}, but an explicit connection to the problem discussed here has not yet been given.

The series (\ref{eq:pts1d})  is \lq asymptotically divergent' \cite{boyd}. It does not converge,
but any partial sum of the series approaches $\lambda_1/\gamma$ as $\varepsilon \rightarrow 0$.
This divergence is caused by the fact that the coefficients $c_l$ increase too rapidly
as the order $l$ increases.  This is also the case for the perturbative expansions
obtained for the  Lyapunov exponents in Refs.~\cite{Meh04,Meh05,Dun05,Wil07}.
These perturbation series, of the form (\ref{eq:ptGamma}), must be resummed.
To give an example we show how to resum the series (\ref{eq:pts1d})
using \lq Borel summation\rq{}.
The coefficients derived from \Eqnref{eq:cl_recursion} have the asymptotic form
\begin{equation}
\label{eq:clasy}
c_l\sim   \frac{(l-1)!}{(2\pi) S^{l}}
\end{equation}
with $S=1/6$.
This implies that the so-called \lq Borel sum\rq{}
\begin{equation}
B(w) \equiv \sum_{l=1}^\infty \frac{c_l}{l!} w^l
\end{equation}
converges for $|w| <  S$.
The sum (\ref{eq:pts1d})  is estimated by
\begin{equation}
\label{eq:borelint}
\frac{\lambda_1}{\gamma} = \frac{1}{\varepsilon^2}
 \int_0^\infty \!\!{\rm d}w \,{\rm e}^{-w/\varepsilon^2} B(w)\,.
\end{equation}
The Borel sum  exhibits poles on the real axis for $w\geq S$.
As a consequence the integral must be interpreted as a principal-value integral.
The integral is performed along
a ray ${\cal C}$ in the upper right quadrant of the $w$-plane. The real part of the resulting expression estimates $\lambda_1/\gamma$.
The imaginary part gives the rate of caustic formation.
This is a consequence of the dispersion relation connecting the Lyapunov
exponent and the rate of caustic formation (Section \ref{sec:liap1d}).
Provided that $B(w)$ is analytic outside $[S,\infty)$ the precise location of ${\cal C}$ does not matter.
But to evaluate the integral, the function $B(w)$ must be continued outside its radius of convergence.
If the only singularities of $B(w)$ are in the interval $[S,\infty)$, analytic continuation
can be achieved by conformally mapping the $w$-plane excluding
$[S,\infty)$ to the unit disk, by the following mapping:
\begin{equation}
\label{eq:cm}
 w = -\frac{4S\xi}{(1-\xi)^2}\,.
\end{equation}
Under this mapping the poles in $[S,\infty)$ are mapped to the boundary
of the unit disk. Fig.~\ref{fig:con}{\bf a},{\bf b} shows how the integration
path ${\cal C}$ is mapped: the image of $\cal C$ is contained in the radius of convergence
of $B(\xi)$ in the $\xi$-plane, so that the integral can be performed.
Fig.~\ref{fig:con}{\bf c} shows the result of this procedure including
coefficients up to $l_{\rm max}=24$. Compared with the exact result (\ref{eq:lyapunov_1d_WN}) one observes excellent agreement.
The approach described here [resummation by the conformal mapping (\ref{eq:cm})] was first used by Leguillou and Zinn-Justin~\cite{ZJ} to compute critical
exponents for the $n$-vector model by resumming asymptotic \lq $\epsilon$-expansions\rq{}.
The approach was also used to resum an asymptotic series for the
correlation dimension (Section \ref{sec:fractal})
of small heavy particles in a two-dimensional random velocity field in the white-noise limit \cite{Wil10b}.

If the analytic structure of $B(w)$ is more difficult to determine,  the analytic continuation can be performed using Pad\'e{} approximants \citep{Ben78}.
This approach was used in Refs.~\cite{Dun05,Wil07} to resum perturbation series of the form (\ref{eq:ptGamma}) for Lyapunov exponents in the white-noise limit. Another possibility is to sum the series to
its \lq optimal order\rq{}, depending on $\varepsilon$. This method is
described in Ref.~\cite{boyd} and was used to compute
the maximal Lyapunov exponent in partially compressible flows \cite{Meh04,Meh05}.

A general difficulty with asymptotic perturbation series is that a given series
is asymptotic to infinitely many different functions
differing by non-analytic terms
that have vanishing Taylor coefficients to all orders. Fig.~\ref{fig:con}{\bf c} shows
that there is no such additional contribution to $\lambda_1$ in one spatial dimension in the white-noise limit.
But for particles in two-dimensional random velocity fields with $\Gamma\neq 1$
(see Section~\ref{sec:rvf} for the definition of $\Gamma$), there is a non-analytic term of the form $N \exp [-1/(6\varepsilon^2)]$
that must be added to the perturbation series in the white-noise limit, see Eq.~(19) in Ref.~\cite{Meh04}.
The following Section explains how such non-analytic terms may arise within asymptotic approximations
to the steady-state solution of Eq.~(\ref{eq:FP_1d}).

\subsubsection{WKB approximation}
\label{sec:WKB}
The \lq{}Wentzel-Kramers-Brillouin\rq{} (WKB) approximation is an asymptotic method that allows to determine possible non-analytic contributions
to the steady-state solution of the Fokker-Planck equation (\ref{eq:FP_1d}).  In its original form it was used to find approximate solutions
to the one-dimensional Schr\"odinger equation \cite{Wen26,Kra26,Bri26}. The method has been used in a wide variety of applications, for example
in fluid mechanics \cite{Monin}, quantum mechanics \cite{Child}, and population dynamics (see \cite{Eriksson2011} and references therein).
It is a common method to find approximate solutions to generalised diffusion
equations when the diffusion constant is small \cite{Frei98} and was used in Refs. \cite{Dun05,Wil07} to compute the rate of caustic formation
in the white-noise limit. The starting point for the WKB approximation to the steady-state solution of the Fokker-Planck equation
(\ref{eq:FP_1d}) is the ansatz
\begin{equation}
P(z)  = N\, {\rm e}^{-\left[S_0(z)+\varepsilon^2 S_2(z)+\cdots\right]/\varepsilon^2}\,
\label{eq:WKB_Ansatz}
\end{equation}
where  $N$ is a normalisation factor and $S_0,S_2,\ldots$ are unknown functions.
Substituting this ansatz into Eq. (\ref{eq:FP_1d}), expanding in powers
of $\varepsilon^2$, and requiring each order to vanish separately, one obtains differential equations for $S_0$, $S_2$, and so forth:
\begin{subequations}
\begin{eqnarray}
0&=&
\label{eq:S0}
\left(\frac{{\rm d}S_0}{{\rm d}z}\right)^2
-(z+z^2) \frac{{\rm d}S_0}{{\rm d}z}\,,\\
0&=&
\label{eq:S2}
\frac{{\rm d}S_2}{{\rm d}z}
\left(2\frac{{\rm d}S_0}{{\rm d}z}-z-z^2\right)-\frac{{\rm d}^2S_0}{{\rm d}z^2}+2z+1\,.
\end{eqnarray}
\end{subequations}
Eq. (\ref{eq:S0}) has two solutions: $S_0^-(z)=a^-$ and  $S_0^+(z)=z^3/3+z^2/2+a^+$, where $a^+$ and $a^-$ are constants of integration. The two corresponding
solutions of Eq. (\ref{eq:S2}) are $S_2^-(z)=-\ln b^-+\ln|z^2+z|$ and $S_2^+(z)=-\ln b^+$ with the
positive constants $b^+$ and $b^-$.
The steady-state solution of (\ref{eq:FP_1d}) is usually obtained by matching linear combinations of the solutions
\begin{subequations}
\label{eq:WKB_Ppm_general}
\begin{eqnarray}
\label{eq:Pm}
P^-&=& \, \frac{b^-}{|z^2+z|}{\rm e}^{-a^-/\varepsilon^2}\,,\\
P^+&=& \, b^+{\rm e}^{-(z^3/3+z^2/2+a^+)/\varepsilon^2}
\label{eq:Pp}
\end{eqnarray}
\end{subequations}
at the classical turning points $z=-1$ and $z=0$. Here, however, we proceed differently.
For small values of $z$ the $z^2$-term on the r.h.s. of (\ref{eq:FP_1d}) is negligible. The corresponding
steady-state solution of (\ref{eq:FP_1d}) is Gaussian
with variance $\varepsilon^2$. This form is obtained from (\ref{eq:Pp}) with $a^+=0$.
For large values of $|z|$ by contrast the $z^2$-term on the r.h.s. of  (\ref{eq:FP_1d}) dominates
and $P(z) \sim |z|^{-2}$.
In this case the appropriate solution is $P^-$. We match the actions $S_0(z)$ of the two solutions 
smoothly at $z=-1$ and continuously at $z=1/2$ with $a^-=1/6$.
The solution $P_+$ is valid when $z>-1$ and the solution $P_-$ is valid when $z<-1$ or when $z>1/2$.
It follows that the probability of observing large values of $|z|$ is exponentially
suppressed  $\propto \exp[-1/(6\varepsilon^2)]$. This explains the non-analytic $\varepsilon$-dependence of
the rate of caustic formation in the white-noise limit, Eq. (\ref{eq:Jc}).
The WKB approximation can be formulated in higher spatial dimensions, too. In this case it is convenient to perform the matching
using Hamilton-Jacobi theory as outlined in Ref. \cite{Frei98}, see also Ref. \cite{Eriksson2011}. This approach rests upon the fact
that Eq. (\ref{eq:S0}) has the form of a Hamilton-Jacobi equation. A variational principle must be invoked in higher dimensions
to determine the functional form of $S_0(z)$. A numerical solution of this variational problem indicates that the
rate of caustic formation in two and three spatial dimensions is also of the asymptotic form $\exp[-1/(6 \varepsilon^2)]$.

This may also explain the non-analytic $\epsilon$-dependence of the Lyapunov exponents
in the white-noise limit, of the form $N \exp [-1/(6\varepsilon^2)]$~\cite{Meh04,Wil07}.
Such non-analytical terms
have vanishing Taylor coefficients to all orders and are not present in the perturbation series (\ref{eq:lwn}).
But the simple procedure outlined above is too crude to yield reliable results for the Lyapunov exponents. Since the amplitude of $P^-$ diverges at $z=-1$, a refined approximation is necessary near this point (commonly referred to as \lq{}uniform approximation\rq{}).

An alternative procedure to extract non-analytical terms from asymptotic perturbation series is to analyse
the asymptotic form of the perturbation coefficients. Coefficients of the form (\ref{eq:clasy})  are expected to give rise to non-analytic terms of the form $\exp(-S/\varepsilon^2)$ as explained
in Ref. \cite{boyd}. In two- and three-dimensional flows with some degree of compressibility, the perturbation coefficients in the white-noise
limit have the asymptotic form $c_l \sim [6 (1-\Gamma)]^{-l} (l-1)!$, apparently encoding a non-analytical
dependence of the form $\exp[-1/(6 |\Gamma-1|\varepsilon^2)]$. Here $\Gamma$ is the compressibility parameter introduced
in Section \ref{sec:Model} (Table~\ref{tab:1}). When $\Gamma$ is not zero this dependence differs from $\exp[-1/(6 \varepsilon^2)]$
that is obtained in numerical simulations in the white-noise limit, and in the WKB approximation outlined above.
The reason for this difference is not understood.

The perturbation methods described in this Section were generalised to compute the
correlation dimension in two \cite{Wil10b} and three \cite{Wil14} spatial dimensions in the white-noise limit $\ku\rightarrow 0$. Also in this case it appears that
there is a non-analytic contribution, possibly of the form $N \exp[-1/(6\varepsilon^2)]$. It is currently not understood how to
compute such contributions to the correlation dimension.

A closely related WKB approximation was used in Ref. \cite{Gus08b} to solve a one-dimensional model for the distribution of relative velocities of
small heavy particles at small separations and at large Stokes numbers where the inertial-range properties of the turbulent velocity fluctuations become important.

\subsection{Finite-$\ku$ approximations}
\label{sec:fkua}
At finite Kubo and Stokes numbers it is convenient to use the dimensionless variables (\ref{eq:dim1}). In this Section all equations are expressed in these variables. For notational convenience the primes are dropped.
\subsubsection{{Expansion around deterministic trajectories}}
\label{sec:texpand}
To generalise the methods described in the previous Section to finite Kubo numbers {and Stokes numbers} is difficult
because the flow velocity $u$ in Eq.~(\ref{eq:langevin_1d_dimless}) is a non-linear function of the particle position $x(t)$.
This renders the coupled equations \eqnref{eq:langevin_1d_dimless} nonlinear, they cannot be explicitly solved.
An approximate solution was obtained in Ref.~\cite{Gus11a} by expanding an implicit solution of (\ref{eq:langevin_1d_dimless}). In the dimensionless variables (\ref{eq:dim1}) this
solution takes the form (dropping the primes):
\begin{eqnarray}
\label{eq:rsolution}
x(t)\!=\underbrace{\!x_0\!+\!\ku\,\st\, v_0\left(1\!-\!{\rm e}^{-t/\st}\right)\!}_{\xd} + \frac{\ku}{\st}\int_{0}^{t}\!\!\! {\rm d}t_1\!\!\int_{0}^{t_1}\!\!\!\!{\rm d}t_2\,{\rm e}^{-(t_1\!-t_2)/\st}u(x(t_2),t_2)\,.
\end{eqnarray}
Here $x_0\equiv x(0)$ is the initial position and $v_0\equiv v(0)$ is the initial velocity
of a particle.
We denote by $\xd$ the deterministic part of the trajectory
(the solution of the equation of motion for $u = 0$, in the absence of turbulent fluctuations).
We define
\begin{equation}
\label{eq:dxt}
\delta x(t)\equiv x(t)-\xd\,,
\end{equation}
the difference between the actual trajectory and the deterministic trajectory.
We note that $\delta x(t)$ is proportional to $\ku$ and can therefore be considered small if $\ku$ is small enough
(how small $\ku$ must be depends on $t$, $\st$, and upon the realisation of the turbulent fluctuations).
Here $\ku$ is used as an expansion parameter to keep track of the order of the stochastic part of the 
solution.
In general, $\ku$ is only a book-keeping parameter, the expansion is in terms of $\delta x(t)$, not in $\ku$.
But Eq. (\ref{eq:dxt}) shows that the expansion becomes a $\ku$-expansion for the dynamics considered here
when $v_0=0$.
Expanding $u(x(t),t)$ in terms of $\delta x(t)=\Ordo(\ku)$ gives:
\begin{equation}
u(x(t),t)=u(\xd,t)+\partial_x u(\xd,t)\,\delta x(t)+\frac{1}{2}\partial_x^2 u(\xd,t)\,\delta x(t)\,\delta x(t)+\dots\,.
\label{eq:uexpansion}
\end{equation}
Since the underlying flow is homogeneous,
the value of $x_0$ does not matter.
Inserting $\delta x$ from Eqs.~\eqnref{eq:dxt} and~\eqnref{eq:rsolution}
into $\eqnref{eq:uexpansion}$
and iterating yields $u(x(t),t)$ in terms of $u$ and its derivatives
evaluated at $\xd$. The result is an expansion of $u(x(t),t)$ around the deterministic dynamics $\xd$. This and related  expansions can
be used to obtain perturbation series for the Lyapunov exponents in powers of $\ku$ as we shall see.

\begin{figure}[t]
\includegraphics[width=\textwidth]{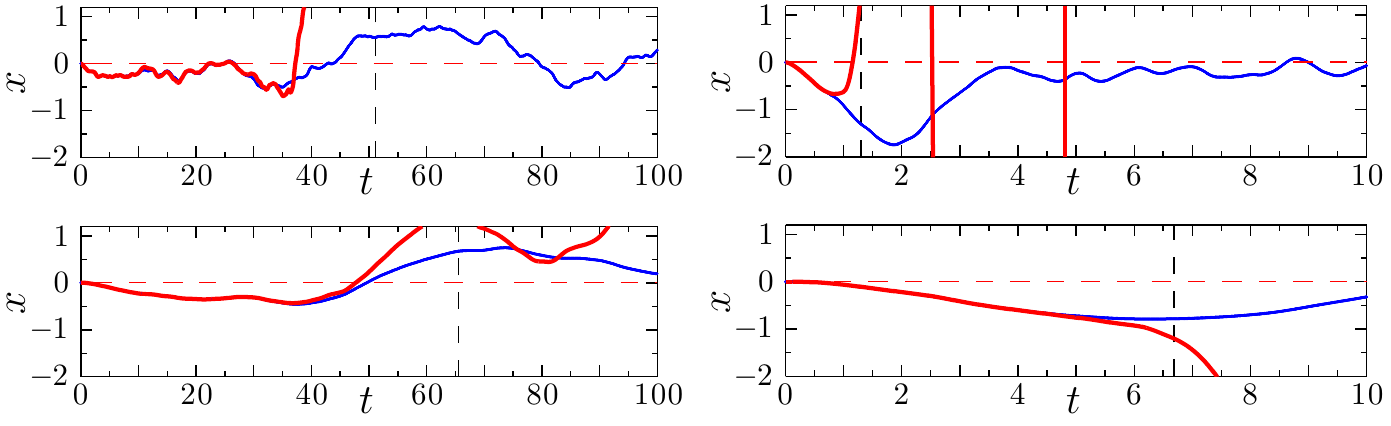}
\caption{\label{fig:trajek} Numerically determined solutions of Eq.~(\ref{eq:langevin_1d_dimless})
for the initial condition $x_0=v_0=0$, for different values of Kubo and Stokes numbers:
$\ku = 0.1$ (left column) and $\ku=1$ (right column), $\st = 0.1$ (upper row), and $\st=10$ (lower row).
Numerically computed trajectories are shown as blue solid lines, the results of
the perturbation expansion described in Section \ref{sec:texpand} to order $\ku^5$
are shown as red solid lines. The initial condition $x_0=0$ is shown as horizontal dashed lines.
Dimensionless units (\ref{eq:dim1}) are used. In all cases the same
realisation of the random velocity {field was} used.
Estimates for the time scale $t_{\rm f}$ at which the trajectory approximation fails, (\ref{eq:fail}), are shown as vertical black-dashed lines. }
\end{figure}

Terms of order $\ku^n$ in the expansion of $u(x(t),t)$ contain $n$ factors of $u$ and spatial derivatives of $u$ evaluated at $\xd$.
In this way $u(x(t),t)$ may in principle be expanded to any order in $\ku$.
Fig.~\ref{fig:trajek} shows numerically computed particle trajectories in the one-dimensional model
at finite Stokes and Kubo numbers. Also shown is the perturbation expansion (\ref{eq:rsolution}) and (\ref{eq:uexpansion})
to order $\ku^5$ for different Stokes and Kubo numbers. The realisation of the fluid-velocity field is the same in all {cases.}
The fluctuations of the particle paths depend strongly on the Stokes number. For large Stokes numbers the particles move almost ballistically, their acceleration is small.
One expects that the expansion yields accurate results provided that the magnitude of $\delta x(t)\equiv x(t)-\xd$ remains smaller than the correlation length.
One way of estimating the largest time $t_{\rm f}$
until which the expansion is valid is to solve
\begin{equation}
\label{eq:fail}
\big\langle \delta x(t)^2|t;v_0,u(\xdo,0),\partial_xu(\xdo,0),\ldots \big\rangle\sim 1
\end{equation}
for $t= t_{\rm f}$.
The average in Eq.~(\ref{eq:fail}) is a finite-time
average of the form discussed later in Section \ref{sec:finite_time}. It is conditional
on the initial values $v_0$, $u(\xdo,0)$, $\partial_xu(\xdo,0)$, and so forth.
Evaluating the average (\ref{eq:fail})
as explained in the following Section
yields estimates for $t_{\rm f}$,
shown in Fig.~\ref{fig:trajek} as dashed lines.

In a similar manner one can expand spatial derivatives of $u$.
The method can also be used for expanding general functionals $F$ of $u(x(t),t)$ and its gradients, provided that $F$  can be implicitly written as 
$F(t)=F^{(\rm d)}(t)+\mathscr{F} [F(t),u(x(t),t),\partial_x u(x(t),t),\ldots]$
where $F^{(\rm d)}$ is the deterministic part of $F$ obtained in the absence of turbulent fluctuations,  and $\mathscr{F}$ is a functional (commonly integral)
of $F(t)$, $u(x(t),t)$, and so forth.

The expansion described above is an expansion around deterministic trajectories, we refer to it as the \lq trajectory expansion\rq{}. In combination
with the averaging method described in the following Section the trajectory expansion allows to compute Lyapunov exponents, fractal dimensions, and averages of other observables in the statistical model.

\subsubsection{Steady-state averages}
\label{ssec:Method_steady_state_averages}
The dynamics of particles suspended in a random velocity field approaches a statistically steady state.
For particles advected in time-uncorrelated random velocity fields
this was first shown in Ref.~\cite{Jan85}. We now show how to compute steady-state averages
in turbulent aerosols at finite Kubo numbers \cite{Gus11a,Gus13a}.
Since the particles may sample the fluid-velocity field and its derivatives preferentially it is necessary to compute the velocity fluctuations
as seen by the particle while it moves through the flow.

Consider an observable $F$ evaluated at the particle position $x(t)$ at time $t$,
such as the fluid-velocity gradient $A(x(t),t)$ or the particle-velocity gradient $z(t)$. The average
of $F$ depends on the path the particle has taken through the flow.
The path depends on the initial conditions $x_0$ and $v_0$, on
the initial values of the flow velocity $u$ and its gradients evaluated at $x_0$ at $t=0$,
and on the fluctuations of the velocity field along the path the particle takes from $x_0$ to $x(t)$.
For the average of $F$ at time $t$ conditional upon the initial conditions
we write
\begin{equation}
\label{eq:avF}
\langle F|t;v_0,u(\xdo,0),\partial_xu(\xdo,0),\ldots\rangle\,,
\end{equation}
making explicit the dependence on time $t$, on the initial particle velocity $v_0$,
and upon the initial flow velocity and its gradients.
The average in \Eqnref{eq:avF} is an average over realisations of the flow conditional on the initial conditions.
The average does not depend on the initial particle position $x_0$, since the flow is assumed to be homogeneous.

As $t\rightarrow \infty$ one expects that a steady state is approached.
In this limit the average (\ref{eq:avF}) must become independent of the initial conditions,
and the corresponding steady-state average is denoted by
\begin{equation}
\langle F\rangle_\infty\equiv \lim_{t\rightarrow\infty}  \langle F|t;v_0,u(\xdo,0),\partial_xu(\xdo,0),\ldots\rangle\,.
\label{eq:timeavg}
\end{equation}
Assuming that all information about the initial conditions is lost in the limit $t\rightarrow \infty$ one may take $v_0=0$ in \Eqnref{eq:timeavg},
as well as
$u(\xdo,0) = 0,\partial_xu(\xdo,0)=0$, and so forth. This substantially simplifies the calculations 
(see below).

Inserting expansions of the form described in Section \ref{sec:texpand} into \eqnref{eq:timeavg} leads to  averages of products of the fluid-velocity field
and its spatial derivatives evaluated at different times:
\begin{equation}
\partial^{n_1}_x u\big(x^{\rm d}(t_1),t_1\big)\partial^{n_2}_x u\big(x^{\rm d}(t_2),t_2\big)\cdots\partial^{n_j}_x u\big(x^{\rm d}(t_j),t_j\big)\,,
\end{equation}
where $n_1,n_2,\dots,n_j$ assume integer values.
Since the flow velocity and its spatial derivatives are Gaussian distributed in the statistical model,
such averages can be evaluated using Wick's formula which expresses an average over a product of Gaussian random functions as a sum over products of their covariances~\cite{Altland}.
The steady-state covariances of $u(\xd,t)$ and its spatial derivatives
are found from Eqs.~\eqnref{eq:velocity_field} and \eqnref{eq:velocity_field_C}.
Setting $v_0=0$ we have that $\xd=x_0$, and the covariances read~\cite{Gus13a}:
\begin{eqnarray}
\left\langle\partial^m_x u(x_0,t_1)\partial^n_x u(x_0,t_2)\right\rangle_\infty= C_{mn} {\rm e}^{-|t_1-t_2|}\,.
\label{eq:correlation_uu_1d}
\end{eqnarray}
Here
\begin{eqnarray}
C_{mn} =
\left\{
\begin{array}{ll}
(-1)^{(n-m)/2}(m+n+1)!! & \mbox{if }m+n\mbox{ even}\,,\cr
0 & \mbox{otherwise\,,}
\end{array}
\right.
\label{eq:cmn}
\end{eqnarray}
and $m,n=0,1,2,\dots$. As mentioned above these steady-state averages are independent of the initial conditions.

As an example consider the steady-state average $\langle A(x(t),t)\rangle_\infty$ of the fluid-velocity gradient $A(x(t),t)$.
The trajectory expansion of $A(x(t),t)$ is obtained as described in Section \ref{sec:texpand}.
Performing the steady-state average (\ref{eq:timeavg}) using (\ref{eq:correlation_uu_1d}) and evaluating
the time integrals over the resulting correlation functions gives \cite{Gus13a}:
\begin{equation}
\label{eq:Abar_1d} \langle A(x(t),t)\rangle_\infty
= \frac{-3\,\ku}{1+\st} + O(\ku^3)\,.
\end{equation}
In the white-noise limit this average tends to zero.
But in general
it is negative because the particles tend to stay longer
in the vicinity of the minima of the \lq potential\rq{} $-\phi(x,t)$ corresponding to the flow velocity $u(x,t)$, Eq.~(\ref{eq:velocity_field_1d}). The result (\ref{eq:Abar_1d}) agrees with the corresponding expression in the telegraph model to lowest order in $\ku$ and $\st$
(Section \ref{sec:othermodels}). The $\st\to 0$ limit  of this expression
was first derived in Ref. \cite{Wil11}.

Other quantities (flow and particle velocities, particle-velocity gradients,\ldots) are averaged analogously.

\subsubsection{Steady-state distributions}
\label{sec:ststd}
Preferential sampling influences the fluctuations of the fluid velocities and their derivatives as seen by the particles.
As an example consider the joint distribution of the fluid velocity $u$ and its gradient
$A \equiv \partial_x u$.
At a given position the joint distribution in the statistical model is Gaussian with zero mean. The variances
are determined by Eq.~(\ref{eq:correlation_uu_1d}):
\begin{eqnarray}
P_0(u,A)=\frac{1}{2\pi\sqrt{3}}{\rm e}^{-u^2/2-A^2/6}\,.
\label{eq:distuA_1d_Ku0}
\end{eqnarray}
But preferential sampling changes this distribution. To find an approximation for the distribution $P(u,A)$ sampled along particle trajectories $x(t)$
we start from the ansatz
\begin{eqnarray}
\label{eq:distz_1d_ansatz}
P(u,A)&=P_0(u,A) \big[a_{00}+\ku (a_{10} u+a_{01} A)+\ku^2(a_{20} u^2+a_{11}uA+a_{02} A^2)\nn\\
&\hspace*{1.2cm}  +\ku^3(a_{30} u^3+a_{21} u^2A+a_{12} uA^2+a_{03} A^3)+\ldots \big]
\,,
\end{eqnarray}
where $P_0(u,A)$ is the distribution of $u$ and $A$ to order  $\ku^0$, Eq.~(\ref{eq:distuA_1d_Ku0}). It corresponds to the joint distribution of $u$ and $A$ at a fixed position in space.
In \Eqnref{eq:distz_1d_ansatz} the order in $\ku$ counts the number of factors $u$ and $A$, just as described in \Secref{sec:texpand}.
The second step is to compute the moments using (\ref{eq:distz_1d_ansatz}). Comparing the resulting expression with the corresponding moments found using the trajectory expansion
yields a linear system of equations for the coefficients $a_{mn}$. Solving this system  one finds
to order $\ku^2$ \cite{Gus13a}:
\begin{equation}
P(u,A)=\frac{1}{2\pi\sqrt{3}}\left[
1 - \frac{\ku\,A}{1+\st} +
\frac{\ku^2(A^2-3u^2)(1+3\st)}{2(1+\st)^2(1+2\st)}\right]{\rm e}^{-A^2/6-u^2/2}\,.
\label{eq:distuA_1d_final}
\end{equation}
Thus  $P_0$ is obtained not only in the limit $\ku\rightarrow 0$, but also in the limit $\st\rightarrow \infty$
where particle trajectories are only weakly affected by the flow.
But in general $P(u,A)$ is not Gaussian.
The result (\ref{eq:distuA_1d_final}) was computed to second order in $\ku$.  In the body of the distribution
it gives good agreement with results from numerical simulations of the statistical model \cite{Gus13a}. One expects that higher-order terms in $\ku$ yield better approximations to the tails. Yet the theory must
fail far in the tails (large fluctuations) because the trajectory expansion is a small-fluctuation expansion.

As a second example consider the distribution of particle-velocity
gradients $z\equiv\partial_x v$.  It was computed in Ref.~\cite{Gus13a}
\begin{eqnarray}
P(z)&=&\sqrt{\frac{1+\st}{6 \pi }}{\rm e}^{-z^2(1+\st)/6}\bigg[1-\ku\frac{2+\st-2\st^2}{(2+\st)(1+2\st)}z
\\
&&-\frac{\ku}{9}\frac{\st(1+\st)(6+9\st+2\st^2)}{(2+\st)(1+2\st)}z^3\bigg]
\nonumber
\label{eq:distz_1d_final}
\end{eqnarray}
to first order in $\ku$. When $\st=0$, this distribution is identical to $P(A)$, as must
be the case in the advective limit.  The white-noise limit of (\ref{eq:distz_1d_final}) is consistent with the distribution
obtained in Ref.~\cite{Wil03},
at least in the body of the distribution. The algebraic tails $P(z) \sim z^{-2}$
that are due to caustic singularities are not reproduced by (\ref{eq:distz_1d_final}). This is due to the fact that the $z^2$-term in Eq.~(\ref{eq:z1d}) is treated perturbatively.
The tails of the distribution can be obtained using WKB approximations.

\subsubsection{Finite-time averages}
\label{sec:finite_time}
\label{sec:Method_finite_time_averages}
To evaluate finite-time averages is more difficult than calculating steady-state averages because finite-time averages
depend upon the initial conditions $v_0,u(\xdo,0),\partial_xu(\xdo,0),\dots$.
Homogeneity of the fluid-velocity field
implies, as mentioned above, that averages cannot depend upon the initial particle position $x_0$.
As an example consider the finite-time average of $A(x(t),t)$.
At finite times one cannot use the steady-state covariances
(\ref{eq:correlation_uu_1d}) but must explicitly account for the initial values
of the flow velocity and its gradients.
The initial flow configuration, at $t=0$, is
defined by the variables $\uxin(\xdo,0)$.
For $t>0$ the dynamics of $\uxi(t)\equiv\uxin(\xd,t)$ is determined by the Ornstein-Uhlenbeck processes
(\ref{eq:OUak}):
\begin{eqnarray}
\delta  \uxi&=-\uxi\delta t +{\delta W_m}\,.
\label{eq:deltauxi}
\end{eqnarray}
The random increments ${\delta W_m}$ satisfy $\langle {\delta W_m} \rangle = 0$ and $\langle {\delta W_m} {\delta W_n}\rangle = 2 C_{mn} \delta t$.
The coefficients $C_{mn}$ are given in Eq.~(\ref{eq:cmn}).
The solution of the Langevin equation (\ref{eq:deltauxi})  can be written as
 \begin{eqnarray}
 \uxi(t)&=\uxi(0) {\rm e}^{-t}+\underbrace{{\rm e}^{-t}\int_0^t{\rm d}t_1{\rm e}^{t_1}c_m(t_1)}_{\equiv \,\Delta u_{x,m}(t)}\,,
 \label{eq:ouu}
 \end{eqnarray}
where $c_m(t_1)$ is white noise such that $\delta W_m\equiv\int_t^{t+\delta t}{\rm d}t_1c_m(t_1)$.
The correlation function of the second term in Eq. (\ref{eq:ouu}), $\Delta u_{x,m}$, is found to be
 \begin{eqnarray}
 \langle\Delta u_{x,m}(t_1)\Delta u_{x,n}(t_2)|t_1,t_2\rangle=C_{mn}\big({\rm e}^{-|t_1-t_2|}-{\rm e}^{-(t_1+t_2)}\big)\,.
 \label{eq:DuDuFinitet}
 \end{eqnarray}
This average is evaluated at fixed times  $t_1$ and $t_2$.
To compute the average of $A(x(t),t)$ at a finite time one inserts (\ref{eq:ouu})
into the trajectory expansion for $A(x(t),t)$ and takes an ensemble average
over the fluctuating part $\Delta u_{x,m}$.
The result depends on the initial fluid-velocity gradients $\uxi(0)$ and upon the initial particle velocity $v_0$.
At large times the average becomes independent of the initial condition, as it must,
and approaches Eq.~(\ref{eq:Abar_1d}). Alternatively one may average the finite-time expression
over the steady-state distributions of the variables determining the initial conditions.
One obtains once more Eq.~(\ref{eq:Abar_1d}).

\subsubsection{Lyapunov exponent in one spatial dimension}
\label{sec:lyapunov_1d}
In order to calculate the maximal Lyapunov exponent $\lambda_1$ at finite Kubo numbers in one spatial
dimension one {may start} from Eq.~(\ref{eq:lambda_1d_def}), expressing the Lypapunov exponent
in terms of the steady-state average of the particle-velocity gradient $z= \partial_x v$.
The dynamics of $z$ is given by Eq.~(\ref{eq:z1d}), and we compute $\lambda_1$ by expanding
an implicit solution of this equation.
In the dimensionless variables (\ref{eq:dim1}) this implicit solution
reads:
\begin{equation}
z(t)=z_0{\rm e}^{-t/\st}+\int_{0}^t\! {\rm d}t_1\,{\rm e}^{-(t-t_1)/\st}[A(x(t_1),t_1)/\st\!-\!\ku\, z(t_1)^2 ]\,.
\label{eq:implicit_z_solution_1d}
\end{equation}
The first term in the integrand in \Eqnref{eq:implicit_z_solution_1d} depends only upon $A(x(t),t)$ and
can be expanded around the deterministic solution $x^{\rm d}(t)$ of Eq. (\ref{eq:langevin_1d_dimless}) in terms of $\ku$ in the same way as $u(x(t),t)$.
The second term in \eqnref{eq:implicit_z_solution_1d} is proportional to $\ku$,  and we expand it in $\ku$
by substituting the r.h.s. of \eqnref{eq:implicit_z_solution_1d} and iterating.
Using computer automation makes it possible to obtain $\lambda_1$ to high orders in $\ku$. Since the distribution of fluid-velocity gradients is symmetric at given values of $x$ and $t$, only
even orders in $\ku$ contribute to the perturbation expansion.
To fourth order one finds:
\begin{eqnarray}
\label{eq:lyapunov_1d_Ci}
\lambda_1\tau&=&
\!-\!{C_{11}}\ku^2
\!+\!\ku^4\frac{{C_{00}C_{22}}(1\!+\!3\st\!+\!2\st^2)\!+\!{C_{11}^2}(1\!-\!\st\!-\!18\st^2\!-\!30\st^3\!-\!10\st^4)}{2(1+\st)^3}\,.
\end{eqnarray}
The coefficients $C_{mm}$, Eq.~(\ref{eq:cmn}),  are given by the variances of the velocity field $u(0,0)$ and
its gradients $\partial^n_xu(0,0)$ evaluated at $x=0$ and $t=0$. From \Eqnref{eq:cmn}
it follows that
\begin{equation}
\label{eq:c2i}
C_{mm}\equiv\langle(\partial^m_xu(x_0,0))^2\rangle=(2m+1)!! \,.
\end{equation}
The coefficients $C_{mm}$ encode
information about the functional form of $u(x(t),t)$. The higher the value of $m$ is, the larger
distances from the deterministic trajectory are taken into account, representing properties of the correlation function $C(X,T)$ at larger spatial scales.
 Higher orders in $\ku$ give coefficients $C_{mm}$ with higher
values of $m$.
These observations  show that the trajectory expansion not only takes into account finite time correlations, it also incorporates the non-linear dynamics on spatial scales below and above $\eta$.
This is in contrast to the white-noise model where the perturbation
expansion for the Lyapunov exponent depends only upon $C_{11}$. This can be seen by multiplying the Lyapunov exponent with $\st$ and taking the limit $\ku\to 0$, $\st\to\infty$ such that $\ku^2\st$ remains constant.
\begin{figure}[tp]
\includegraphics[height=5.5cm,clip]{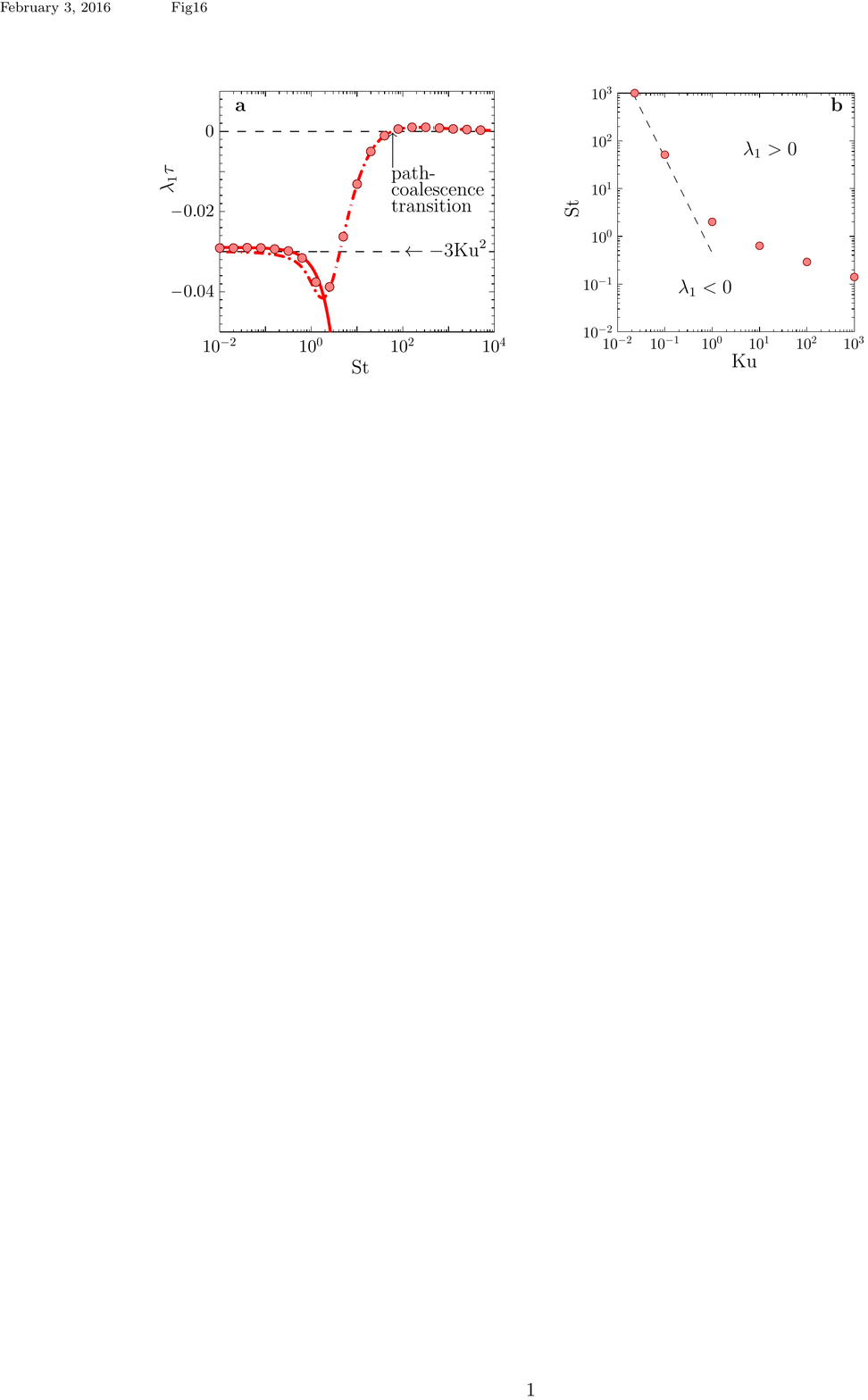}
\caption{\label{fig:lyapunov_1d} {\bf a} Maximal Lyapunov exponent $\lambda_1$
for $\ku=0.1$ as a function of $\st$ in one spatial dimension.
Symbols (red filled $\circ$) show results of numerical simulations of the statistical model. Also shown is the white-noise result (\ref{eq:lyapunov_1d_WN}),
dash-dotted line, as well as perturbation theory in $\ku$, Eq. (\ref{eq:lyapunov_1d_final}), extended to order $\ku^6$, solid line.  
Arrows indicate the location of the path-coalescence transition, and the limiting value of $\lambda_1$ as $\st\to 0$ [Eq.~(\ref{eq:lyapunov_1d_final})].
{\bf b} Shows location of the path-coalescence transition in one spatial dimension. Results of numerical
simulations of the statistical model are shown as red filled $\circ$, the white-noise prediction
$\varepsilon^2 = 3 \ku^2 \st = 1.77$ is shown as a dashed line.
}
\end{figure}

Inserting (\ref{eq:c2i}) into  (\ref{eq:lyapunov_1d_Ci}) yields:
\begin{eqnarray}
\label{eq:lyapunov_1d_final}
\lambda_1\tau&=&
-3\ku^2
+3\ku^4\frac{4+6\st-22\st^2-45\st^3-15\st^4}{(1+\st)^3}\,.
\end{eqnarray}
We note that the $\ku^2$-term in this expression is identical to the corresponding $\ku^2$-asymptote of the telegraph model~\cite{Fal07}.
But the $\ku^4$-terms differ between the two models.

The trajectory expansion for $\lambda_1$ extended to order $\ku^6$ is
compared with results of numerical simulations of the one-dimensional statistical
model in Fig.~\ref{fig:lyapunov_1d}{\bf a}. The Figure shows that the expansion around deterministic trajectories
works well up to $\st\sim 1$ for $\ku=0.1$, this is where caustics become frequent.
But we observe that the result to order $\ku^6$ is not precise enough to capture the path-coalescence transition where $\lambda_1$ changes sign.  An example where the  trajectory expansion describes the path-coalescence transition is given in Ref.~\cite{Gus13b}, describing the path-coalescence transition of particles advected in two-dimensional compressible flows.  In this case the series for $\lambda_1$ can be computed to order $\ku^8$. Resummation yields the precise location of the path-coalescence transition in this case (Fig.~2{\bf b} in Ref.~\cite{Gus13b}).

Fig.~\ref{fig:lyapunov_1d}{\bf b}
shows the location of the path-coalescence transition in the $\ku$-$\st$-plane.
At small $\st$ and $\ku$ the maximal Lyapunov exponent is negative (path coalescence). But when
inertial effects are sufficiently large then $\lambda_1$ turns positive \cite{Meh09}.
The one-dimensional system has interesting single-particle dynamics in the upper {right}-hand
corner of the phase-diagram Fig.~\ref{fig:lyapunov_1d}{\bf b}. When $\st \gg 1$ and {$\ku^2\gg \st$} the system exhibits
non-Gaussian velocity fluctuations and anomalous diffusion. These results were obtained by mapping the problem onto a quantum-mechanical problem
with a staggered-ladder spectrum and non-linear annihilation and creation operators \cite{Arv06,Bez06}. We are aware of only two other
physical {systems} that exhibit ladder spectra: the quantum harmonic oscillator and the Zeeman-splitting Hamiltonian.

The expansion (\ref{eq:lyapunov_1d_final}) was derived for general values of $\st$. It is necessary to discuss possible reasons that cause the perturbation theory to fail at large values of $\st$.
There are at least two ways in which the series expansion in $\ku$ may fail
at large Stokes numbers.
First, it is likely that
the series expansion (\ref{eq:lyapunov_1d_final}) is asymptotically divergent, just as
the white-noise expansion. In order to obtain accurate results at
larger values of $\st$ it is then necessary to resum the series. This requires high-order coefficients. The resummation
of the white-noise series shown in Fig.~\ref{fig:con} uses the first $24$ coefficients.
Second, it may be necessary to supplement the expansion with non-analytical terms, as explained
in Section \ref{sec:pt}. In one spatial dimension there is no additional contribution to the perturbation expansion for $\lambda_1$ in the white-noise limit. But this does not rule out that such a term
may exist at finite Kubo numbers.

Fig.~\ref{fig:lyapunov_1d}{\bf a} also shows the white-noise approximation (\ref{eq:lyapunov_1d_WN}).
We note that the lowest-order contribution to $\lambda_1$ in Eq.~(\ref{eq:lyapunov_1d_final}),
of order $\ku^2$, is independent of $\st$.
This explains why the white-noise theory works comparatively well.
Since the $\ku^2$-term is independent of $\st$, the white-noise approximation,
exact as $\st\to\infty$, yields the correct small-$\st$ behavior to lowest order in $\ku$, namely $\lambda_1\tau = -3\ku^2$. It is interesting to note that
the coloured-noise approximation (Section \ref{sec:othermodels}) yields
the incorrect result $\lambda_1\tau = -3\ku^2 \st/(1+\st)$ to {second} order in $\ku$.
That the coloured-noise approximation predicts a wrong factor $\st/(1+\st)$ is
explained by the fact that this approximation neglects preferential effects, while keeping
time correlations finite. We infer that preferential effects
must give a contribution  containing a factor $1/(1+\st)$:
\begin{equation}
\lambda_1\tau =  -3\ku^2  = -3 \ku^2\, \Big(\!\!\underbrace{\frac{1}{1+\st}}_{\stackrel{\mathrm{\scriptstyle preferential}}{\mathrm{sampling}}} + \underbrace{\frac{\st}{1+\st}}_{\stackrel{\mathrm{\scriptstyle time}}{\mathrm{correlations}}}\!\!\Big)\,.
\end{equation}
This equation exhibits, by way of example, which contributions are due to preferential sampling, and which
contributions are due to finite time correlations (and appear in models that take such correlations
into account but neglect preferential effects).

At small Stokes numbers and finite Kubo numbers higher orders in $\ku$ matter.
Therefore (\ref{eq:lyapunov_1d_final}) yields a better description in this regime than the white-noise result (Fig.~\ref{fig:lyapunov_1d}{\bf a}).

\subsubsection{Lyapunov exponents in $d$ dimensions.}
\label{sec:ddim}
The one-dimensional model discussed in the previous Section is special:
a one-dimensional velocity
field can always be written as the gradient of a potential, Eq.~(\ref{eq:velocity_field_1d}).
The one-dimensional velocity field
is thus compressible. As a result the Lyapunov exponent $\lambda_1$ can be negative.
In incompressible flows, by contrast, the maximal Lyapunov exponent
must always be positive, but clustering can nevertheless be substantial.
In this Section we explain how to calculate the Lyapunov exponents in general $d$-dimensional flows.
The calculation is analogous
to the one-dimensional case. The infinitesimal distance $\mathscr{R}_t$,
area $\mathscr{A}_t$, and volume $\mathscr{V}_t$ in Eq.~(\ref{eq: 2.4})
are computed by following the motion of $d$ particles close to a test particle at $\ve x(t)$.
The small separations between the $d$ particles and the test particle are denoted by
$\ve X_\mu$, $\mu=1,\dots,d$. We use Greek indices to label particles and Roman indices
to label spatial vector and matrix components.
The separations $\ve X_\mu$ and relative velocities $\ve V_\mu$ follow the linearised equation of motion:
\begin{eqnarray}
\label{eq:lind}
\frac{{\rm d}\ve X_\mu}{{\rm d}t} &=& \ku \ve V_\mu\,,\quad
\frac{{\rm d}\ve V_\mu}{{\rm d}t} = \frac{1}{\st}(\ma A({\ve x}(t),t)\ve X_\mu-\ve V_\mu)
\end{eqnarray}
in the dimensionless variables (\ref{eq:dim1}).
The dynamics of $\ve X_\mu$ determines the Lyapunov exponents.
To compute the exponents it is convenient to set up
a time-dependent coordinate system $\hat{\ve  n}_{\mu}$ such that $\hat {\ve n}_{1}$ points in the direction of $\ve X_1$, $\hat{\ve  n}_{2}$ is orthonormal to $\hat {\ve n}_{1}$,
 $\hat {\ve n}_{1}$ and $\hat{\ve  n}_{2}$ span the plane formed by $\ve X_1$ and $\ve X_2$, and so on.
 One writes
 \begin{equation}
 \ve X_{\mu}=\sum_{\nu=1}^d W_{\mu\nu}\hat {\ve n}_{\nu}\,.
 \label{eq:Xmu_decomposition}
 \end{equation}
 The tensor $\ma W$ with elements $W_{\mu\nu}$ is lower diagonal.
 Multiplying the equation of motion (\ref{eq:lind}) for $\ve X_\mu$
with $\ma W^{-1}$, linearising $\ve V_\mu\approx \ma Z\ve X_\mu$, and using Eq.~\eqnref{eq:Xmu_decomposition} one finds an equation for $\hat {\ve n}_\mu$:
 \begin{eqnarray}
 \frac{{\rm d}\hat{\ve n}_{\mu}}{{\rm d}t}
&=&\ku\Big(\ma Z\hat{\ve n}_{\mu}
-\sum_{\nu=1}^{\mu}(2-\delta_{\mu\nu})\,(\hat{\ve n}_{\mu}\cdot \ma Z\hat{\ve n}_{\nu})\hat{\ve n}_{\nu}\Big)\,.
 \label{eq:langevin_Zmatrix_nhat}
 \end{eqnarray}
The Lyapunov exponents
are given by the steady-state average of $Z_{\mu\nu}'\equiv {\hat {\ve n}}_{\mu}\cdot \ma Z{\hat{\ve  n}}_{\mu}$.
This can be seen as follows.
First, $\hat{\ve n}_{1}$ is the direction of the separation vector between two particles.
The maximal Lyapunov exponent $\lambda_{1}$ is thus given by the average of
\begin{equation}
\frac{\rm d}{{\rm d}t} \log|\ve X_1|=\ku\,\hat{\ve n}_{1}\cdot\ma Z\hat{\ve n}_{1}\,.
\end{equation}
Second, the vector $\ve n_{2}$ is orthogonal to $\ve n_{1}$. It determines  the direction between two particles
projected onto the subspace orthogonal to  $\ve n_{1}$.
The maximal Lyapunov exponent in this subspace is $\lambda_{2}$, it is given by the average of
$\hat{\ve n}_{2}\cdot\ma Z\hat{\ve n}_{2}$, and so forth. Note that the unit vectors $\hat{\ve n}_\mu $
are orthogonal to all $\hat{\ve n}_\nu$ for  $\nu<\mu$ by construction, $\ma W$ is chosen
to be a lower-diagonal tensor. In short
 the Lyapunov exponents are given by
  \begin{equation}
\label{eq:lmu}
  \lambda_\mu = \ku\, \langle {\hat {\ve n}}_{\mu}\cdot \ma Z{\hat{\ve  n}}_{\mu}\rangle_\infty\,.
  \end{equation}
The steady-state expectation values (\ref{eq:lmu}) are evaluated using the techniques
described in Sections \ref{sec:texpand} and \ref{ssec:Method_steady_state_averages}.
Writing $ \delta \ve x(t)\equiv \ve x(t)-\xxd$ one has
 \begin{eqnarray}
 \label{eq:x_z_implicit}
  \delta \ve x(t) &\equiv&\ve x(t)-\xxd =\frac{\ku}{\st}\int_{0}^{t}\!\! {\rm d}t_1\int_{0}^{t_1}\!\!\!{\rm d}t_2\,{\rm e}^{-(t_1-t_2)/\st}\ve u(\ve x(t_2),t_2)\,
\end{eqnarray}
with $\xxd =\ve x_0+\ku\,\st(1-{\rm e}^{-t/\st}) \ve v_0$.
Expanding $\ve u(\ve x(t),t)$ around $\xxd$ for $\ve v_0=0$ yields:
\begin{eqnarray}
\label{eq:uexpansion_general_dim}
u_i(\ve x(t),t)&=&u_i(\ve x_0,t)+\sum_j\frac{\partial u_i}{\partial x_j}(\ve x_0,t)\,\delta x_j(t)\\
&+&\frac{1}{2}\sum_{jk}\frac{\partial^2 u_i}{\partial x_j\partial x_k}(\ve x_0,t)\,\delta x_j(t)\,\delta x_k(t)+\dots\,,
\nonumber
\end{eqnarray}
analogous to \eqnref{eq:uexpansion}.
Next one inserts $\delta\ve x$ from (\ref{eq:x_z_implicit})
into \eqnref{eq:uexpansion_general_dim}.
Iterating and collecting terms in powers of $\ku$ gives the trajectory expansion
of $\ve u(\ve x(t),t)$ analogous to the one-dimensional case.
As in the one-dimensional case the parameter $\ku$ is used as a book-keeping parameter to group terms in powers of $\ku$.
A corresponding expansion is obtained for $\ma A(\ve x(t),t)$.
In order to compute the Lyapunov exponents from (\ref{eq:lmu}) one must also expand implicit solutions of
Eqs.~(\ref{eq:langevin_Zmatrix_nhat}) and (\ref{eq:langevin_Zmatrix_0}).
In the dimensionless variables (\ref{eq:dim1}) Eq.~(\ref{eq:langevin_Zmatrix_0}) reads:
\begin{eqnarray}
\dot{ \ma Z}
&=&\frac{1}{\st}(\ma A-\ma Z)-\ku\, {\ma Z}^2\,.
\label{eq:langevin_Zmatrix}
\end{eqnarray}
We consider the implicit solution of (\ref{eq:langevin_Zmatrix}) that is of the same form as (\ref{eq:implicit_z_solution_1d}), but matrix-valued:
\begin{align}
&\ma Z(t)= \ma Z_0 \,{\rm e}^{-t/\st}+\int_{0}^t\!\!\!{\rm d}t_1\, {\rm e}^{(t_1-t)/\st}\big [ \frac{1}{\st}\ma A(\ve x(t_1),t_1)-\ku\, \ma Z^2(t_1)
\big]\,.
\label{eq:impmaZ}
\end{align}
Eq.~(\ref{eq:impmaZ}) is expanded in $-\ku\,\ma Z^2$.
The corresponding implicit solution of Eq.~(\ref{eq:langevin_Zmatrix_nhat})
\begin{align}
 \hat{\ve  n}_\mu(t)&\!=\!\hat{\ve  n}_{\mu,0}+\ku\int_{0}^{t}\!\!\!{\rm d}t_1
\Big[ \ma Z(t_1)\hat{\ve  n}_\mu(t_1)
\!-\!\sum_{\nu=1}^{\mu}(2-\delta_{\mu\nu})\,
\big(\hat{\ve n}_{\mu}(t_1)\cdot \ma Z(t_1)\hat{\ve n}_{\nu}(t_1)\big)
\hat{\ve n}_{\nu}(t_1)\Big]
\label{eq:impn}
\end{align}
is expanded in deviations $\delta \hat{\ve n}_\mu$ from the initial orientation, $\delta \hat{\ve n}_\mu\equiv \hat{\ve n}_\mu(t)-\hat{\ve n}_{\mu,0}$.
One obtains a perturbation expansion in powers of $\ku$,
a sum of integrals of multi-time correlation functions of the
flow-velocity field and its gradients,
\begin{equation}
\label{eq:CCij}
\CC{_{ij}}{_{i_1i_2\dots i_nj_1,j_2\dots j_n}}(\ve X,T)
\equiv \left\langle\frac{\partial^{n}u_{i}(\ve r,t)}{\partial x_{i_1}\cdots\partial x_{i_n}}\frac{\partial^{n}u_{j}(\ve r + \ve X,t+T)}{\partial x_{j_1}\cdots\partial x_{j_n}}\right
\rangle_\infty\,.
\end{equation}
These correlation functions are determined by Eq.~(\ref{eq:velocity_field_C}). Since the fluid-velocity field is
homogeneous, isotropic, and time reversible they are functions of $\ve X$ and $|T|$ only.
In order to compute the Lyapunov exponents it is sufficient to consider
the limit $\mathscr{R}=|\ve X|\to 0$:
\begin{subequations}
\label{eq:allcc}
\begin{equation}
\mbox{}\hspace*{-6.3cm}\CC{_{ij}}{}(0,T)={\cal N}_d^2C_{00}{\rm e}^{-|T|}(d-1+\beta^2)\delta_{ij}\,,\\[-0.8cm]
\end{equation}
\begin{equation}
\label{eq:CC4_R0}
\CC{_{ij}}{_{kl}}(0,T)
=\frac{{\cal N}_d^2C_{11}}{3}{\rm e}^{-|T|}\left[
(d+1+\beta^2)\delta_{ij}\delta_{kl}
+(\beta^2-1)(\delta_{ik}\delta_{jl}+\delta_{il}\delta_{jk}) \right]\,,\\[-0.8cm]
\end{equation}
\begin{eqnarray}
\mbox{}\hspace*{-1.3cm}\CC{_{ij}}{_{klmn}}(0,T)
&=&\frac{{\cal N}_d^2C_{22}}{15}{\rm e}^{-|T|}\Big[
(d+3+\beta^2)\delta_{ij}[\delta_{lm}\delta_{kn}+\delta_{km}\delta_{ln}+\delta_{kl}\delta_{mn}]\nn\\
&&+(\beta^2-1)\delta_{ik}(\delta_{jl}\delta_{mn}+\delta_{jm}\delta_{ln}+\delta_{jn}\delta_{lm})\nn\\
&&\hspace*{0.5mm}+(\beta^2-1)\delta_{il}(\delta_{jk}\delta_{mn}+\delta_{jm}\delta_{kn}+\delta_{jn}\delta_{km})\nn\\
&&+(\beta^2-1)\delta_{im}(\delta_{jk}\delta_{ln}+\delta_{jl}\delta_{kn}+\delta_{jn}\delta_{kl})\nn\\
&&+(\beta^2-1)\delta_{in}(\delta_{jk}\delta_{lm}+\delta_{jl}\delta_{km}+\delta_{jm}\delta_{kl})
\Big]\,,
\end{eqnarray}
\end{subequations}
and so forth. The normalisation ${\cal N}_d$ is given in Eq.~(\ref{eq:normalisation}).
Expanding $\langle \ve n_\mu \cdot \ma Z \ve n_\mu\rangle_\infty$
and averaging using the correlations \eqnref{eq:allcc} yields
the Lyapunov exponents.  To order $\ku^2$ one finds:
\begin{equation}
\lambda_\mu\tau=\ku^2\frac{d(d+1-2\mu)+\beta^2(d-4\mu)}{d(d-1+\beta^2)}\,.
\end{equation}
This result is independent of $\st$ and thus equal to Eq.~(\ref{eq:ladv}),
an equation that was obtained in the advective limit $\st \rightarrow 0$.

Higher-order expressions for $\lambda_\mu\tau$ are lengthy.
Here we quote only the result for incompressible fluid-velocity fields
to order $\ku^4$:
\begin{eqnarray}
\label{eq:lyapunov_spectrum_Ku4}
\lambda_\mu\tau&=&\ku^2\frac{d(1+d-2\mu)}{d(d-1)}\\
&&\hspace*{-8mm}+\ku^4\frac{1}{d^2(d-1)^2}\frac{1}{(1+\st)^3} \big\{-d^4(2+8\st+12\st^2+9\st^3+3\st^4)\nn\\
&&\hspace*{-8mm} +d^3(1\!+\!\st)(-5\!-\!14\st\!-\!12\st^2\!-\!6\st^3\!+\!2\mu(2\!+\!10\st\!+\!14\st^2\!+\!7\st^3)) \nn\\
&&\hspace*{-8mm}-d^2(1\!+\!5\st\!+\!14\st^2\!+\!21\st^3\!+\!7\st^4\!+\!12\mu^2\st(1\!+\!\st)^3
\!-\!2\mu(3\!+\!13\st\!+\!19\st^2\!+\!12\st^3\!+\!4\st^4)) \nn\\
&&\hspace*{-8mm}{+2d(1\!+\!3\st\!-\!6\st^3\!-\!2\st^4\!+\!\mu(-2\!-\!6\st\!-\!\st^2\!+\!9\st^3\!+\!3\st^4))}
{+4\mu\st^2(1\!+\!3\st\!+\!\st^2)}\nn\big\}\,.
\end{eqnarray}
 This result generalises the white-noise result (\ref{eq:lwn}) to finite Kubo numbers.
Taking the  white-noise limit of Eq.~(\ref{eq:lyapunov_spectrum_Ku4}) results in Eq.~(\ref{eq:lwn}), that was derived using diffusion approximations.

It follows from Eq.~(\ref{eq:lyapunov_spectrum_Ku4}) that the
middle exponent $\lambda_2\rightarrow 0$ in three spatial dimensions as $\st\to 0$.
This is a consequence of the time-reversal symmetry
of the statistical model.

Inserting series expansions such as Eq.~(\ref{eq:lyapunov_spectrum_Ku4})
into the Kaplan-Yorke formula (\ref{eq:dL1}) yields the Lyapunov dimension. Eq.~(\ref{eq:DeltaLd}), for example, is obtained by extending the series (\ref{eq:lyapunov_spectrum_Ku4}) to order $\ku^6$.

\subsubsection{Small-$\st$ expansion}
The trajectory expansion described in Sections \ref{sec:texpand} to \ref{sec:finite_time} allows to quantify clustering
in terms of the steady-state average $\langle\tr \ma Z\rangle_\infty$. This perturbative method corresponds to an expansion
in the Kubo number $\ku$. It is instructive to contrast this expansion with a systematic expansion in $\st$. This shows how it comes about that fractal clustering
at very small Stokes numbers is determined by instantaneous fluid-velocity gradients, their history does not  matter in this limit.

The perturbation series in $\st$ is obtained by systematically expanding
$\ma A(\ve x(t),t)$ and $\ma Z(t)$ as series in $\st$:
\begin{equation}
\label{eq:ZAexp}
\ma Z=\sum_{i=0}^\infty \ma Z_{(i)}\st^i\,,\quad\mbox{and}\quad \ma A=\sum_{i}\ma A_{(i)}\st^i\,,\quad \mbox{with}\quad
\ma A_{(i)}=\frac{1}{i!}\left.\frac{\partial^i\ma A}{\partial\st^i}\right|_{\st=0}.
\end{equation}
Note that $\ma A(\ve x(t),t)$ depends upon the Stokes number because  $\ve x(t)$ depends upon $\st$.
The coefficients $\ma Z_{(i)}$ are determined by inserting the expansion (\ref{eq:ZAexp}) into  the equation of motion (\ref{eq:langevin_Zmatrix_0}) for the matrix of
particle-velocity gradients.  Collecting powers  of $\st$ yields:
\begin{eqnarray}
\label{eq:ZsmallStExpansion}
\ma Z_{(0)}&=&\ma A_{(0)}\,,\quad
\ma Z_{(1)}=-\frac{{\rm d}\ma A_{(0)}}{{\rm d}t} +\ma A_{(1)}-\ku \ma A_{(0)}^2\,,\\
\ma Z_{(2)}
&=&\frac{{\rm d}^2\ma A_{(0)}}{{\rm d}t^2}-\frac{{\rm d \ma A_{(1)}}}{{\rm d}t}+\ma A_{(2)}
+\ku\Big(2\frac{{\rm d}\ma A_{(0)}^2}{{\rm d}t}-\ma A_{(0)}\ma A_{(1)}+\ma A_{(1)}\ma A_{(0)}\Big)+2\ku^2 \ma A_{(0)}^3\,.\nn
\end{eqnarray}
The next step is to take the trace and average. The calculation simplifies since the fluid-velocity field is assumed to
be homogeneous. The final result is
\begin{eqnarray}
\label{eq:nablav}
\langle{\nabla\cdot\ve v}\rangle_\infty=\langle{\tr\,\ma Z}\rangle_\infty=-\ku\,\st^2\frac{\partial}{\partial\st}\langle {\tr\,\ma A^2}\rangle_\infty\Big|_{\st=0}+\Ordo(\st^3)\,.
\end{eqnarray}
All quantities are expressed in the dimensionless variables (\ref{eq:dim1}). Eq.~(\ref{eq:nablav}) 
is identical to \Eqnref{eq:nablav0} discussed in Section \ref{sec:expadv}.  
It quantifies how preferential sampling of straining regions contributes to spatial clustering.
The contribution is small because it scales as $\st^2$ (Eq.~(\ref{eq:nablav}) was derived for small values of $\st$). 
It turns out that the $\st^3$-terms in this expansion contain averages of combinations of $\ma A$ and time derivatives of $\ma A$. 
This indicates that clustering is not an instantaneous effect in general, it depends on the history of fluid-velocity gradients experienced by the particles.

\section{Conclusions}
\label{sec:conc}
We have reviewed how statistical-model calculations allow us to understand
the mechanisms that cause heavy small particles in incompressible turbulent flows to form spatial patterns. The model approximates the turbulent fluctuations by Gaussian random functions with appropriate spatial and temporal correlations, it is amenable to mathematical analysis, and accounts for the fundamental mechanisms at play. The analysis requires advanced methods of Mathematical Physics (high-order perturbation theory, resummation of divergent series, and multi-dimensional WKB approximations) and yields, in certain cases, asymptotically exact results that explain the mechanisms at work.
In this review we have shown that the statistical-model analysis makes it possible to qualitatively
understand and to quantitatively describe small-scale clustering observed in numerical simulations  of heavy small particles in turbulence.

Apart from approximating the turbulent velocity fluctuations, the statistical model assumes a simplified equation of motion for the particles that applies to dilute suspensions of heavy small spherical particles, and disregards possible direct interactions such as electrostatic forces. In the simplest case it is assumed that all particles are identical. In this case
the statistical model is determined by three dimensionless parameters. The Stokes number $\st$ quantifies the importance of inertia. When $\st=0$ then the particles are advected by the flow, inertia plays no role. At large Stokes numbers, by contrast, inertial effects are important.
The Kubo number $\ku$ is a dimensionless measure of
the correlation time of the flow. When $\ku$ is small or $\st$ is large the dynamics can be
computed by means of diffusion approximations as explained in this review. The third parameter,
 ${\rm F}$, measures the importance of settling due to gravity.

The statistical model calculations in combination with results of numerical simulations of
particles in turbulence give rise to the following picture. Small-scale clustering is determined by the multiplicative process of random contractions and expansions of a small cloud of particles (multiplicative amplification). The strength of the resulting fractal clustering is
quantified by the spatial Lyapunov exponents of the particle dynamics.
In general, the Lyapunov exponents are determined by the history of fluid-velocity gradients
that the particles encountered in the past. The particle paths may
be biased due to preferential sampling. Any theory of fractal clustering must take this into account: what matters is the time-series of fluid-velocity gradients that the particle experienced in the past as it moved through the flow.

In certain limits the picture simplifies. For very small Stokes numbers
the particle-velocity gradients are determined entirely by the instantaneous
fluid velocity gradients. In this limit clustering occurs in straining regions of the flow.
Particles gather in the straining regions because their inertia allows them to centrifuge from vortical regions. This mechanism (Maxey's centrifuge) is an example of preferential sampling.
In the white-noise limit, by contrast, there is no preferential sampling. In this limit small-scale clustering bears no relation to the instantaneous fluid-velocity gradients at the particle positions, it is entirely determined by the history of fluid-velocity gradients.

At finite Stokes and Kubo numbers preferential sampling indirectly affects fractal clustering by biasing the history of fluid-velocity gradients seen by the particles. Recent results on the dynamics of particle separations in turbulence~\cite{Bra15a} confirm this picture.
We conclude that the statistical model qualitatively explains the mechanisms that cause small-scale clustering and preferential sampling (on scales of the order of the Kolmogorov length), resolving the dichotomy mentioned in the Introduction.

In some cases the statistical model even makes quantitative predictions. One example is seen in Fig.~\ref{fig:compare},
illustrating how the degree of small-scale clustering depends on the Stokes number. The agreement is remarkable
considering that the statistical model is a highly simplified model of the turbulent velocity fluctuations. One fundamental difference is that turbulence breaks time-reversal invariance while the statistical model does not. This does not appear
to have much influence on small-scale clustering, and we infer that the dynamics of heavy small spherical particles in the dissipative range of turbulence is
in the first place determined by the kinematic equation (\ref{eq:eom}), and is less sensitive to 
the exact nature of the turbulent driving at small scales.
The fact that the Gaussian statistical model explains the small-scale spatial clustering observed in fully developed turbulence indicates that the clustering mechanisms are universal and apply more generally in incompressible unsteady mixing flows.
We note, however, that the tumbling of non-spherical particles in turbulence at small Stokes numbers \cite{Par12} is quite sensitive to the breaking of time-reversal invariance~\cite{Gus14b}.

The results described in this review pertain to a particular limit of the inertial-particle problem, given either by Eq.~(\ref{eq:stokes}) or by (\ref{eq:eom}), depending on whether gravitational
settling matters or not. Most results described in this review were obtained for ${\rm F}=0$,
but the effect of gravitational settling was also discussed. Settling may either decrease
or increase the strength of fractal clustering, depending on the value of the Stokes number.

We did not consider buoyancy forces or \lq added-mass\rq~forces due to acceleration of surrounding fluid
in this review. These forces could be treated using the methods described in this review, allowing
to investigate the dynamics of light particles  in turbulent flows \cite{Vol08,Cal08,Cal08b,Cal12,Pra12}.
The added-mass force is an inertial effect, due to unsteady fluid inertia.
It remains an open problem to find a particle equation of motion that
consistently takes into account fluid-inertia effects in turbulence \cite{Candelier2016}.

As far as heavy-particle dynamics at finite Kubo numbers is concerned a number of important problems remain to be solved.
First, other measures of particle clustering need to be analysed. One important question is to compute the correlation
dimension at finite Kubo and Stokes numbers.
Up to now analytical model calculations of the correlation dimension have only been performed
at small Stokes numbers and in the white-noise limit.
A possible approach is suggested in Refs.~\cite{Zai07,Bra15a,Zaichik}. It would be of interest to use the trajectory expansion to analytically evaluate the \lq drift velocity of particle separations\rq{} entering the constitutive equation for the pair correlation function in this approach~\cite{Zai07,Bra15a,Zaichik}.
Second, an important open problem concerning the white-noise
perturbation series for the Lyapuonv exponents and the correlation dimension is this:
we have not yet found a general scheme to compute possible
non-analytical contributions to the perturbation series.
These non-analytic contributions are believed to be closely related to the formation of caustics.
Third, a natural way to characterise the importance of preferential sampling
are finite-time Lyapunov exponents. Their fluctuations relate different fractal dimensions. It is possible to use
the trajectory expansion described in Section \ref{sec:Method} to compute finite-time Lyapunov exponents.
This may make it possible to compute other fractal dimensions.
Fourth, in this review we have considered a statistical model that is homogeneous and isotropic. In many applications
one or both symmetries may be broken. It would be of interest to formulate a statistical model that
describes the dynamics of heavy particles in stratified flows.
Fifth, in some applications (grain dynamics in circumstellar
accretion disks for example \cite{Wil08}) molecular diffusion is important for very small particles, smoothing out clustering
on the smallest spatial scales.
The effect of molecular diffusion has been studied by a number of authors (see for example Ref.~\cite{Bal01}),
but we have not reviewed these results in this paper.

The list of interesting problems that can be attacked with the methods described in this review does not end here: it is important to analyse the dynamics of
non-spherical particles \cite{Par12,Gus14b,Byr15}, of active particles \cite{Cen13,Zha14,Berglund2016}, of spherical particles with different Stokes numbers, and to compute fluctuations of relative velocities between particles in turbulent aerosols. There is recent progress concerning these problems, but many open questions remain.  We expect that the methods reviewed here will make further progress possible.

 {\em Acknowledgements}.  
We thank J. Meibohm for his comments on the manuscript.
This work was supported by Vetenskapsr\aa{}det [grant number 2013-3992], Formas [grant number 2013-3992], by the G\"oran Gustafsson Foundation for Research in Natural Sciences and Medicine, and by the grant {\em Bottlenecks for particle growth in turbulent aerosols} from the Knut and Alice Wallenberg Foundation, grant number 2014.0048.  BM is grateful to NORDITA for funding a four-week stay at their Institute in Stockholm where parts of this review were written.
KG acknowledges funding from the European Research Council under the European Union's Seventh Framework Programme, ERC Grant Agreement No 339032 during his stay at the University of Rome, Tor Vergata. The numerical computations were performed using resources provided by C3SE and SNIC.

\vfill\eject

\end{document}